\documentclass{aa}
\usepackage{ifthen}
\usepackage{graphicx}
\usepackage{amssymb}
\usepackage{aalongtable}
\usepackage{txfonts}

\begin{document}

\newcommand{\water}{H$_2$O}
\newcommand{\micron}{$\mu$m}
\newcommand{\kms}{km\,s$^{-1}$}
\newcommand{\recta}[4]{$\alpha=#1^{\rm{h}}\,#2^{\rm{m}}\,#3\fs#4$}
\newcommand{\rectabf}[4]{$\mathbf{\alpha=#1^{\rm{h}}\,#2^{\rm{m}}\,#3\fs#4}$}
\newcommand{\dec}[3]{$\delta=#1\degr\,#2\arcmin\,#3\arcsec$}
\newcommand{\decli}[4]{$\delta=#1\degr\,#2\arcmin\,#3\farcs#4$}
\newcommand{\declibf}[4]{$\mathbf{\delta=#1\degr\,#2\arcmin\,#3\farcs#4}$}
\newcommand{\powerten}[1]{$10^{#1}$}
\newcommand{\vc}{$v_{\rm{c}}$}
\newcommand{\ve}{$v_{\rm{e}}$}
\newcommand{\vch}{$v_{\rm{c}}^{\rm{H_2O}}$}
\newcommand{\veh}{$v_{\rm{e}}^{\rm{H_2O}}$}
\newcommand{\vp}{$v_{\rm{p}}$}
\newcommand{\vcoh}{$v_{\rm{c}}^{\rm{OH}}$}
\newcommand{\vchzweio}{$v_{\rm{c}}^{\rm{H}_2\rm{O}}$}
\newcommand{\dvoh}{$\Delta\rm{v}^{\rm{OH}}$}
\newcommand{\dvhzweio}{$\Delta\rm{v}^{\rm{H}_2\rm{O}}$}
\newcommand{\Sp}{$S_{\rm{p}}$}
\newcommand{\SI}{$S_{\rm{I}}$}
\newcommand{\smax}{$S_{\rm{max}}$}
\newcommand{\sir}{$S_{25}$}
\newcommand{\Lhzweio}{${\cal L}^{\rm{H_2O}}$}
\newcommand{\Lsun}{$L_{\sun}$}
\newcommand{\Myr}{$M_{\sun}$\,yr$^{-1}$}
\newcommand{\mdot}{$\dot{\rm{M}}$}
\newcommand{\Teff}{$T_{\mathrm{eff}}$}
\newcommand{\gm}{$\;\;\;\;\;$}
\newcommand{\vgm}{$\;\;\;\;\;\;\;$}

\newcommand{\pad}{.\hskip-2pt$^\circ$}
\newcommand{\pam}{.\hskip-2pt$^\prime$}
\newcommand{\pas}{.\hskip-2pt$^{\prime\prime}$}
\newcommand{\gsim}{\;\lower.6ex\hbox{$\sim$}\kern-7.75pt\raise.65ex\hbox{$>$}\;}

\newcommand{\lsim}{\;\lower.6ex\hbox{$\sim$}\kern-7.75pt\raise.65ex\hbox{$<$}\;}
\newcommand{\rpx}{$\,\,$}
\newcommand{\spx}{$\:\:$}
\newcommand{\gpx}{\spx\spx}
\newcommand{\hpx}{\npx\spx}
\newcommand{\ipx}{\npx\npx}
\newcommand{\jpx}{\npx\opx}
\newcommand{\mpx}{$\;\;\:$}
\newcommand{\npx}{$\;\;$}

\newcommand{\sevendash}{---------------------}
\newcommand{\fivedash}{---------------}
\newcommand{\threedash}{---------}
\newcommand{\twodash}{------}

\title {Water vapour masers in long-period variable stars } \subtitle{I.
RX Boo and SV Peg}

\titlerunning{Water vapour masers in long-period variable stars. I}

\author {A.~Winnberg \inst{1}
        \and D.~Engels \inst{2}
        \and J.~Brand \inst{3}
        \and L.~Baldacci \inst{3,}  \inst{4}
        \and C.M.~Walmsley \inst{5}}

\offprints{A.~Winnberg}

\institute{Onsala Rymdobservatorium, Observatoriev\"{a}gen,
           S--43992 Onsala, Sweden\\
           \email{anders.winnberg@chalmers.se}
      \and Hamburger Sternwarte, Gojenbergsweg 112,
           D--21029 Hamburg, Germany
      \and INAF - Istituto di Radioastronomia, Via P. Gobetti 101,
           I--40129 Bologna, Italy
      \and Dipartimento di Astronomia, Universita` di Bologna, 
              Via Ranzani 1, I-40127 Bologna, Italy
      \and INAF - Osservatorio Astrofisico di Arcetri,
           Largo E. Fermi 5,
           I--50125 Florence, Italy}

\date{Received date; accepted date}

\abstract {Water vapour maser emission from late-type stars characterises
them as asymptotic-giant-branch stars with oxygen-rich chemistry
that are losing mass at a substantial rate.  Further conclusions on
the properties of the stars, however, are hampered by the strong
variability of the emission.}
{We wish to understand the reasons for the strong variability of
\water\ masers in circumstellar shells of late-type stars.  In this
paper we study RX~Bootis and SV~Pegasi as representatives of semiregular
variable stars (SRVs).}
{We monitored RX~Boo and SV~Peg in the 22-GHz maser line of water
vapour with single-dish telescopes.  The monitoring period covered two
decades for RX~Boo (1987 -- 2007) and 12 years for SV~Peg
(1990 -- 1995, 2000 -- 2007).  In addition, maps were obtained of RX~Boo
with the Very Large Array over several years.}
{We find that most of the emission in the circumstellar shell of
RX~Boo is located in an incomplete ring with an inner radius of 91~mas
(15~AU).  A velocity gradient is found in a NW--SE direction.  The
maser region can be modelled as a shell with a thickness of 22~AU,
which is only partially filled.  The gas crossing time is 16.5 years.
The ring-like structure and the velocity gradient remained stable for
at least 11 years, while the maser line profiles varied strongly.
This suggests that the spatial asymmetry is not accidental, so that
either the mass loss process or the maser excitation conditions in
RX~Boo are not spherically symmetric.  The strong variability of the
maser spectral features is mainly due to incoherent intensity
fluctuations of maser emission spots, which have lifetimes of the
order of 1 year.  We found no correlation between the optical and the
maser variability in either star.  The variability properties of the
SV~Peg masers do not differ substantially from those of RX~Boo.  There
were fewer spectral features present, and the range of variations was
narrower.  The maser was active on the $>$10-Jy level only 1990 -- 1992
and 2006/2007.  At other times the maser was either absent ($<$1~Jy) or
barely detectable.}
{The variability of \water\ masers in the SRVs RX~Boo and SV~Peg is due
to the emergence and disappearance of maser clouds with lifetimes of
$\sim$1 year.  The emission regions do not evenly fill the shell of
RX~Boo leading to asymmetry in the spatial distribution, which
persists at least an order of magnitude longer.}

\maketitle

\section{\label{intro} Introduction}
Water vapour maser emission of the $6_{16} - 5_{23}$ transition
at\linebreak 1.35-cm wavelength (22~GHz) is often found in the
circumstellar shells of oxygen-rich stars on the asymptotic giant
branch (AGB) and in several red supergiants (RSGs).  The AGB stars
have intermediate masses ($M \approx 1 - 8$~M$_\odot$) and are usually
long-period variables.  They encompass semi-regular variables (SRVs),
Mira variables and OH/IR stars.  The classification of SRVs and Mira
variables is made in the optical regime and they are distinguished by
the amplitude of their intensity variations.  Due to their higher
mass-loss rates OH/IR stars are optically faint and their variability
is observed in the infrared (see review by Habing \cite{Habing96}).

SRV, Mira and OH/IR stars form a sequence of increasing mass-loss
rates, but it is not clear if this is also an evolutionary sequence or
a sequence of increasing progenitor masses.  RSGs are stars with
masses $M >$ 8~M$_\odot$ in a short evolutionary phase, where
mass-loss rates are high and they develop circumstellar dust and gas
shells as AGB stars do.

Immediately after the discovery of \water\ masers in the envelope of
the RSG VY~CMa, variability was noticed (Knowles et
al. \cite{Knowles69}; Buhl et al.  \cite{Buhl69}).  From monitoring
observations of AGB stars and RSGs (Schwartz et al. \cite{Schwartz74})
evidence was found that the observed maser spectra varied in peak and
integrated flux, in line profiles and possibly in line-of-sight
velocity\footnote{In this paper we use the expression `line-of-sight
velocity' instead of `radial velocity' because the latter could be
confused with the velocity radially from the star, i.e. the expansion
velocity of the stellar envelope.  All velocities in this paper are
relative to the local standard of rest (LSR).} (Dickinson et
al. \cite{Dickinson73}).  The earliest studies of long-period
variables found that the maser variability was well correlated with
variations in the optical and in the infrared (Schwartz et
al. \cite{Schwartz74}), but later observations showed that the
regularity of the radio variations had been overestimated (Engels et
al. \cite{Engels88}; Cohen \cite{Cohen89}). The first interferometric
maps of the Mira variables R~Aql and RR~Aql taken at two epochs
separated by 15 months also showed dramatic changes (Johnston et
al. \cite{Johnston85}).  Knowledge of the properties of the
variability and understanding of the reasons for it, however, are
prerequisites for the interpretation of the observations and for the
derivation of basic stellar properties.

Water vapour masers are potentially very sensitive tools to probe the
physics and dynamics in parts of the circumstellar shells, whose radii
range between $\sim$5 and 50~AU in case of Mira variables and SRVs
(Bowers et al. \cite{Bowers93}; Bowers \& Johnston \cite{Bowers94};
Colomer et al. \cite{Colomer00}; Bains et al. \cite{Bains03}; Imai et
al. \cite{Imai03}).  High-resolution mapping with interferometers
shows that the integrated maser emission is composed of a multitude of
individual clouds of sizes $\sim$2 -- 4~AU and a low filling factor of
$\sim$0.01 (Bains et al. \cite{Bains03}).  In conjunction with the SiO
masers located closer to the central star and with the OH masers in
the outer parts of the envelope, \water\ masers allow us to map the
shell that is created by the strong mass loss ($>$\powerten{-7}~\Myr)
on the upper AGB.  Information on the velocities and on the
distribution of the maser spots can be used to reconstruct the
geometry of the mass-loss process and sometimes even its history.

Monitoring of \water\ maser emission with single-dish radio
telescopes, even covering more than a decade, provides only limited
insight.  The spectra mostly consist of many components, which often
overlap in velocity.  The profiles and intensities of \water\ maser
spectra therefore depend not only on temperature variations due to the
pulsating star but also on other factors, including random intensity
fluctuations of individual clouds (Engels et al. \cite{Engels88}).
Early programmes followed only the strongest lines.  Three years of
monitoring led Cox \& Parker (\cite{Cox79}) to the conclusion that the
masers are not stable for more than a few pulsation cycles.  Later
Little-Marenin and co-workers monitored a few Mira variables over a
number of years and found complex variations in intensity with many
components varying independently from one another (e.g. Little-Marenin
et al. \cite{Little91}).  Usually a phase lag of $\sim$0.3 (30 -- 100
days) with respect to the visual maximum was observed (Little-Marenin
et al. \cite{Little93}).  A Russian group monitored a number of stars
with strong \water\ masers for up to twenty years.  They found that by
and large the maser flux correlates with the visual light curve, and
confirmed the presence of a phase lag.  They attributed the maser
variations and the phase lag to the passage of periodic shocks (Lekht
et al.  \cite{Lekht05}, and references therein).

\begin{table}
\caption[]{\label{tab:epochs} The VLA observing schedule}
\begin{flushleft}
\begin{tabular}{lcl}
\hline\noalign{\smallskip}
Date & Conf. & Sources \\
\noalign{\smallskip}
\hline\noalign{\smallskip}
1990 Feb. 26 &   A & RX~Boo, U~Her, OH~39.7+1.5, \\
             &     & OH~83.4$-$0.9, VX Sgr \\
1990 June 3  &   A & RX~Boo, U~Her, OH~39.7+1.5, \\
             &     & OH~83.4$-$0.9, VX Sgr \\
1991 Aug. 23 &   A & OH~39.7+1.5 \\
1991 Oct. 20 & BnA & RX~Boo, U~Her \\
1992 Jan. 10 &   B & VX~Sgr, NML~Cyg \\
1992 Dec. 28 &   A & RX~Boo, U~Her, OH~39.7+1.5 \\
1995 Sept. 5 &   A & RX~Boo \\
1996 Dec. 10 &   A & OH~26.5+0.6 \\
1998 May 31  &   A & OH~26.5+0.6 \\
\noalign{\smallskip}
\hline
\end{tabular}
\end{flushleft}
\end{table}

In order to improve the understanding of the properties of maser
variability for different types of late-type stars, we started in 1987
a monitoring programme of several such stars using the Effelsberg and
Medicina radio telescopes.  The sample included SRVs, Mira variables,
OH/IR stars and RSGs.  In order to break the spatial degeneracy, which
usually makes the interpretation of single-dish maser spectra
difficult, we added interferometric observations using the Very Large
Array (VLA).  Of each object type we selected prototypical stars,
which were observed during several epochs (see Table
\ref{tab:epochs}).  In this paper we present the results for two SRVs:
RX~Bootis (RX~Boo) and SV~Pegasi (SV~Peg).  In subsequent papers we
will present results for the other types of late-type stars.

\object{RX~Boo} is a 9 -- 11 mag SRV with a mean spectral type of M7.5 and a
period varying between 340 and 400 days (Kukarkin et
al. \cite{Kukarkin71}).  Olofsson et al.  (\cite{Olofsson02}) in their
model fit involving several CO transitions estimated a mass-loss rate
of 6\, $\times$ \powerten{-7}~\Myr\ and an expansion velocity of
9.3~\kms\,, whereas Teyssier et al. (\cite{Teyssier06}), as a result
of another model fit, obtained 2\, $\times$ \powerten{-7}~\Myr\ and
7.5~\kms, respectively.  The distance is 160$\pm$25~pc (The Hipparcos
Catalogue; Perryman et al. \cite{Perryman97}).  Dyck et
al. (\cite{Dyck95}), using an IR interferometer, measured the stellar
diameter at 2.2~ \micron\ and found a uniform-disk (UD) diameter of
19$\pm$1~mas.  They calculated an effective temperature of 3000 $\pm$
100~K.  Chagnon et al. (\cite{Chagnon02}), observing in a band centred
at 3.8~\micron, found a UD diameter of 21.0 $\pm$ 0.3~mas.

The \water\ maser of RX~Boo is one of the first masers of this kind
detected in late-type stars (Schwartz \& Barrett \cite{Schwartz71};
Dickinson et al. \cite{Dickinson73}), and hence it is one of the best
studied.  Single-dish monitoring was performed by Schwartz et
al. (\cite{Schwartz74}), Cox \& Parker (\cite{Cox79}), Berulis et
al. (\cite{Berulis83}), and Hedden et al.  (\cite{Hedden91}) who all
found a weak correlation of maser intensity with optical flux.  Maps
taken with the VLA show an overall extent of the maser shell of 150 --
230~mas (Bowers et al. \cite{Bowers93}; Colomer et
al. \cite{Colomer00}), giving an outer shell radius of 12 -- 18~AU or
9 -- 14 stellar radii.

\object{SV~Peg} also varies with $\sim$2~mag amplitude (9 -- 11~mag)
and has a spectral type of M7 and a mean period of 144 days (Kukarkin
et al. \cite{Kukarkin71}).  The mass-loss rate is 3\, $\times$
\powerten{-7}~\Myr\ at an expansion velocity of 7.5~\kms\ (Olofsson et
al. \cite{Olofsson02}) and its distance is 205$\pm$45~pc (The
Hipparcos Catalogue; Perryman et al. \cite{Perryman97}).  The \water\
maser was discovered by Dickinson (\cite{Dickinson76}) and was only
occasionally observed before our monitoring efforts.  No mapping has
been done as yet.

\section {\label{observations} Observations}
Single-dish observations of the \water\ maser line at 22235.08~MHz
were made with the Effelsberg\footnote{The 100-m Effelsberg telescope
is operated by the Max-Planck-Institut f\"ur Radioastronomie (MPIfR)
at Bonn, Germany.} 100-m and Medicina\footnote{The 32-m Medicina
telescope is operated by the Istituto di Radioastronomia (IRA) at
Bologna, Italy.} 32-m telescopes at typical intervals of a few months.
RX~Boo was monitored over twenty years between 1987 and 2007, and
SV~Peg in 1990 -- 1995 and 2000 -- 2007.  VLA\footnote{The Very Large
Array is operated by National Radio Astronomy Observatory (NRAO) from
its subsidiary `Array Operations Center' (AOC) at Socorro, New Mexico,
USA.} observations were made of RX~Boo on four occasions in the period
1990 -- 1992 and later in 1995 after a strong flare event (see Table
\ref{tab:epochs}).  This event will be discussed in a separate paper.

\subsection{\label{obs_Effelsberg} Effelsberg observations}
With the 100-m telescope observations were made of RX~Boo between 1990
February and 2002 June and of SV~Peg between 1990 February and 1995
June.  We used the 18 -- 26-GHz receiver to observe the $6_{16}
\rightarrow 5_{23}$ transition of the water molecule.  The receiver
had a cooled maser as pre-amplifier and passed left-hand circularly
polarised radiation.  As circumstellar water masers are found to be
unpolarised to limits of a few percent (Barvainis \& Deguchi
\cite{Barvainis89}), significant flux losses due to the polarization
properties of the receiver are not expected.  At 1.3-cm wavelength the
beamwidth is $\sim$40\arcsec\ (FWHM).  We observed in total power mode
integrating ON and OFF the source for 5~min each. `ON-source' the
telescope was positioned on RX~Boo at the coordinates
\recta{14}{24}{11}{6}, \dec{+25}{42}{13} (J2000) and on SV~Peg at
\recta{22}{05}{42}{1}, \dec{+35}{20}{54} (J2000), while the
`OFF-source' position was displaced 3\arcmin\ to the east of the
source.  Typical system temperatures were 60 -- 120~K depending on
weather conditions and elevation.  The typical \textsl{rms}
noise-level in the spectra is $\sim$0.2~Jy.

The backend consisted of a 1024-channel autocorrelator.  Observations
were made with a bandwidth of 6.25~MHz, centred on the systemic
line-of-sight velocity of the star.  The velocity coverage was
$\sim$80~\kms\ and the velocity resolution 0.08~\kms.

Standard procedures were used to reduce the spectra.  First a baseline
was fitted using the channels not affected by emission, then the
calibration was applied.  Finally, spectra from several integrations
of the same source were averaged, if they were taken during the same
day. Hanning smoothing to 0.16~\kms\, was applied only to noisy
spectra.

The flux calibrations were made against NGC 7027 (5.9~Jy) and Mon R2
(6~Jy), and corrections for atmospheric attenuation and elevation were
applied using a standard gain curve determined under good weather
conditions.  Deviations from the standard gain curve were less
than 10\%, except at low elevations where deviations up to 30\% were
measured.  Losses due to pointing are at most secondary effects, as
pointing checks and corrections were regularly carried out on strong
continuum sources, showing deviations from the nominal position of at
most 10\arcsec.

\subsection{\label{obs_Medicina} Medicina observations}
Observations at Medicina were carried out between 1987 March and 2007
January on RX~Boo and between 1990 February and 1995 January and again
between 2000 October and 2007 January on SV~Peg.  The Medicina 32-m
telescope (HPBW at 22~GHz $\sim1$\pam9) is set up primarily to be used
for VLBI measurements, and therefore the front-end is typical for such
work.  For line observations at 22~GHz only the right-hand circular
polarization output from the receiver is available.

Over the years the system has undergone various improvements.  In 1989
realignement of the antenna surface resulted in an improvement of the
efficiency at 22~GHz from 17 to 38\%.  The backend was a 512-channel
digital autocorrelator; in 1991 the number of channels was increased
to 1024.  In the same year a HEMT amplifier replaced the GaAs FET
front-end, reducing the system temperature (120~K at zenith under good
weather conditions).  Also in 1991 an active sub-reflector control
increased the gain at low and high elevations.  The available
bandwidth varied between 3.125~MHz and 25~MHz.  Up to 1997 the
spectral resolutions used corresponded to 0.082~\kms\ or 0.165~\kms.
In 1997 the VLA-1 chip correlator was replaced by one based on the
NFRA correlator chips and boards (Bos \cite{Bos91}); in the standard
configuration the latter has 2048 channels and a maximum bandwidth of
160~MHz.  However, to maintain consistency with the older observations
we have only used a 10~MHz band and 1024 channels, giving a velocity
resolution of 0.132~\kms.

The telescope pointing model is typically updated a few times per
year, and is quickly checked every few weeks by observing strong maser
sources (e.g. W3~OH, Orion-KL, W49~N, and/or Sgr~B2 and W51).  The
pointing accuracy was always better than 25\arcsec; from about mid
2004 the \textsl{rms} residuals from the pointing model were of the
order of $8$\arcsec $-10$\arcsec.  The antenna gain as a function of
elevation was determined by daily observations of the continuum source
DR~21 at a range of elevations; we assume a flux density of 16.4~Jy
(after scaling the value of 17.04~Jy given by Ott et al. (\cite{Ott94}) for the
ratio of the source size to the Medicina beam\footnote{In previous works and up
to 2003 April a flux density of 18.8~Jy was adopted for DR~21, following Dent
(\cite{Dent72}).  In this paper, all data have been recalibrated using the value
given by Ott et al. (\cite{Ott94}).}).  The daily gain curve was determined by
fitting a polynomial to the data, which was then used to convert antenna
temperature to flux density for all spectra taken that day.  From the dispersion
of the single measurements around the curve we find the typical calibration
uncertainty to be 19\%.  On the few days that no separate gain curve was
measured, we have applied the one closest in date, and estimate a corresponding
calibration uncertainty of 7\%.  The average overall calibration uncertainty is
therefore estimated at $\sim$20\%.

Observations were taken in total power mode, with both ON and OFF scans of 5-min
duration.  The OFF position was taken 1\degr~W of the source position.  
Typically, 2 ON/OFF pairs were taken. The typical \textsl{rms} noise-level in
the spectra is $\sim$1.5~Jy.  Finally, no observations were taken between 1996
April and 1997 February, due to extensive maintenance work on the telescope.

For the purpose of studying the long-term variability of the maser
emission, spectra of the same source taken less than 4~days apart were
averaged.  During the period that the sources were also monitored at
Effelsberg it may happen that we have spectra from both telescopes
within 4~days.  In these cases we always used the Effelsberg spectra
because of their much lower noise level.  With very few exceptions
both spectra showed the same line profile and flux densities (within
10\%) of the individual emission components, providing an {\it a posteriori}
check on the correctness of the calibration procedures of both telescopes.

\subsection{\label{obs_VLA} VLA observations}

\begin{table}
\caption[]{\label{tab:VLAspec} VLA map specifications for RX~Boo}
\begin{flushleft}
\begin{tabular}{lrrrrr}
\hline\noalign{\smallskip}
Date &  \multicolumn{3}{c}{HPBW} & rms & $S/N$ \\
 & maj.a.  & min.a. & p. a.  &  & \\
 & (\arcsec) & (\arcsec) & (\degr) & (Jy/b.) & \\
\noalign{\smallskip} \hline\noalign{\smallskip}
1990 Feb. & 0.090 & 0.077 & 86.23 & 0.015 & 3700 \\
1990 June & 0.082 & 0.078 & 40.12 & 0.014 & 1700 \\
1991 Oct. & 0.305 & 0.092 & $-76.62$ & 0.025 & 190 \\
1992 Dec. & 0.077 & 0.072 & $-23.67$ & 0.027 & 130 \\
1995 Sep. & 0.127 & 0.084 & $-67.27$ & 0.009 & 580 \\
\noalign{\smallskip} \hline
\end{tabular}
\end{flushleft}
\emph{maj.a.}: half-power beam width (HPBW) for the major axis of
the best-fit three-dimensional Gaussian component to the synthesised beam \\
\emph{min.a.}: HPBW for the minor axis \\
\emph{p.a.}: position angle of the major axis (E of N)\\
\emph{rms}: the root-mean-square noise fluctuations in signal-free
channels in units of Jansky per beam area \\
\emph{S/N}: signal-to-noise ratio or `dynamic range' in the
channel with strongest signal
\end{table}

RX~Boo was observed with the VLA on four occasions between 1990 February and
1992 December (Table \ref{tab:epochs}).  An additional observation was made in
1995 September after the occurrence of a strong flare, which will be discussed
in a separate paper.  All 27 antennas were used yielding synthesised beamwidths
as specified in Table \ref{tab:VLAspec}.  These beamwidths depend mainly on the
configuration, i.e. the extent of the array.  For three of the four
epochs we used the largest extent (A; see Table \ref{tab:epochs}), while the
October 1991 observations were carried out with a hybrid configuration (BnA).

We chose a receiver bandwidth of 3.125~MHz to obtain a total velocity range of
42~\kms\ and the bandwidth was split into 64 channels, yielding a velocity
resolution of 0.66~\kms.  Data from the right and left circular polarization
modes were averaged.  Typical integration times were 30~min on the star and
12~min on the phase calibrator J1407+284 with a sampling time of 30~sec.  Flux
calibration was obtained relative to 3C286 that was assumed to have a flux
density of 2.55~Jy and 3C84 was used to correct for the bandpass
shapes.

\begin{figure*}
\centering
\resizebox{16cm}{!}{\includegraphics{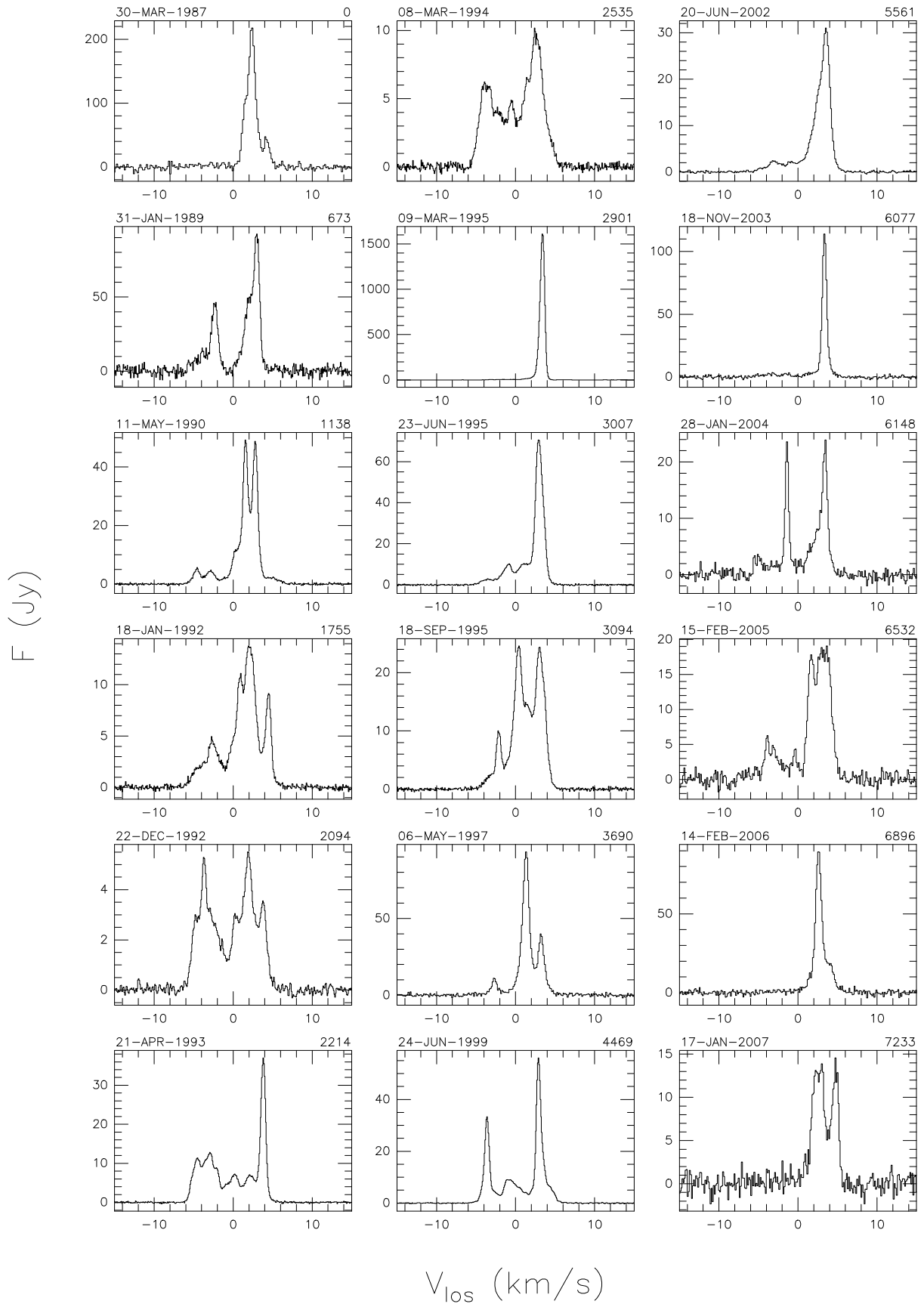}}
\caption{Sample \water\ maser spectra of RX~Boo showing typical profiles
starting in 1987 and ending in early 2007.  Note the varying flux density 
scale.}
\label{fig:sample_spectra_rxboo}
\end{figure*}

The data were reduced with AIPS (Astronomical Image Processing
System\footnote{http://www.aoc.nrao.edu/aips/}) using the spectral line data
routines.  To improve the sensitivity as well as the dynamic range,
self-calibration was applied.  Care was taken to select an unresolved feature as
phase reference.  It turned out that bright features often are extended and we
made several trials to find the best unresolved feature.  Usually we used a
point source model for the first iteration.  The final sensitivity reached on
line-free channels was 9 -- 27~mJy/beam depending on weather conditions and
integration time (see Table \ref{tab:VLAspec}).

\section{\label{SingleDishData} Single dish data of RX~Boo}

\subsection{\label{sdd_Def} Definitions}
To interpret the single dish maser profiles we will use several terms, whose
definitions are given here.  The integrated maser emission from a circumstellar
shell is a superposition of emissions arising from many physical entities ({\it
maser clouds}).  These emissions are in the form of narrow {\it maser lines},
which emit at the velocity of the cloud projected on the line-of-sight. 
Observed with a single radio telescope all the maser lines together form a {\it
maser spectrum} with usually several peaks.  The shape of the often contiguous
spectrum is named {\it maser profile}.  The data analysis aims at a
decomposition of the spectrum into individual maser lines.  However, due to
blending in velocity space, the analysis rarely results in individual lines but
in {\it maser features}, which are composed of two or more maser lines.  These
maser features might be traced over some period of time in several spectra and
may reappear at later times perhaps with a slightly different velocity ($\la0.5$
\kms).  They might be identified with real maser clouds persisting over this
period of time.  However, most likely the features are composed of several real
clouds with velocities close to one another but spatially separated.  We will
use the interferometric maps to identify these (groups of) maser clouds, which
we name more generally {\it maser spatial components}.  For accounting purposes
all spectral features were grouped according to their velocities into {\it maser
spectral components}.  At different times within the monitoring period of 20
years different spectral features contribute to a spectral component, reflecting
the asynchronous formation and dissolution of the contributing maser clouds.

\subsection{\label{sdd_MaserSpec} \water\ maser spectra}
\begin{figure}
\centering
\includegraphics[width=8cm]{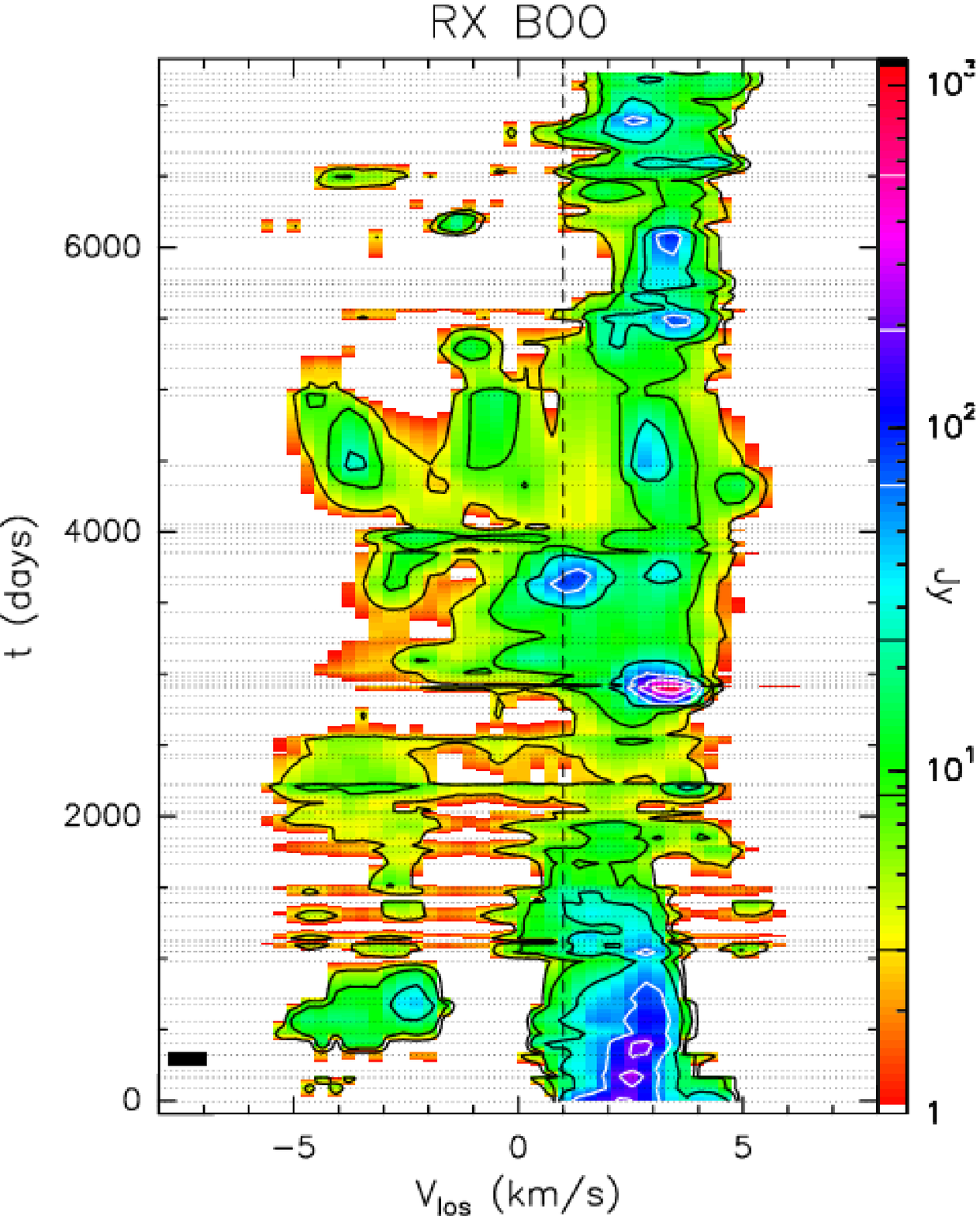}
\caption{Flux density of the \water\ masers of RX~Boo as a function of
velocity and time.  The time axis covers the years 1987 -- 2007.  The
time stamp $t=0$ corresponds to 1987 March 30 (J.D. 2446885).  The
dotted horizontal lines mark the actual observations averaged in 4-day
intervals.  Flux densities in between the observations were interpolated
linearly.  The spectra were resampled to a resolution of 0.3~\kms.  The dashed
vertical line marks the stellar systemic velocity (1~\kms) as given by Teyssier
et al. (\cite{Teyssier06}).  Contours are drawn at levels 3.00, 8.49, 24.03,
68.02, 192.51, 544.85~Jy (logarithmic scale). These six levels are indicated by
black and white lines in the flux scale along the right side of the map. 
(For a colour version of this figure, see the electronic edition.)}
\label{fig:v_t_diagram_rxboo}
\end{figure}

Our monitoring of RX~Boo covers almost 21 years, from 1987 March to
2007 January with several (typically 4 -- 5) observations per year.
Depending on telescope (i.e. Effelsberg or Medicina), date of
observation and integration time the \textsl{rms} sensitivity of the
observations was very inhomogeneous ranging from 0.1 to 6~Jy.  All
spectra taken are shown in the Appendix (Fig. A1).  Sample spectra
showing typical profiles are given in Fig.
\ref{fig:sample_spectra_rxboo}.  A general view on the properties of
the masers and their variations is given in
Fig. \ref{fig:v_t_diagram_rxboo}.  Here the spectra are arranged along
a time axis and the flux densities in between observations are
interpolated to obtain an approximated continuous variability profile.

The \water\ masers of RX~Boo covered a constant velocity interval
between --5 and +5~\kms, although at particular velocities the maser
emission could drop below the detection threshold for extended periods
of time (months to even years).  In the time range 1987 -- 1989 ($t \la
1000$ in Fig. \ref{fig:v_t_diagram_rxboo}) RX~Boo displayed a strong
maser feature at 2.5~\kms\ reaching flux density levels of a few
hundred Janskys.  In 1995 ($t \sim 2900$) a flare was observed at a
velocity of 3.4~\kms\ with a peak flux density of $ >$1500~Jy.  During
the rest of the time typical flux density levels were $ <$100~Jy.  The
maser never completely disappeared, although occasionally the flux
density of the strongest emission peak was only a few Janskys.

Although strong flux variations were observed, leading to a great
variety of spectral profiles, they did not occur at random velocities
within the observed velocity interval.  With a few exceptions the
emission at $v_{\rm los} > 0$~\kms (observed line-of-sight velocity)
was always stronger than that at $v_{\rm los} < 0$~\kms, with the
strongest peak occurring between 1.5 and 3.5~\kms.

\subsection{\label{sdd_LineProfAna} Line profile analysis}
The \water\ maser spectra of RX~Boo were decomposed into separate
features by fitting multiple Gaussian line profiles.  We generally
used an iterative procedure in which we first fitted and subtracted
the strongest features in each spectrum and then extracted the weaker
features from the residual spectrum.  The number of extracted features
depended on the quality of the spectra as well as the number of
distinct peaks.  Blending of maser lines is a major obstacle and the
extracted features may consist of two or more lines.  Depending on the
strength of the contributing lines the central velocity of the
extracted feature may vary.  As a rule of thumb, the FWHM of strong
features (visible as distinct peaks in the spectra) is $\sim$1~\kms,
and therefore features with FWHM $\ga$ 2~\kms\ are most probably
blends.  Comparison with adjacent spectra often shows that blended
features split when the relative intensities of the contributing lines
change.  In complex situations when Gaussian fitting failed, features
were identified by eye and velocities and flux densities measured by
hand.  We assumed that maser features in adjacent spectra with
velocity differences $\la$0.5~\kms\ belong to a unique spatial maser
component in the shell of RX~Boo, which varied in intensity with time.
All features were accordingly assigned to 11 maser spectral components
labeled A, B,..., K.  These components are listed in Tables
\ref{tab:compRXBooA-E} and \ref{tab:compRXBooF-K} in the Appendix.
For each spectrum we give the Gregorian and Julian dates, its
\textsl{rms} noise level in Jansky, the \SI\ in
\powerten{-22}~W~m$^{-2}$, and the velocities (\vp) and peak flux
densities (\Sp) for each component.  Occasionally, features were
considered as blends and were assigned to two or more
components. Values marked by ``:'' and upper limits were measured with
the cursor on the computer screen.

\subsection {\label{sdd_LineFitRes} Line fitting results}
Tables \ref{tab:compRXBooA-E} and \ref{tab:compRXBooF-K} show that the
maser features at $v_{\rm los}<$~0~\kms\ were weaker than those at
$v_{\rm los} >$ 0~\kms\ over the full monitoring period.  One spectral
component, C, at --2.7~\kms, was detected at almost every epoch,
whilst A, B, D and E were only detected sporadically.  This may be
largely a sensitivity effect, because many spectra did not have a
sufficient signal-to-noise ratio to allow detection of the fainter
features on the level of a few Janskys.  In the red part of the maser
profile ($v_{\rm los} >$ 0~\kms) heavier blending of lines made the
unambiguous assignment of features to the components F -- K a
non-trivial task (Table \ref{tab:compRXBooF-K}).  The features
assigned to component I, for example, show a systematic velocity shift
with \vp $=+2.7\pm0.3$~\kms\ until 1992 and $+3.2\pm0.3$~\kms\ after
1994.  Most likely the emission contributing to component I came from
(at least) two maser clouds, of which one of them dominated the
emission during the first time range and the other one during the
second time range.  Similar shifts are observed among the other
spectral components.  The direction of the shifts does not show any
systematics so that we regard these shifts generally as a result of
blending by two or more maser lines contributing to a spectral
component.  Some components show a rather large scatter of velocities,
as for example component G with 0.4 $\le$ \vp $\le$ 1.9~\kms.  These
``shifts'' are almost random and are also not well explained as
movements of a physical maser cloud, but rather as a result of
blending of several maser lines in the velocity interval 0 -- 2~\kms\
and varying relative intensities.  The spectral components F and G
show therefore some overlap in velocity, indicating that they are not
well defined.  Spatial resolution of the maser emission is required to
overcome the limited interpretational power of the single dish maser
profiles.

However, it is obvious from Fig. \ref{fig:v_t_diagram_rxboo} as well
as from the run of velocities in individual spectral components
(Tables \ref{tab:compRXBooA-E} and \ref{tab:compRXBooF-K}) that the
lifetimes of individual components are limited on the scale of months
up to a few years.  The lifetimes will be addressed in more detail in
Sect.  \ref{CrossCorrel-InterfData}.

\begin{figure*}
\centering
\includegraphics[width=16cm]{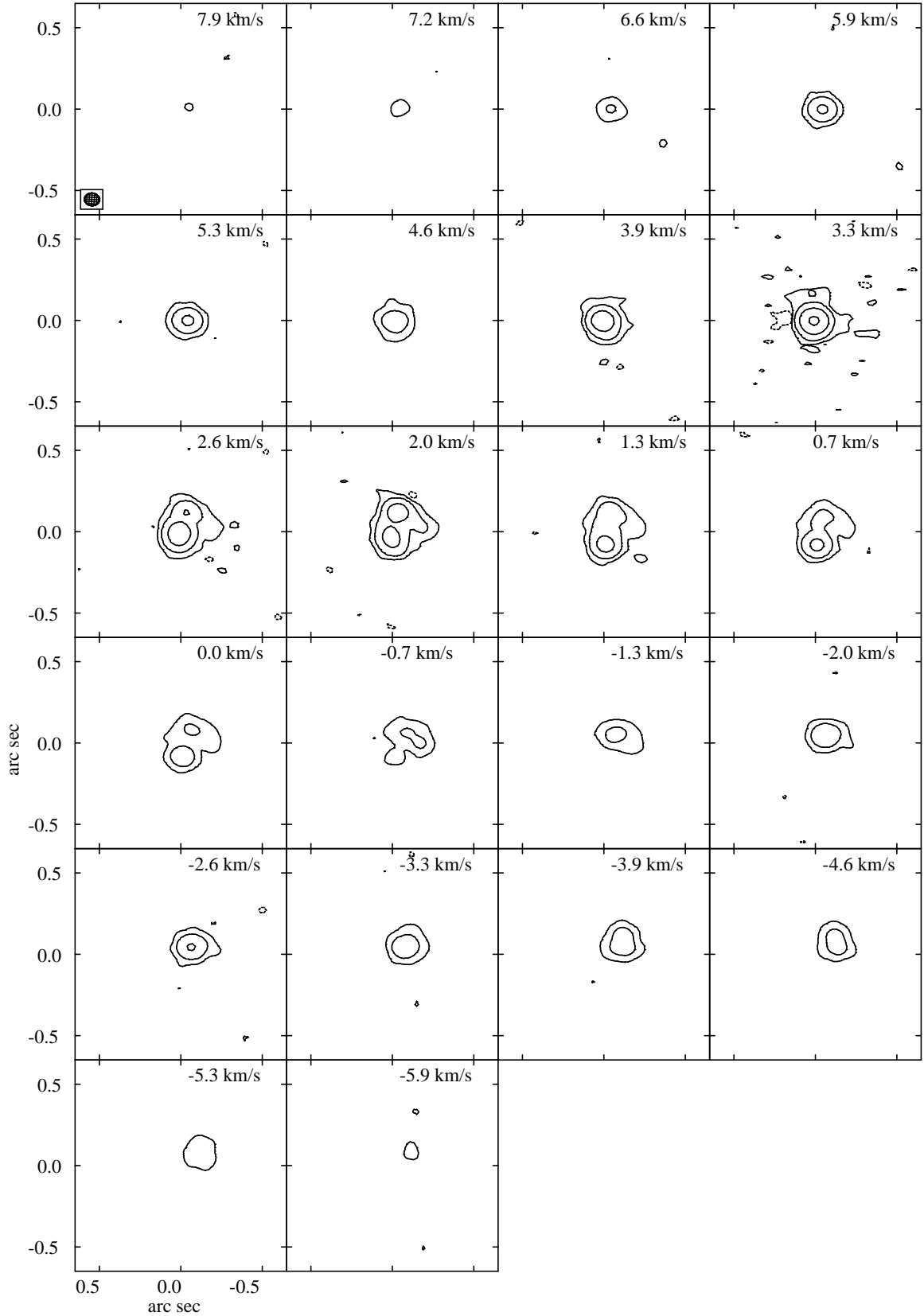}
\caption{Sample \water\ maser images of RX~Boo from 1990 February.
Each image is separated from its neighbours by the line-of-sight
velocity interval 0.658~\kms.  The synthesised FWHM beam size (major
axis: 0\pas09; minor axis: 0\pas08; position angle of the major axis:
86\degr) is shown in the uppermost left image (7.9~\kms).  The images
are oriented along right ascension and declination and the angular
scales are relative to the position \recta{14}{24}{11}{6},
\dec{+25}{42}{14} (J2000).  Brightness contours are -0.05 (dashed
contour), 0.05, 0.5, 5, 50~Jy per beam area (1.45 \powerten{-13}~sr).}
\label{fig:channel_maps}
\end{figure*}

\section{\label{InterfData} Interferometric data of RX~Boo}
Interferometric observations of the maser emission aim at mapping the
maser clouds and breaking the spatial degeneracy.  Due to the limited
spatial resolution of the instrument this may be only partially
successful, and the data analysis of the images will yield {\it maser
spatial components}, which will be superpositions of several maser
clouds close to each other in space as well as in velocity.

\subsection{\label{id_ChanMapAna} Channel-map analysis}
The VLA interferometric data consist of one data cube for each of the four
observations, containing 63 maps each.  The maps are separated in velocity by
0.658~\kms.  An example is given in Fig. \ref{fig:channel_maps} where maps of
the 22 channels with detected signal are displayed from the first VLA epoch
(1990 February).  Except for the central velocity range the maser emission
consists of point-like and unresolved spots.  In order to single out maser
components the data cubes were analyzed within AIPS in a three step process. 
First, the individual channel maps were analyzed one by one, then components
were identified by comparing neighbouring channel maps, and finally these
components were verified in velocity space.

The analysis of the channel maps was made by fitting up to four
two-dimensional Gaussian components with elliptical cross sections to
the emission spots.  The results of this analysis tend to be insecure
as soon as spots are blended spatially as well as in velocity.  The
positions of the fitted Gaussian components are obtained with an
accuracy of $\sim$0.02 pixel (0.4~mas) for strong emission spots
($>$1~Jy).  Comparing the positions of such spots extending over
several channels yields deviations $\leq$0.2 pixel ($\leq$4~mas)
between the positions of the same spot in different channel maps.  The
positions of fainter emission spots have errors up to 0.5 pixel
(0\pas01).  Very faint spots ($\leq$300~mJy) can be localised by
summing over several channels, usually three.  This is guided by the
experience that spatial components extend in velocity by not more than
$\sim$3 channels ($\sim$2~\kms).  Often the Gaussian components fitted
to weak spots from merged channel maps tend to be extended, indicating
that these spots are actually blends of several weak emission spots.
Due to high signal-to-noise ratios the positional accuracy obtained
from Gaussian fits is much better than the resolution of the
interferometer of $\sim$4 pixel (FWHM $\sim$ 0\pas 08), allowing us to
determine the relative spatial separation of spots down to 4 --
10~mas.

\begin{table}
\caption[]{\label{tab:components} Spatial and spectral components of RX~Boo}
\begin{tabular}{rrrrrrr}
\hline\noalign{\smallskip}
Spat. & Date & Vel. & Flux & Xoff & Yoff & Spec. \\
C. &      & [km/s] & [Jy] & [mas] & [mas] & C. \\
\noalign{\smallskip}
\hline\noalign{\smallskip}
A1 & Feb. 90 &$-$4.5 &  1.5 &$-$40 &   +84 & A  \\
A1 & Jun. 90 &$-$4.5 &  0.9 &$-$37 &   +68 &    \\
A1 & Oct. 91 &$-$5.1 &  0.7 &$-$38 &   +84 &    \\
A1 & Dec. 92 &$-$4.8 &  0.6 &$-$62 &   +66 &    \\[0.1cm]
A2 & Feb. 90 &$-$4.5 &  1.5 &$-$27 &  +151 &    \\
A2 & Jun. 90 &$-$4.5 &  1.5 &$-$33 &  +140 &    \\
A2 & Oct. 91 &$-$4.5 &  1.6 &$-$34 &  +146 &    \\
A2 & Dec. 92 &$-$4.2 &  0.5 &$-$58 &  +126 &    \\[0.1cm]
B1 & Dec. 92 &$-$3.8 &  3.5 & $-$6 &   +90 & B  \\[0.1cm]
C1 & Feb. 90 &$-$3.2 &  6.0 &  +28 &   +81 & C  \\
C1 & Jun. 90 &$-$3.2 &  2.0 &  +21 &   +72 &    \\[0.1cm]
C2 & Feb. 90 &$-$3.2 &  0.5 &$-$36 &  +105 &    \\
C2 & Jun. 90 &$-$3.8 &  0.3 &$-$41 &  +106 &    \\[0.1cm]
C3 & Oct. 91 &$-$2.5 &  4.2 &$-$18 &   +92 &    \\
C3 & Dec. 92 &$-$2.5 &  1.0 &$-$42 &   +92 &    \\[0.1cm]
C4 & Dec. 92 &$-$2.5 &  0.7 &  +20 &   +60 &    \\[0.1cm]
D1 & Dec. 92 &$-$2.2 &  0.7 &$-$86 &   +16 & D  \\[0.1cm]
D2 & Jun. 90 &$-$1.9 &  0.6 & $-$9 &   +90 &    \\[0.1cm]
E1 & Oct. 91 &$-$0.9 &  1.5 &$-$12 &   +94 & E  \\
E1 & Dec. 92 &$-$1.2 &  0.8 &  +48 &   +64 &    \\[0.1cm]
F1 & Feb. 90 &$-$0.5 &  1.0 &$-$80 & $-$35 & F  \\
F1 & Jun. 90 &$-$0.5 &  0.3 &$-$77 &   +62 &    \\[0.1cm]
F2 & Feb. 90 &  +0.1 &  1.0 &  +29 &  +116 &    \\
F2 & Jun. 90 &  +0.1 &  3.0 &  +39 &  +134 &    \\
F2 & Dec. 92 &$-$0.2 &  0.4 &   +6 &  +106 &    \\[0.1cm]
F3 & Dec. 92 &  +0.4 &  1.5 &  +78 &   +88 &    \\[0.1cm]
G1 & Feb. 90 &  +0.1 & 12.0 &  +81 & $-$46 & G  \\
G1 & Jun. 90 &  +0.1 &  5.0 &  +83 & $-$55 &    \\
G1 & Oct. 91 &  +0.8 &  5.2 &  +86 & $-$55 &    \\
G1 & Dec. 92 &  +0.8 &  0.4 &  +84 & $-$65 &    \\[0.1cm]
G2 & Oct. 91 &  +1.1 &  4.0 &   +4 &   +92 &    \\[0.1cm]
H1 & Feb. 90 &  +1.4 & 18.0 &  +58 &  +154 & H  \\
H1 & Jun. 90 &  +1.4 & 18.0 &  +55 &  +148 &    \\
H1 & Oct. 91 &  +2.1 &  5.8 &  +66 &  +158 &    \\
H1 & Dec. 92 &  +2.1 &  2.8 &  +68 &  +168 &    \\[0.1cm]
H2 & Feb. 90 &  +1.4 &  0.5 &$-$80 &   +74 &    \\
H2 & Jun. 90 &  +1.4 &  1.3 &$-$81 &   +64 &    \\[0.1cm]
H3 & Feb. 90 &  +2.1 & 10.0 &  +97 &   +10 &    \\
H3 & Jun. 90 &  +1.4 &  8.0 &  +99 &  $-$7 &    \\
H3 & Dec. 92 &  +2.1 &  1.2 &  +70 &   +58 &    \\[0.1cm]
H4 & Dec. 92 &  +2.1 &  0.2 &  +80 & $-$59 &    \\[0.1cm]
I1 & Feb. 90 &  +2.8 & 70.0 & +100 &   +34 & I  \\
I1 & Jun. 90 &  +2.8 & 26.0 &  +99 &   +26 &    \\[0.1cm]
I2 & Feb. 90 &  +3.4 &  5.0 &  +88 &   +14 &    \\
I2 & Oct. 91 &  +2.4 &  3.8 &  +88 &   +18 &    \\[0.1cm]
I3 & Dec. 92 &  +3.4 &  0.1 &$-$44 &  +164 &    \\[0.1cm]
J1 & Oct. 91 &  +4.4 &  5.3 &  +66 &   +50 & J  \\
J1 & Dec. 92 &  +3.7 &  2.3 &  +76 &   +58 &    \\[0.1cm]
K1 & Feb. 90 &  +4.7 &  8.5 &  +59 &   +99 & K  \\
K1 & Jun. 90 &  +4.7 &  1.4 &  +51 &   +30 &    \\[0.1cm]
K2 & Jun. 90 &  +6.0 &  0.6 &  +47 &   +34 &    \\
\noalign{\smallskip}
\hline
\end{tabular}
\end{table}

\subsection{Spatial component identification}
Subsequently, neighbouring channel maps were compared and spatial
maser components were identified by looking for spatially coincident
Gaussian components.  Spatial maser components were accepted as real,
if they had Gaussian components in at least two neighbouring channel
maps.  Gaussian components present in only one map are always weak,
and are not considered further.  They might be real maser components
but cannot be verified.  At last we made sure that the list of
identified spatial maser components can be identified also in velocity
space.  For this purpose `slices' along the velocity axis through the
cube were obtained at the spatial positions of the components.
Usually the components stand out as distinct peaks in these `slices',
confirming their reality.  Spatial components blended in velocity by
another strong component, reveal themselves by adding considerable
asymmetry to the velocity profile of the major component.  Asymmetric
velocity profiles of components is a strong indication of blending and
may reveal new components that are separated marginally in velocity
but not spatially within the resolution of the maps.  The final list
of components consists of all components verified spatially as well as
in velocity

Typically 10 to 15 spatial maser components were identified in each VLA
observing epoch.  These components are listed in Table \ref{tab:components}.  
For the epoch 1990 June the components can be compared to those found by Colomer
et al. (\cite{Colomer00}), who used the VLA to observe RX~Boo one day apart from
our observation.  To determine the spatial components they used a 3-D Gaussian
fitting programme developed to analyze directly the spectral data cubes.  They
find about the same number of components (15 vs. 14), and their Gaussian-fitted
intensities and the overall spatial distribution of their components are
compatible with our results.  The \textsl{rms} scatter in intensities and
positions for components in common are 2~Jy and 30~mas, respectively.  The
positional deviations are a factor $>$3 higher than the relative positional
accurary of 4 -- 10~mas derived above.  A possible cause for the scatter
is the extensions of the maser clouds.  Bains et al. (\cite{Bains03}) found
sizes of 2 -- 4~AU for the Mira stars U~Ori and U~Her.  Adopting these sizes for
RX~Boo, angular sizes of $\sim$40~mas are predicted, much larger than the
implied positional accuracies from the Gaussian fits.  We suspect that the two
fitting methods treat the effect of blending of the extended maser clouds
differently, spatially as well as kinematically.

To search for continuum emission of the star we formed the average of
all channels without line emission, yielding \textsl{rms} fluctuations
of the order of 5~mJy/beam.  The non-detection is in agreement with
the expected blackbody flux density of $\sim$0.3~mJy from a star with
a diameter of 19~mas and a photospheric temperature of 3000~K.

\subsection{\label{id_Alignment} Alignment of the maps}
In order to identify spatial components present over two or more
observing epochs we aligned the four maps to a common origin.  The
maps of 1990 were very similar showing the majority of components to
be present in both maps and making the alignment trivial.  Several
spatial components in common were found also in the 1991 and 1992
maps, however, the components in common with the 1990 maps were
scarce.  Finally, we chose the strong spatial components G1, H1 and
I1/I2 (see Table \ref{tab:components}) to obtain an alignment of all
four maps.  After alignment these components scattered in position by
$\sim$10~mas.

Spatial coincidences among the weaker components were searched for in the
aligned maps.  After subtraction of coincident components we ended up with 25
different spatial components of which 64\% were present in at least two maps.  
Positional deviations between maps were typically of the order of 30~mas,
although in a few ambiguous cases we accepted as coincidences also components
with deviations up to 60~mas (cf. E1, H3, K1 in Table \ref{tab:components}).  
The velocity and spatial resolutions were clearly insufficient to avoid such
ambiguities.

According to their velocity, the merged spatial components were assigned to the
eleven spectral components A...K.  The merged components are listed in Table
\ref{tab:components}, which gives the spatial component (Spat. C.), the VLA
observing epoch, the line-of-sight velocity and peak flux of the spatial
component identified, the spatial offsets (Xoff and Yoff) from the adopted map
centre (see below) and the associated spectral component (Spec. C.) from Tables
\ref{tab:compRXBooA-E} and \ref{tab:compRXBooF-K} in the Appendix.

\begin{figure*}
\centering
\includegraphics[width=13.5cm]{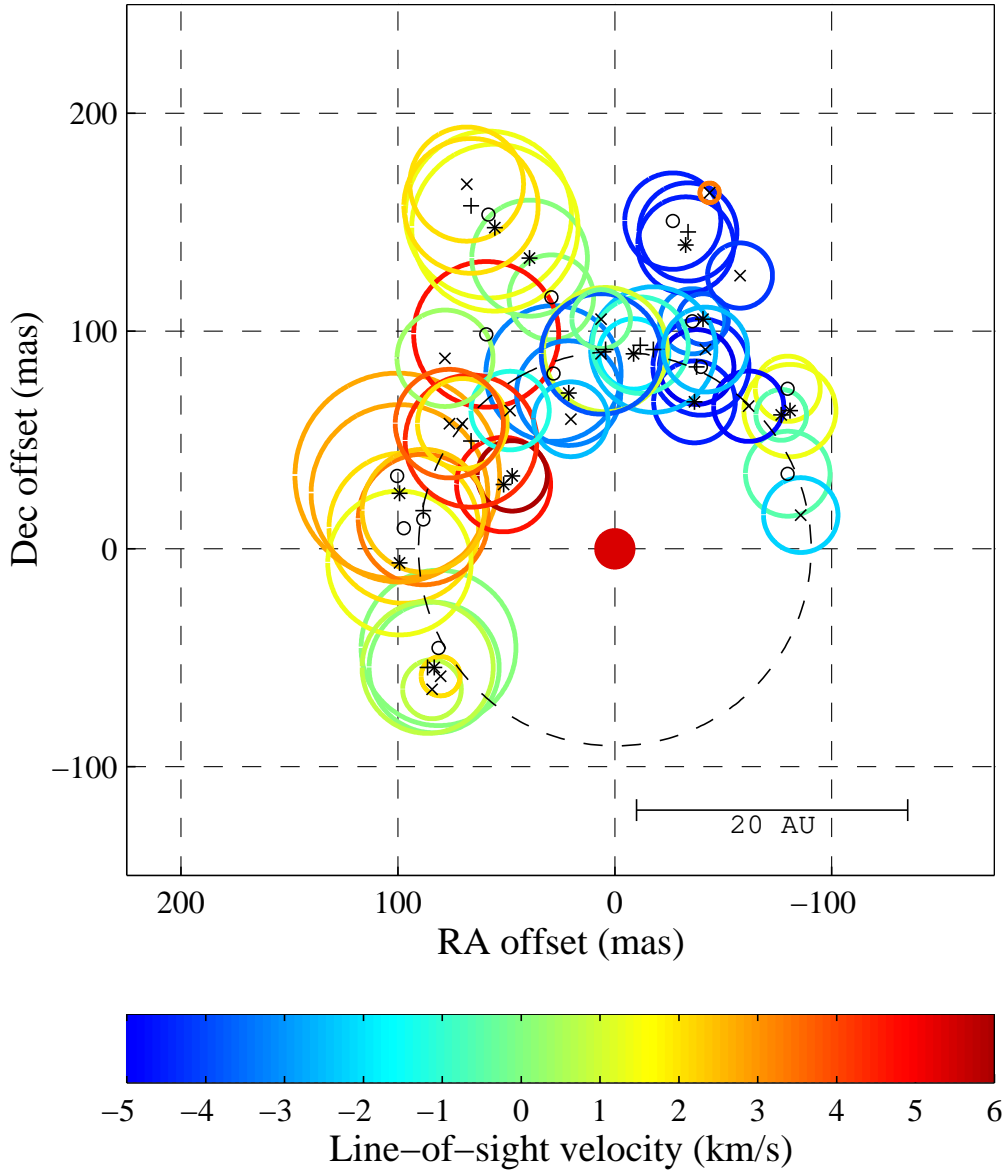}
\caption{All the maser components listed in Table \ref{tab:components}
plotted on the sky.  Each component is represented by a symbol and a
circle whose diameter is logarithmically proportional to the flux
density of the component (see text for the mathematical expression).
The epochs are represented by different symbols: 1990 February by
small circles; 1990 June by asterisks; 1991 October by plus signs;
1992 December by crosses.  The circles around the components are
colour coded according to the line-of-sight velocity of the component
($v_{\rm los}$ in the text).  (For a colour version of this figure,
see the electronic edition.)  A circle (dashed) has been fitted to the
components on the arc (see text for a detailed list of components) and
the origin has been moved to the centre of this circle (Table
\ref{tab:components}).  The filled dark red circle at the origin
symbolises the central star with a diameter of 19~mas.}
\label{fig:aligned_maps}
\end{figure*}

\subsection{\label{id_MapsCompsId} Distribution of spatial components}
The distribution of spatial components from all maps after alignment
is shown in Fig. \ref{fig:aligned_maps}.  The flux density of the
spatial components is represented by the sizes of circles centred on
the position of the components.  Due to the large range in flux (see
the third column of Table \ref{tab:components}) we have chosen a
logarithmic relation:

\begin{equation}
d = 20\,{\rm log}\,S  + 26
\end{equation}

\noindent where $d$ is the diameter of the circle in milliarcseconds
and $S$ is the flux density in Janskys.

In this map the most prominent feature is the incomplete ring-like
structure where most of the spatial components reside.  The
south-western part of the ring contained no emission during all four
observing epochs covering 3 years.  The second obvious feature is a
velocity gradient with most of the redshifted emission ($v_{\rm los} >
2$ \kms) located in the south-eastern part and most of the blueshifted
emission ($v_{\rm los} < -1$ \kms) in the north-western part of the
shell.  Although containing fewer spatial components, also the VLA map
taken in 1995 September shows the velocity gradient and is compatible
with the incomplete ring geometry.

We interpret the ring-like structure as outlining the inner boundary
of tangential masers.  Therefore, the position of the central star
should be close to the centre of the circle that fits best to the
components in this ring-like structure.  Consequently, we have made a
least-squares fit of a circle to all components except A2, F2, H1 and
I3 (of all epochs), which are situated off the ring-like structure at
about 11 o'clock and 1 o'clock.  The radius of this circle is 91 $\pm$
6~mas and the statistical errors of the centre position are of the
same order of magnitude (6 -- 7~mas).  We have then moved the origin
of the distribution from an earlier arbitrary position to this centre.
The spatial offsets Xoff and Yoff listed in Table \ref{tab:components}
are given relative to this centre.  The choice of the stellar position
is discussed further in Sect. \ref{shell size}.

\subsection{\label{CrossCorrel-InterfData} Cross correlation of single-dish
and interferometric data 1990-1992} As suspected already in
Sect. \ref{sdd_LineFitRes} the spectral components are actually blends
of two or more spatial components.  During 1990 -- 1992 many spectral
components were made up of coeval spatial components (e.g. A1 and A2
in Table \ref{tab:components}).  At later times the same spectral
component may come from emission of entirely different spatial
components.  The drift in velocity seen in component C, for example,
is accompanied by a change of the emission region within the shell
(see spatial components C1 -- C4 in Table \ref{tab:components}).  Some
components, as for example the most redshifted emission (spectral
components J and K), are only temporarily present.

While the spatial components detected in the 1990 February map were
mostly redetected in 1990 June, many of them had disappeared in 1991
October.  Also there were fewer coincidences of spatial components
between the 1991 and 1992 maps, compared to those between the 1990
maps.  This indicates that the typical lifetime of individual maser
clouds is in the range 0.5 -- 1 year.  However, spectral components J
and K could be detected in the single-dish spectra over 1.3 and 2.5
years, respectively (Table \ref{tab:compRXBooF-K}).  We conclude
therefore that there is a range of lifetimes of individual maser
clouds of 0.5 -- 3 years.  The regions within the shell able to
sustain maser emission at least from time to time, seem to have a
longer lifetime, otherwise the stability of the distribution of maser
components over $\sim$3 years cannot be understood.  As the maser
components we detected with the VLA are probably made up of several
maser clouds, the quoted lifetimes are certainly upper limits for the
lifetimes of single maser clouds.  Indeed VLBA maps taken in 1995 by
Marvel (\cite{Marvel96}) show a much larger number of spatial
components, which would have been impossible to resolve with the VLA.

\subsection{\label{RXBooPropMot} Proper motion}
Over the period of the VLA observations (2.8 years) an individual
cloud with the maximum possible velocity perpendicular to the line of
sight, i.e.  8.4~\kms\ (see Sect. \ref{3-dCSE}), would have moved
$\sim$30~mas.  With much smaller line-of-sight velocities and
lifetimes mostly considerably shorter than 3 years no detectable
proper motions of spots within the \water\ maser shell were expected.
However, the proper motion of the whole system was likely to be
detected.

To test for the presence of proper motion of RX~Boo between the 1990
and 1992 VLA observations we searched for spatial maser components
present on both occasions in maps obtained from the calibrated
dataset, in which phase calibration was made with the use of the phase
calibrator J1407+284, i.e. before self calibration was applied.  Only
spatial components G1 (0.2 -- 0.8~\kms) and H1 (1.4 -- 2.1~\kms) could
be identified unambiguously in the 1990 and 1992 maps, while they
could not be recovered in the 1991 October maps, due to decreased
resolution (BnA configuration) and lower signal-to-noise ratio.  To
measure the positions of the components we used the AIPS task IMFIT.
Positional differences between the epochs were averaged for both
components yielding $\Delta \alpha = +49 \pm 11$~mas and $\Delta
\delta = -159 \pm 6$~mas.  These differences are attributed to a
common movement of maser sites and central star.  The proper motion of
RX~Boo was then $\Delta \alpha = +18 \pm 4$~mas/yr and $\Delta \delta
= -59 \pm 2$~mas/yr.  For comparison, the proper motion determined by
Hipparcos is $\Delta \alpha = +21.74 \pm 0.75$~mas/yr and $\Delta
\delta = -49.70 \pm 0.69$~mas/yr (Perryman et al. \cite{Perryman97}).
Using the geometry suggested in Fig. \ref{fig:aligned_maps} we
determined the likely position of RX~Boo for 1990 February:
\recta{14}{24}{11}{52} $\pm$ 0\fs01, \decli{+25}{42}{13}{7} $\pm$
0\farcs1 (J2000).

\section{\label{CSE} The circumstellar envelope of RX~Boo}
\subsection{\label{2-dCSE} The projected structure}
In an isotropically expanding envelope the relationship between the
line-of-sight velocity and the projected distance from the star is an
elliptical one:

\begin{equation}
\left( \frac{r_{\rm p}}{r}\right) ^2 + \left( \frac{v_{\rm los} - v_*}{v}\right)
^2 = 1
\end{equation} 

\noindent where $r$ is the distance, $r_{\rm p}$ is the projected
distance, $v$ is the expansion velocity, $v_{\rm los}$ is the
line-of-sight velocity and $v_*$ is the line-of-sight velocity of the
star (+1 \kms).  Therefore, when plotting ($v_{\rm los} - v_*$)
against $r_{\rm p}$ for spatial components in such an envelope where,
in addition, all the masers are situated in a thin spherical shell
(like, for example, OH masers in an OH/IR star), the points will be
close to (half) an ellipse with its centre at the origin and its axes
coinciding with the coordinate axes.  However, if the components are
spread out within the envelope, as is obviously the case for the
\water\ masers in RX~Boo, the plot will become a scatter diagram
without any discernable ellipse.  On the other hand if, for example,
there is a minimum distance from the star where masers can exist and
the component density is high, there will be a void in the lower part
of the diagram with a boundary that will fit well to an ellipse.

\begin{figure}
\centering
\includegraphics[width=8.75cm]{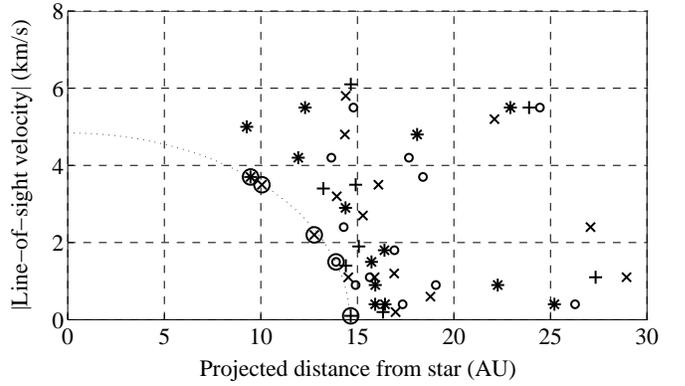}
\caption{The absolute value of the line-of-sight velocity relative to
that of the star (1~\kms) plotted against the projected distance of
all the spatial components listed in Table \ref{tab:components} from
the origin (the central star).  The epochs are represented by the same
symbols as in Fig.  \ref{fig:aligned_maps}.  Five of the components
(marked by surrounding circles) fit accurately to an ellipse (dotted
line) centred on the origin and with its axes along the coordinate
axes of the diagram.  If the expansion of the stellar envelope is
isotropic, this suggests that there exists an inner distance (14 --
16~AU) at a certain expansion velociy (4 -- 6~\kms) where water vapour
masers first appear in the expanding gas.}
\label{fig:aligned_CSE}
\end{figure}

In order to test this method of analysis on RX~Boo we plotted the
absolute value of the line-of-sight velocity of the spatial components
in Table \ref{tab:components} versus the projected distance
($\sqrt{\rm{Xoff}^{2} + \rm{Yoff}^{2}}$) from the star
(Fig. \ref{fig:aligned_CSE}).  By choosing $\vert v_{\rm los} -
v_*\vert$, the density of points in the plot is doubled and the chance
of seeing a possible boundary is increased.  The fact that no masers
are seen close to the star in Fig. \ref{fig:aligned_maps} and no
spatial components are found at projected distances smaller than
$\sim$9~AU in Fig. \ref{fig:aligned_CSE} are solid proof that there
are no masers radiating in the radial direction (`radial masers') in
the circumstellar envelope (CSE) of RX~Boo.  Most of the masers in
fact are radiating in directions that probably are close to tangential
(`tangential masers').  They can be seen forming a small cluster
around about 15~AU (94~mas) from the star at velocities close to zero
in Fig. \ref{fig:aligned_CSE}, corresponding to several of the
components along the dashed circle in Fig. \ref{fig:aligned_maps} with
radius of 91~mas.  However, there is a mixture of line-of-sight
velocities among the components close to this circle, which is clearly
seen in Fig.  \ref{fig:aligned_CSE}, namely the roughly vertical
extension of the above mentioned cluster towards higher (in an
absolute sense) line-of-sight velocities.  There are about 10
components spread out at larger distances ($r_{\rm p} \gtrsim 20$~AU)
from the star visible in both figures.  They must be located in the
gas that has been carried out to these distances in the outflow.  The
fact that the line-of-sight velocities are not very high suggests that
these masers too are more or less tangential masers.

In Fig. \ref{fig:aligned_CSE} a quarter of an ellipse has been drawn.
This represents a least-squares fit to five of the data points in the
diagram, marked by circles around the symbols.  The best-fit values of
its axes are $14.6 \pm 1.4$~AU and $4.8 \pm 0.9$~\kms.  We interpret
this well-fitting ellipse as an indication of the existence of a
minimum distance for water vapour masers of 14 -- 16~AU in the CSE of
RX~Boo.  At this distance the expansion velocity probably is in the
range 4 -- 6~\kms.

\begin{figure}
\centering
\includegraphics[width=8.75cm]{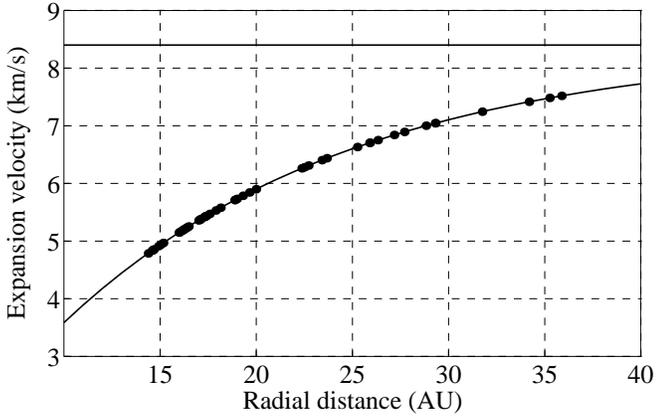}
\caption{The model expansion velocity as a function of the distance from the
star (Eq. \ref{eq:exp_model}).  It approaches the terminal velocity (8.4~\kms)
asymptotically (horizontal straight line).  The dots along the expansion law
curve symbolise the distances and velocities of the maser components.}
\label{fig:expansion_law}
\end{figure}

\begin{figure}
\centering
\includegraphics[width=6.75cm]{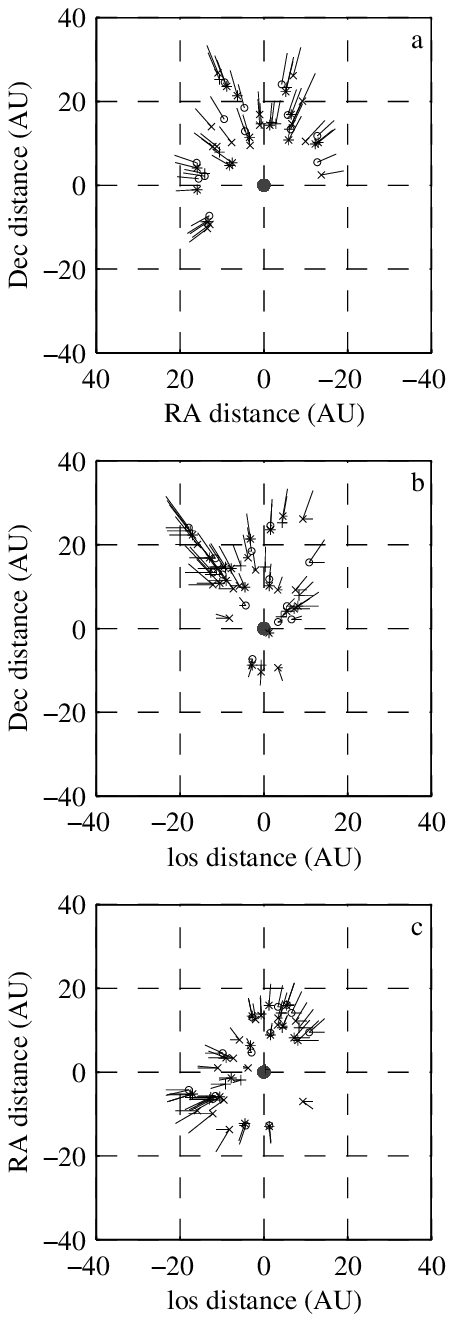}
\caption{The three-dimensional distribution of maser components and
their associated velocities according to the expansion model given by
Eq.  (\ref{eq:exp_model}).  The distribution is seen from three
cardinal directions: on the sky (a), `from the side' (b) and `from
above' (c).  The maser components are represented by the same symbols
as earlier.  In addition the projected expansion velocities and the
line-of-sight velocities are shown in the same scale in units of \kms\
as the distance scale in units of AU.}
\label{fig:envelope_3D}
\end{figure}

\subsection{\label{3-dCSE} The 3-D structure}
By assuming a plausible expansion law it is possible to get a rough
idea of the three-dimensional structure of the envelope.  Supported by
theoretical models of radiation-pressure-driven mass loss from
late-type stars (cf.  Goldreich \& Scoville \cite{Goldreich76} and
Bowen \cite{Bowen88}) we assume that the expansion law is exponential
in nature and leading asymptotically to the final velocity, $v_{\rm
f}$:

\begin{equation}
v = v_{\rm f} \left\lbrace  1 - {\rm exp}\,\left[ -k (r - r_0)\right]
\right\rbrace 
\label{eq:exp_model}
\end{equation} 

\noindent where $k$ is a scaling factor and $r_0$ is a radial shift.

Eq. (\ref{eq:exp_model}) thus contains three constant parameters
($v_{\rm f}$, $k$ and $r_0$) and in order to find plausible values for
them we need to estimate the coordinate values for three points on the
expansion law.  The value of $v_{\rm f}$ (the expansion velocity at $r
= \infty$) can be estimated from (sub)-millimetre CO data, because it
is believed that CO emission originates from gas throughout the CSE to
very large distances from the star.  We have two estimates of the
terminal expansion velocity available: 9.3~\kms\ (Olofsson et
al. \cite{Olofsson02}) and 7.5~\kms\ (Teyssier et al.
\cite{Teyssier06}).  Both these values are the results of complicated
model fits involving several rotational CO lines.  Therefore we are
unable to judge the correctness of these results and we simply assume
that they reflect the uncertainty inherent in the two models that give
rise to them.  Therefore we adopt the mean value of 8.4~\kms\ for the
value of the final expansion velocity.

In addition, we have a rough idea of the expansion velocity at one
distance from the star (4.8~\kms\ at 14.6~AU) from the analysis of the
diagram in Fig. \ref{fig:aligned_CSE}.

Finally, we assume that the expansion law is purely exponential all
the way down to the photosphere, which is stationary, and where the
expansion velocity is zero.

Thus the values of the parameters $k$ and $r_0$ can be determined from
the condition that the stellar wind passes the positions $(r_1,v_1) =
(14.6,4.8)$ and $(r_2,v_2) = (1.5,0)$:

\begin{equation}
k = \frac{{\rm ln}\,v_{\rm f} - {\rm ln}\,(v_{\rm f} - v_1)}{r_1 - r_0}
\end{equation} 

\noindent and

\begin{equation}
r_0 = r_2 = r_*
\end{equation} 

\noindent resulting in $k = 0.065\,{\rm AU}^{-1}$ and $r_0 = 1.5\,{\rm AU}$.

Now, the observables are the line-of-sight velocity relative to the star
$(v_{\rm los} - v_*)$ and the projected radial distance $r_{\rm p}$.  Their
relations to the physical expansion velocity $v$ and radial distance $r$ are:

\begin{equation}
v_{\rm los} - v_* = v\,{\rm sin}\,\theta
\label{eq:v_los}
\end{equation} 

\noindent and

\begin{equation}
r_{\rm p} = r\,{\rm cos}\,\theta
\label{eq:r_p}
\end{equation} 

\noindent where $\theta$ is the angle between the normal to the line
of sight and the radius from the star to the maser.  Therefore, for
every maser component the angle $\theta$ can be determined from the
equations (\ref{eq:exp_model}), (\ref{eq:v_los}) and (\ref{eq:r_p}).

In Fig. \ref{fig:expansion_law} the expansion law is shown graphically
together with the maser component values determined by the model and
represented by filled circles.  They are shown mainly as an indication
of the distance range for the maser components ($\sim$14 to
$\sim$36~AU).

The 3D-distribution of the maser components and their associated
velocities is shown in Fig. \ref{fig:envelope_3D} as seen from three
different directions: on the sky (a; similar to
Fig. \ref{fig:aligned_maps} but with the projected expansion
velocities added), `from the side' (b) and `from above' (c).  The
dominance of tangential masers is seen clearly in
Fig. \ref{fig:envelope_3D} (b) and (c).  The obliquity of the `maser
arc' seen in Fig.  \ref{fig:envelope_3D} (c) is in accordance with the
preference of negative line-of-sight velocities on the lower-RA side
of the star and of positive line-of-sight velocities on the higher-RA
side, which is evident in Fig.  \ref{fig:aligned_maps}.

However, one should keep in mind that an observer seeing RX~Boo from
directions deviating considerably from the geocentric direction would
see a different set of maser components.  Thus,
Fig. \ref{fig:envelope_3D} is merely showing the three-dimensional
positions of the maser components that we see from earth.  Most of
them probably would be too weak to be detected from the directions
represented by Figs. \ref{fig:envelope_3D} (b) and (c).  Of course a
similar situation prevails for our present point of observation: there
may be maser spots that {\it we} do not see because their emission may
happen to be beamed in the wrong direction from our vantage point.

Having adopted an expansion law it is of interest to investigate the
associated time scale for the gas to travel through the envelope.
Thus, we have integrated the function $1/v$, where $v$ is described by
Eq. (\ref{eq:exp_model}), along the radius $r$, avoiding $r = r_0 =
1.5$~AU where $v = 0$.  We find that it takes $\sim$16.5 years for gas
to travel through the zone (14 -- 36~AU) where the \water\ masers
reside.

\section{\label{SdDataSVPeg} Single dish data of SV~Peg}
\begin{figure}[h]

\begin{minipage}[t]{8cm}
\includegraphics[width=9.5cm]{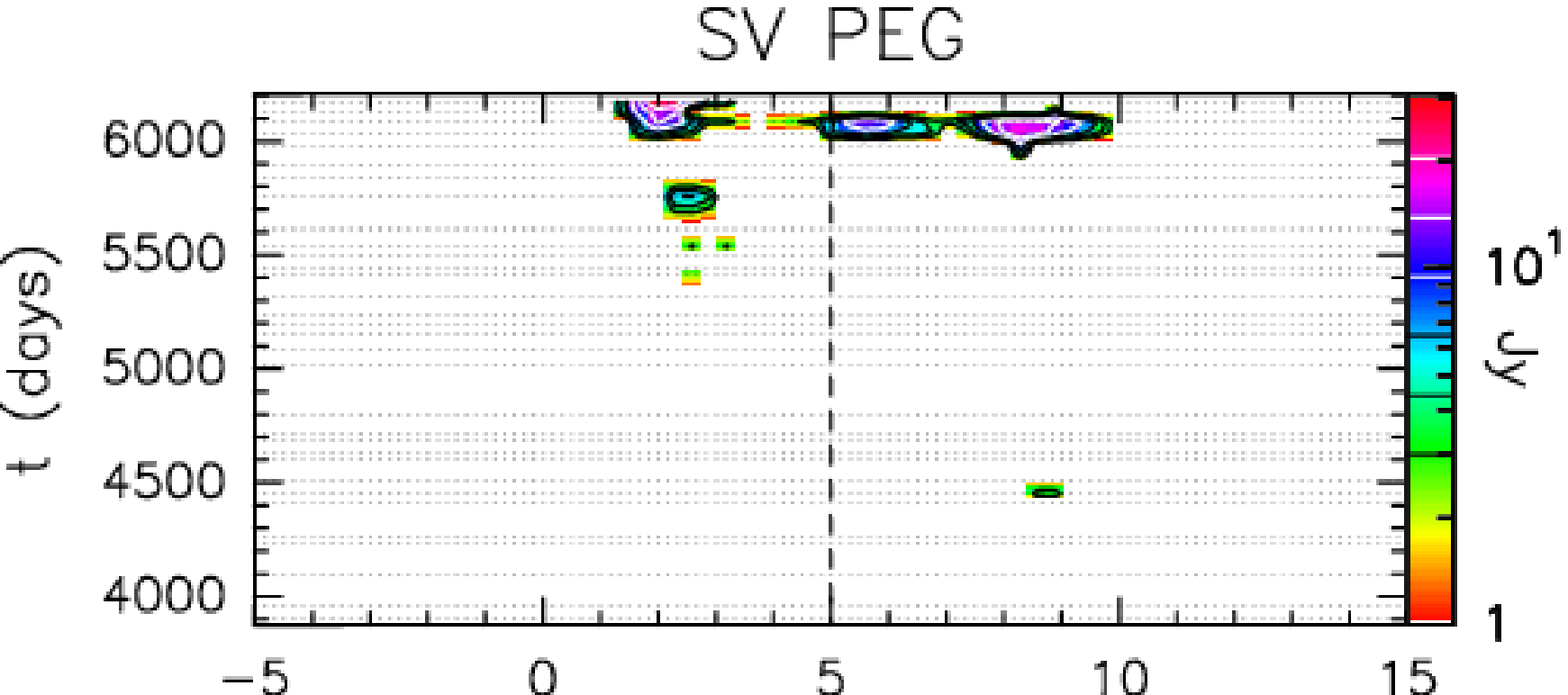}  
\includegraphics[width=9.5cm]{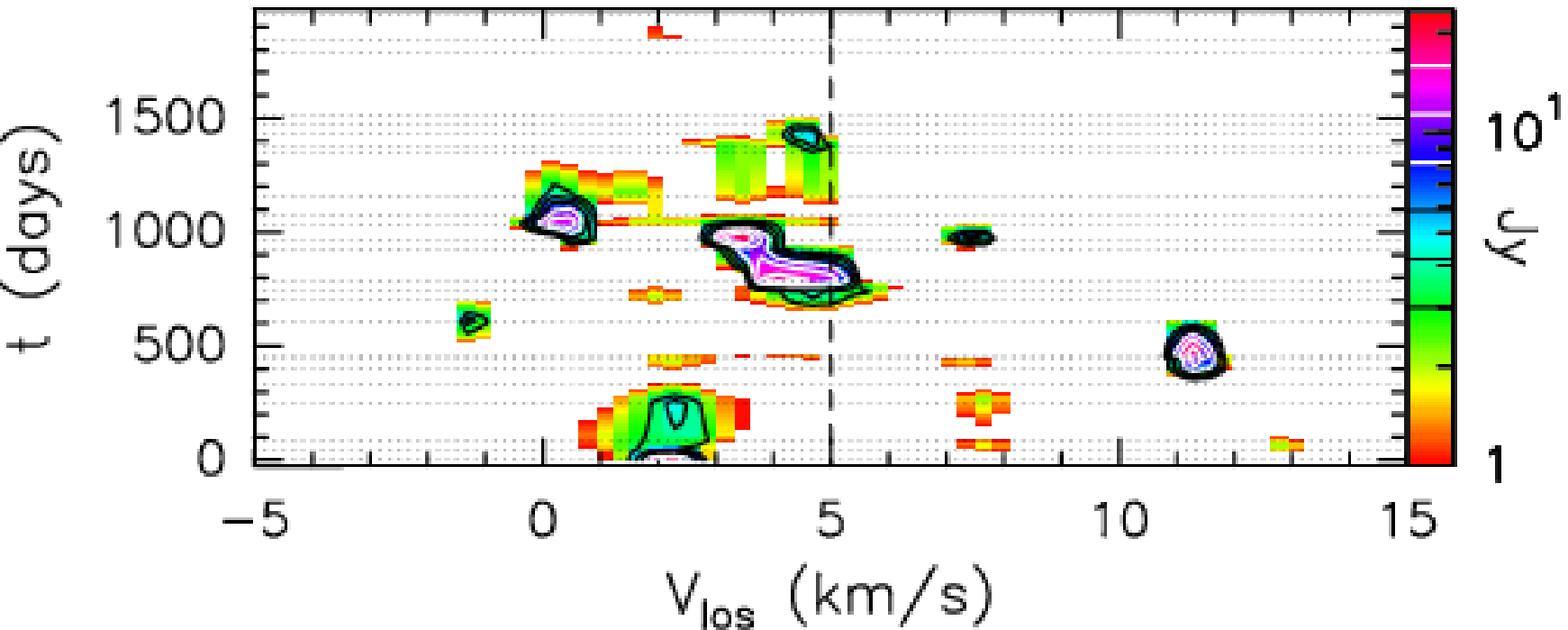}  
\end{minipage}

\caption{Flux density of the \water\ maser of SV~Peg as a function of
velocity and time.  The time axis covers the years 1990 -- 1995
(bottom) and 2000 -- 2007 (top).  The time stamp $t=0$ corresponds to
1990 February 17 (J.D. 2447933).  The dotted horizontal lines mark the
actual observations averaged in 4-day intervals.  Flux densities in
between the observations were interpolated linearly.  The dashed
vertical line marks the stellar systemic velocity (5~\kms) as given by
Kerschbaum \& Olofsson (\cite{Kerschbaum99}).  The spectra were
resampled to a resolution of 0.3~\kms.  Contours in the lower panel
are drawn at levels 3.00, 4.19, 5.86, 8.18, 11.43, 15.97~Jy
(logarithmic scale) and in the upper panel at 3.00, 4.40, 6.45, 9.45,
13.86, 20.32~Jy. These levels are indicated by black and white lines
in the flux scales along the right side of the two maps.  (For a
colour version of this figure, see the electronic edition.)}
\label{fig:v_t_diagram_svpeg}
\end{figure}

The SRV SV~Peg was selected for monitoring because of the detection of
several maser features in 1990 not seen in previous spectra.  We
suspected that this star may show different variability properties
compared to RX~Boo.  The \water\ maser was detected for the first time
in 1974 by Dickinson (\cite{Dickinson76}), showing a strong line of
$\sim$60~Jy at a line-of-sight velocity of 2~\kms.  The maser was
re-detected by Engels et al. (\cite{Engels88}) in 1984 showing two
lines at 2 and 9~\kms\ with similar flux densities of $\sim$12~Jy.
The development of the maser emission since 1990 is shown in Fig.
\ref{fig:v_t_diagram_svpeg}.  The maser was observed also by Benson \&
Little-Marenin (\cite{Benson96}) in 1990 and by Takaba et
al. (\cite{Takaba94}) in 1991 yielding line profiles that were similar
to ours.

\begin{figure*}
\centering
\resizebox{16cm}{!}{\includegraphics{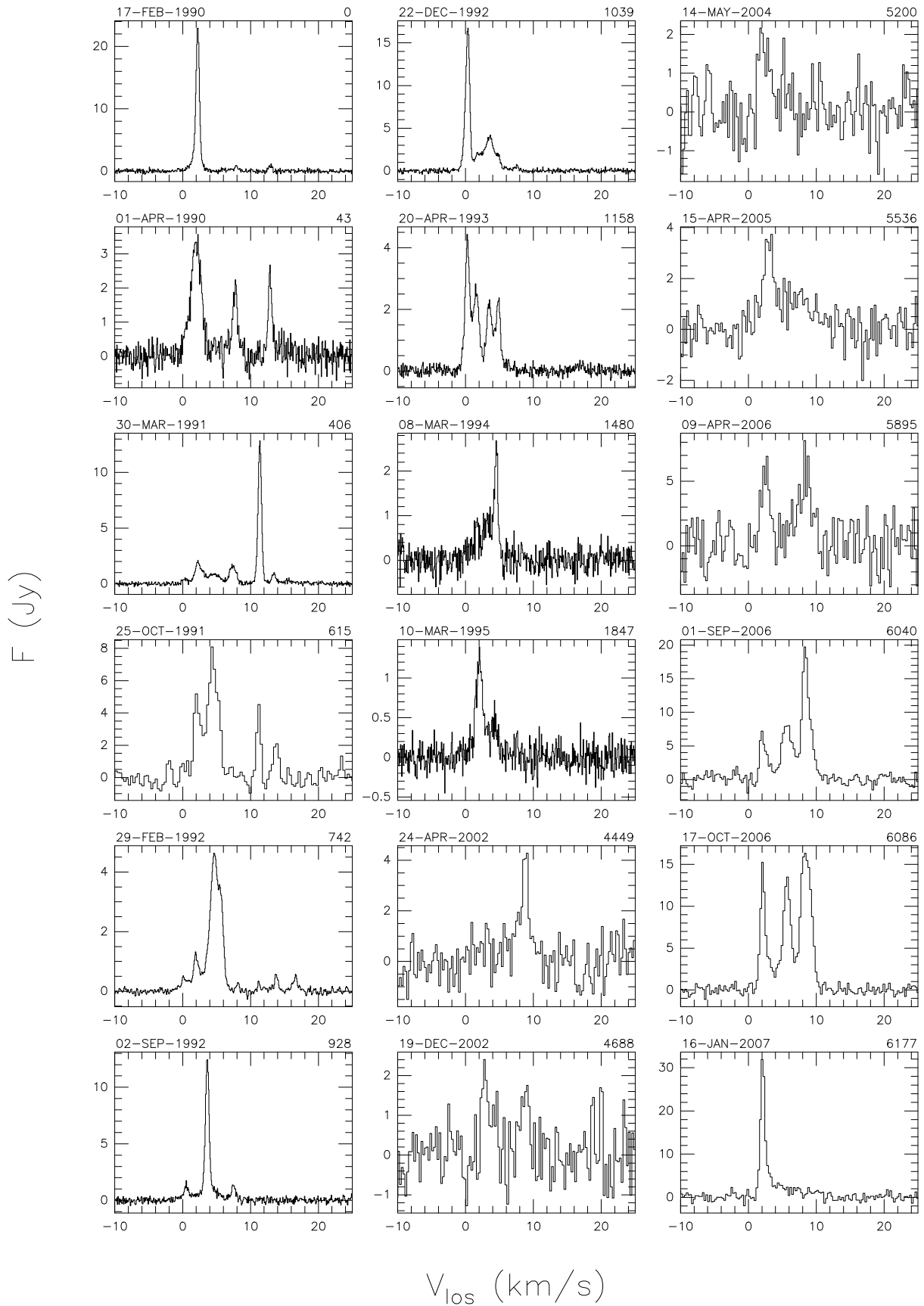}}
\caption{Sample \water\ maser spectra of SV~Peg showing typical profiles
starting in 1990 and ending in early 2007.  Note the varying flux density
scale.}
\label{fig:sample_spectra_svpeg}
\end{figure*}

Our monitoring of SV~Peg covered two periods 1990 -- 1995 and 2000 --
2007.  Sample spectra are shown in
Fig. \ref{fig:sample_spectra_svpeg}.  Until 1992 the integrated maser
emission was dominated by different single maser features with flux
densities $>$10~Jy.  Such features were not observed after 1992 and
the integrated maser emission correspondingly declined abruptly during
the first period.  Also after the monitoring restart in the year 2000
the maser emission remained weak or absent until mid 2006.  All
spectra taken are shown in the Appendix
(Figs. \ref{fig:spectra_svpeg_1} -- \ref{fig:spectra_svpeg_4}).

The \water\ masers of SV~Peg appeared non-simultaneously over a
velocity interval of 0 to 16~\kms.  Actually, maser emission in the
outer parts of this interval ($v_{\rm los} < 2$ and $>$10 \kms) was
detected only during 1990 -- 1993.  The velocity range of the maser in
general would be severely underestimated, if its determination were
based on individual spectra alone.

Compared to RX~Boo the spectral profiles of SV~Peg are different.
While for RX~Boo generally maser emission was detected over the full
velocity range and the individual features are strongly blended, the
SV~Peg maser features are usually well separated.  This is in part due
to the weakness of the SV~Peg features, of which mostly only the
brightest were detected.  But even during the bright phases around
1991 and near the end of 2006 most features are well separated
suggesting that the number of individual maser features is lower in
SV~Peg compared to RX~Boo.

Like we did for RX~Boo (cf. Sect. \ref{sdd_LineProfAna}) we made a
line profile analysis for SV~Peg by fitting multiple Gaussian
components to the maser profiles.  As before, we assumed that maser
features in adjacent spectra with velocity differences $\la$0.5~\kms\
belong to a unique physical maser component in the shell of SV~Peg.
All features were assigned to 9 maser spectral components labelled A,
B,..., H.  These components are listed in Tables
\ref{tab:compSVPegA-E} and \ref{tab:compSVPegF-I} in the Appendix. For
a description of the tables we refer to Sect. \ref{sdd_LineProfAna}.

Except for features B and C we find that all spectral components are
fairly well separated in velocity.  Therefore the central velocities
show in general a small scatter ($\la$0.5~\kms).  Larger drifts in
velocity ($\la$1.5~\kms) are seen in component B indicating the
presence of blends.  Almost all features dominated the spectral
profile at some epochs.  The typical timescales for major intensity
variations of individual components were 6 -- 12 months.  A
particularly well defined example for the temporary appearance of a
feature is component F ($v = 11.3$~\kms\ in
Fig. \ref{fig:v_t_diagram_svpeg}; Fig.
\ref{fig:sample_spectra_svpeg}, Date: 1991 March 30), which was
detected for 11 months only.  Noticeable is also the spectral feature
H at +16.4~\kms, which was observed only twice in the beginning of
1992 (cf. Fig \ref{fig:sample_spectra_svpeg}, Date: 1992 February
29). We consider this feature as real, although it was detected at the
$2-3 \sigma$ level only.

Summarizing, the case of SV~Peg shows that the \water\ maser is
composed of at least 8 components, where each of them dominated the
profile for several months.  At any given epoch, maser emission was only
seen for a limited fraction (a few \kms) of the whole velocity span liable to
exhibit a maser.  Thus the non-detection of \water\ masers in many SRVs might
be merely a temporary effect, and the detection rate of 20\% obtained by
Szymczak \& Engels (\cite{Szymczak95}) among SRVs using two observing epochs
separated by four months might be raised considerably with observations
separated by years.

\begin{figure}
\centering
\includegraphics[width=7.5cm,angle=-90]{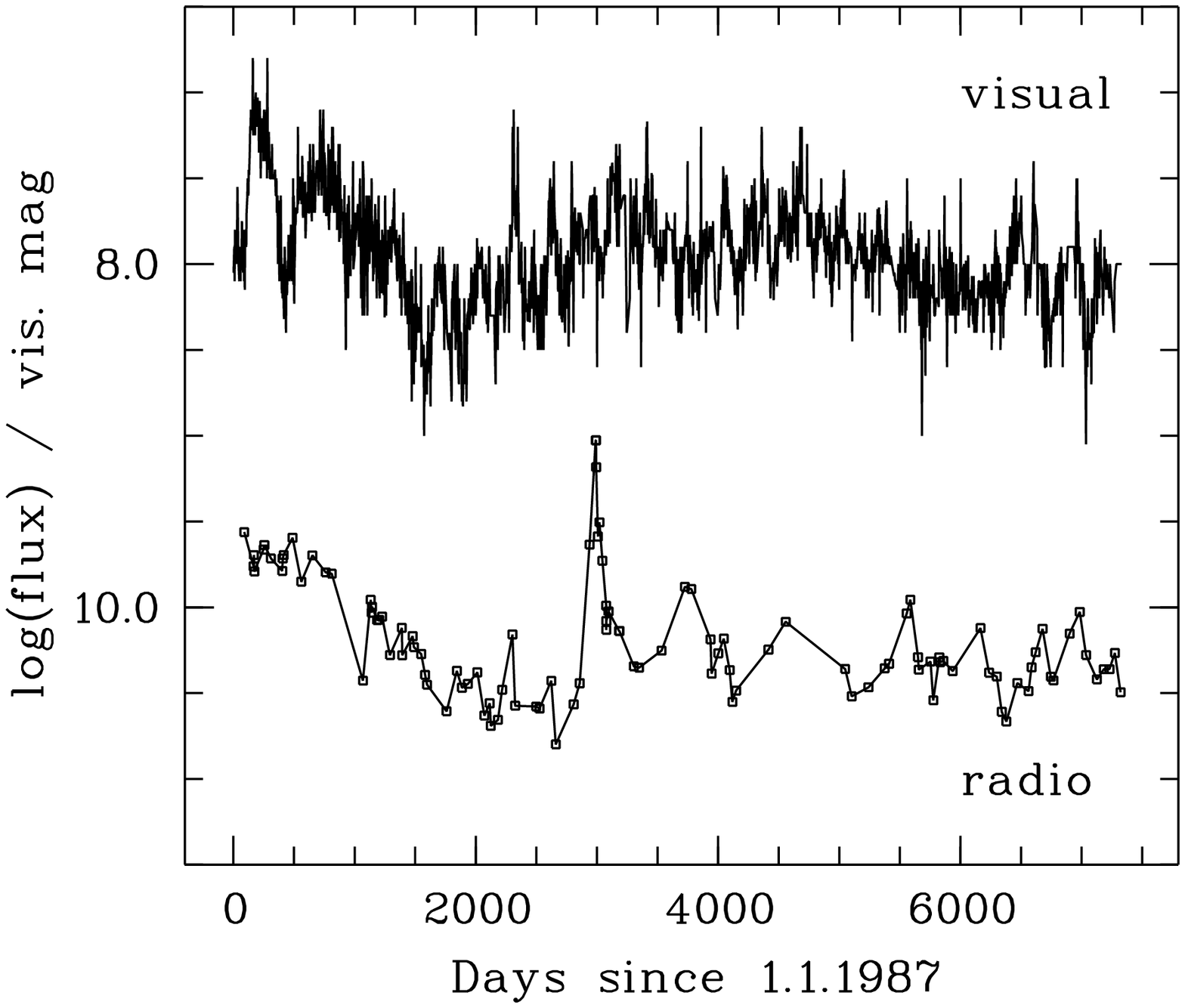}
\caption{Optical AAVSO lightcurve and the lightcurve of the integrated \water\
maser emission for RX~Boo during the years 1987 -- 2007.  The optical data are
visual magnitudes and the integrated radio flux has been offset by an arbitrary
amount.}
\label{fig:lightcurve}
\end{figure}

\section{\label{correlation} Radio and optical variability in SRVs}
It has been reported that the variability of the \water\ maser
emission of RX~Boo is weakly correlated with the optical lightcurve,
and that different maser features react to optical variations with
different time delays (Hedden et al. \cite{Hedden91}).  The optical
and the radio lightcurve during the period 1987 -- 2007 is shown in
Fig. \ref{fig:lightcurve}.  The optical data given as visual
magnitudes were obtained from the American Association of Variable
Star Observers (AAVSO\footnote{http://www.aavso.org/}; Hedden
\cite{Hedden07}).  We rebinned the data into three-day intervals and
interpolated over intervals without observations.  The integrated
radio fluxes were taken from Table \ref{tab:compRXBooA-E} in the
Appendix.

A cross correlation between the optical and the radio lightcurves has
been performed with delays of up to 2000 days ($\sim$6 years)
following the suggestion of Lekht et al. (\cite{Lekht05}) that the
maser emission may respond to the passage of shock waves formed during
a previous bright phase of the star.  No significant signal has been
found for any delay.  However, the course of the lightcurves in Fig.
\ref{fig:lightcurve} gives the impression that a weak correlation
without significant delay is present with a timescale of several
years.  In particular, both lightcurves show a minimum at 2000 days
(1992 -- 1993).

We searched also for particularly bright phases in the optical
lightcurve before 1995, which could have caused the strong radio flare
in 1995 March ($t \approx 3000$).  Such optically
bright phases occurred irregularly every few years, while the strength of the
flare was unique during our observing period.  It was therefore not possible to
link the radio flare with a particular part of the optical lightcurve.

We made a similar analysis of the correlation between optical and
maser variations for SV~Peg with negative result.  Lekht et
al. (\cite{Lekht99}) observed the maser variations of another SRV
RT~Vir over the period 1985 -- 1998 and found large amplitude variations
including two flares with flux densities $>$1000 Jy.  Although they do
not discuss the optical behaviour of the star during this time,
available AAVSO optical data do not show any magnitude variations
larger than the usual $\pm0.5$~mag around the mean $m_{\rm vis}
\approx 8.9$.  We conclude therefore that there is basically no
correlation between \water\ maser radio data and optical data for
SRVs.

\section{Discussion}
The long-term monitoring programme of the two SRVs RX~Boo and SV~Peg
allows us to draw a few general conclusions on the variability
properties of their \water\ masers.

\begin{itemize}
\item Maser emission occurs over a fixed velocity range $\Delta v$,
but not necessarily at all velocities simultaneously.  There is no
evidence that the velocity range varied systematically with time.
Occasionally, maser emission outside the regular velocity interval may
appear.
\item The individual maser features vary irregularly and any
correlation with the optical variability of the stars is negligible
compared to these irregular variations.  The intensity of a particular
feature characterises this feature alone and is not representative for
the maser emission as a whole.  The integrated intensity obtained over
the velocity range $\Delta v$ is a more suitable measure to describe
the maser ``activity'' at a given epoch.
\item The lifetime of individual maser components is of the order 1
year, much less (5 -- 10\%) than the time needed for a mass element to
cross the \water\ maser shell.
\item The distribution of the maser spatial components in the
circumstellar shell of RX~Boo has been found to be remarkably stable
over the range of 6 years (1990 -- 1995).  The distribution showed a
persistent asymmetry, with emission coming always from the NE part and
never from the SW part of the shell. The likelihood to find a new
maser component in the shell, is higher in those parts, where a maser
has been observed previously.
\end{itemize}

\subsection{RX~Boo velocity range}
During the monitor programme the \water\ maser velocity range of
RX~Boo never exceeded the interval $-5 \la v_{\rm los} \la +5$~\kms\
(Fig. \ref{fig:v_t_diagram_rxboo}).  This is well within the
velocity range defined by CO emission extending from --8.3 to
+10.3~\kms\ according to Olofsson et al.  (\cite{Olofsson02}) or --6.5
to +8.5~\kms\ according to Teyssier et al.  (\cite{Teyssier06}),
although the \water\ interval is not symmetric with respect to the
stellar velocity of $v_*= + 1$~\kms\ as defined by the CO lines (see
also Sect. \ref{asymmetry}).  Larger velocity ranges have been
observed in the past (Engels et al. \cite{Engels93}).  Before 1983 the
strongest feature was always detected at $v_{\rm los} \approx 6$~\kms\
(see the compilation in Engels et al.  \cite{Engels88}), which is well
beyond the red edge of the current velocity interval.  On one
occasion, in 1982.2, emission was detected even over an interval $-11
< v_{\rm los} < +9$~\kms\ (Johnston et al. \cite{Johnston85}; de Vegt
et al. \cite{deVegt87}), almost twice the range observed in 1980
(Dickinson \& Dinger \cite{Dickinson82}) and after 1990.

Except for this one occasion, the larger maser velocity range can be
explained within our model (cf. Sect. \ref{CSE}), as line-of-sight
velocities $\vert v_{\rm los} - v_*\vert$ up to the adopted final
outflow velocity $v_{\rm f} = 8.4$~\kms\ are permitted.  According to
Eq. (\ref{eq:v_los}) a larger $\vert v_{\rm los} - v_*\vert$ would
require a larger outflow velocity $v$ at the maser site and/or a
larger $\theta$.  Following Eq. (\ref{eq:r_p}) the latter implies a
smaller projected distance $r_{\rm p}$ of the maser site.  Because the
outflow velocity increases outwards, larger $\vert v_{\rm los} -
v_*\vert$ are achieved at larger radial distances from the star and
RX~Boo may have at some time sustained \water\ maser emission in the
outer reaches of the shell. On the other hand, in the case of a
smaller $r_{\rm p}$, maser components may have been present within the
inner projected ring drawn in Fig. \ref{fig:aligned_maps}, in which no
emission was detected after 1983.

Maser emission at larger radial distances might have been possible due
to a temporal increase in luminosity of the star allowing the rise of
the temperature in the shell or due to the passage of a subshell with
increased density.  In the former case the minimum temperature
required to excite \water\ masers will be achieved at larger radii
than normally.  However, the mean optical brightness of the star was
not brighter before 1983 than afterwards, making this explanation
unlikely.  In the latter case the size of the shell having the minimum
density required to excite \water\ masers might have been increased
for some time.

On the other hand, maser sites with smaller $r_{\rm p}$ could have
been triggered by local enhancements of the gas density or by
temporary increases of gain lengths.  Such sites could be located at
any distance from the star, but we would be seeing a maser that has
become more 'radial'.  All these possible causes can be probed during
future periods, if RX~Boo should again display an extended velocity
range, by searching for an increase of the outer shell radius, or by
searching for maser components within the inner ring.

At odds with our hypothesis is the 1982.2 observation, where the
blueshifted emission exceeded the blue edge of the CO velocity range
by 3 -- 4~\kms.  This observation apparently has recorded a faster than
usual outflow, crossing the \water\ maser shell by 1982.

\subsection{SV Peg velocity range}
The observed velocity interval in SV~Peg is 0 to +16.5~\kms.  This is
to be compared to the CO emission velocity interval --2.5 to
+12.5~\kms\ (Olofsson et al. \cite{Olofsson02}), which implies a
stellar radial velocity $v_* = + 5$~\kms\ and a final outflow velocity
of 7.5~\kms.  \water\ maser emission at the most blue-shifted
velocities $v_{\rm los} < 0$~\kms\ was never observed, while emission
at the redshifted final outflow velocity was seen in 1990 -- 1992
(spectral components F and G).  This implies that the final outflow
velocity is already reached within the \water\ maser shell.

As in the case of RX~Boo before, also SV~Peg showed emission outside
the CO velocity range.  Spectral feature H at +16.4~\kms\ (outflow
velocity 11.4~\kms), which was seen only in 1992, might have traced an
outflowing mass segment with temporarily higher speed than is normally
achieved in the CSE of SV~Peg.

\subsection{\label{shell size} \water\ maser shell size of RX~Boo}
Unless there are 100\% tangentially beaming maser clouds at the outer
periphery of the shell, the observed diameter $\phi_{\rm p}$ of the
\water\ maser shell will be smaller than the true diameter $\phi$
because of projection effects.  As the most distant maser regions are
usually weak, the observed diameter is certainly depending also on the
sensitivity of the observations.  Finally the diameter will depend on
the assumptions made with respect to the location of the star relative
to the observed distribution of maser clouds.

For 1982 Johnston et al. (\cite{Johnston85}) give a projected diameter
$\phi_{\rm p} = 200$~mas compared to $\phi_{\rm p} = 140$~mas in 1983
(Lane et al. \cite{Lane87}) and $\phi_{\rm p} = 150$~mas in 1985
(Bowers et al.  \cite{Bowers93}).  This apparently confirms the case,
that the large velocity range in 1982 was due (at least in part) to a
more extended \water\ maser shell.  However, Colomer et
al. (\cite{Colomer00}) determined $\phi_{\rm p} = 230$~mas for 1990
June and we fitted $\phi_{\rm p} = 180$~mas for the bulk of the
emission features observed in 1990 -- 1992, however, omitting outlying
components at radial distances $100 < r_{\rm p} < 160$~mas for the
adopted stellar position.  Usually the stellar position is adopted in
the centre of the maser spot distribution, which leads to the smallest
projected diameters possible.  Our proposal for the position of the
star relative to the masers is offset by $\sim $50~mas from the
centroid of the maser emission, which leads to an unprecedented high
$\phi_{\rm p} = 320$~mas.  Thus the implied projected shell diameter
is 51~AU.  According to our model the component with the highest
radial distance from the star has $r \sim 36$~AU, which implies that
the true angular diameter of the shell might be as high as $\phi =
440$~mas.  So far, a variation of the \water\ maser shell size of
RX~Boo cannot be determined, because the differences in the diameters
derived by different authors are still dominated by data quality and
model assumptions.

\subsection{The connection of maser variations with stellar variations}
Long-term monitoring of \water\ maser emission has been made by Lekht
et al.  (\cite{Lekht05} and references therein) for various SRVs, Mira
variables, and RSGs.  For the objects with large-amplitude variations
they find that the integrated \water\ maser emission is correlated
with the optical emission of the star.  The radio emission lags behind
the optical one by a few tenths of the period.  They generally favour
a model, in which the variations of the \water\ masers are due to
passages of shock waves generated on the stellar surface.  The long
travelling times ($>$10 years), however, make it difficult to find
correlations between the optical and radio lightcurves over such
delays.

However, the luminosity variations of the star certainly affect the
excitation conditions in the whole maser shell.  The response of
\water\ masers to the light variations in general is well documented
for Mira variables (e.g. Lekht et al. \cite{Lekht01}; Brand et
al. \cite{Brand02}) and OH/IR stars (Engels et al.  \cite{Engels86}).
For the SRVs studied here the variations of the integrated radio flux
are largely caused by brightness variations of single maser features,
and not by variations at all velocities simultaneously.  We argue
therefore that the \water\ maser variations in SRVs are dominated by
local effects, such as variations of the velocity coherence length due
to turbulence.  The influence of the stellar luminosity variations on
the maser variation is negligible in these stars.

\subsection{The lifetime of individual maser clouds}
The changes seen in the VLA maps as well as the velocity shifts seen
over time for spectral components determine a typical lifetime of
maser clouds in the range 0.5 -- 3 years
(Sect. \ref{CrossCorrel-InterfData}).  This is in excellent agreement
with the estimates derived from \water\ maser cloud imaging of four
AGB stars of Bains et al. (\cite{Bains03}), who argue that instability
will destroy the clouds on such time scales.  They derived sound
crossing times of 1.5 -- 3 years for the observed sizes of the clouds.

Based on observations with the VLBA Imai et al. (\cite{Imai03})
derived for RT~Vir clouds considerably shorter lifetimes of 1 -- 2
months.  This order of magnitude shorter lifetime compared to ours is
probably due to different spatial resolutions.  The number of clouds
identified in a particular maser shell is significantly increased when
the spatial resolution is increased.  For example, Marvel
(\cite{Marvel96}) distinguished $>$50 maser clouds in VLBA RX~Boo
\water\ maser maps, while we were restricted to $<$15.  Thus, the
clouds identified here for RX~Boo are certainly composed of several
sub-components, where each of them might survive on the timescale of
months only.  Our larger lifetimes apply then to spatially closely
located groups of maser clouds (our {\it maser spatial components}),
of which the spectral components in single-dish spectra are made.  The
dependence of the lifetimes on the size of the maser clouds is also
evident from observations of RSGs, where the clouds have
physical diameters a magnitude larger than are observed in AGB stars,
and lifetimes of the order of 10 years (Richards et al. \cite{Richards98} and
\cite{Richards99}).

The typical gas crossing time through the \water\ maser supporting
shell in RX~Boo (thickness $\sim$22~AU) is $\sim$16.5 years
(Sect. \ref{CSE}).  The maser clouds most probably do not survive a
full crossing of this shell.  Due to the limited lifetimes of the
maser clouds proper motion measurements require therefore an observing
frequency high enough to sample a given cloud at least twice.  At the
highest resolutions achievable today, recurrent observations with
intervals less than a month are reasonable (see Imai et al. \cite{Imai03}).

\subsection{\label{aspect angle} The beaming directions of masers
in the RX~Boo shell} The distribution of maser components in
Fig. \ref{fig:aligned_maps} tells us that the individual masers are
radiating non-isotropically in the sense that their maximum output is
in a direction that does not deviate very much from the tangential
direction.  Our model of the shell expansion has given us values of
the angle $\theta$ (the {\it aspect angle}) for every maser
component. The distribution function of these values has a maximum at
zero degrees and an extent from about --60\degr\ to +60\degr.  In
other words, the masers are not all in a plane perpendicular to the
line of sight [see Figs.  \ref{fig:envelope_3D}(b) and (c))].

Now, imagine for a moment that all masers were radiating in the
tangential direction and that we therefore see many of them from an
angle that deviates from the direction of maximum amplification.
Furthermore, assume that all the masers have the same intrinsic
strength.  Then by plotting the observed flux density against the
aspect angle we would get the `beam shape' of the `unit maser'.  Of
course there is probably a spread in the direction of maximum
amplification from the tangential direction.  Also there is certainly
a considerable spread in the intrinsic strength of the masers.  These
spreads would destroy to a large extent the relation between flux
density and aspect angle.  However, it is conceivable that a weak
correlation would remain.

\begin{figure}[h]
\includegraphics[width=9cm]{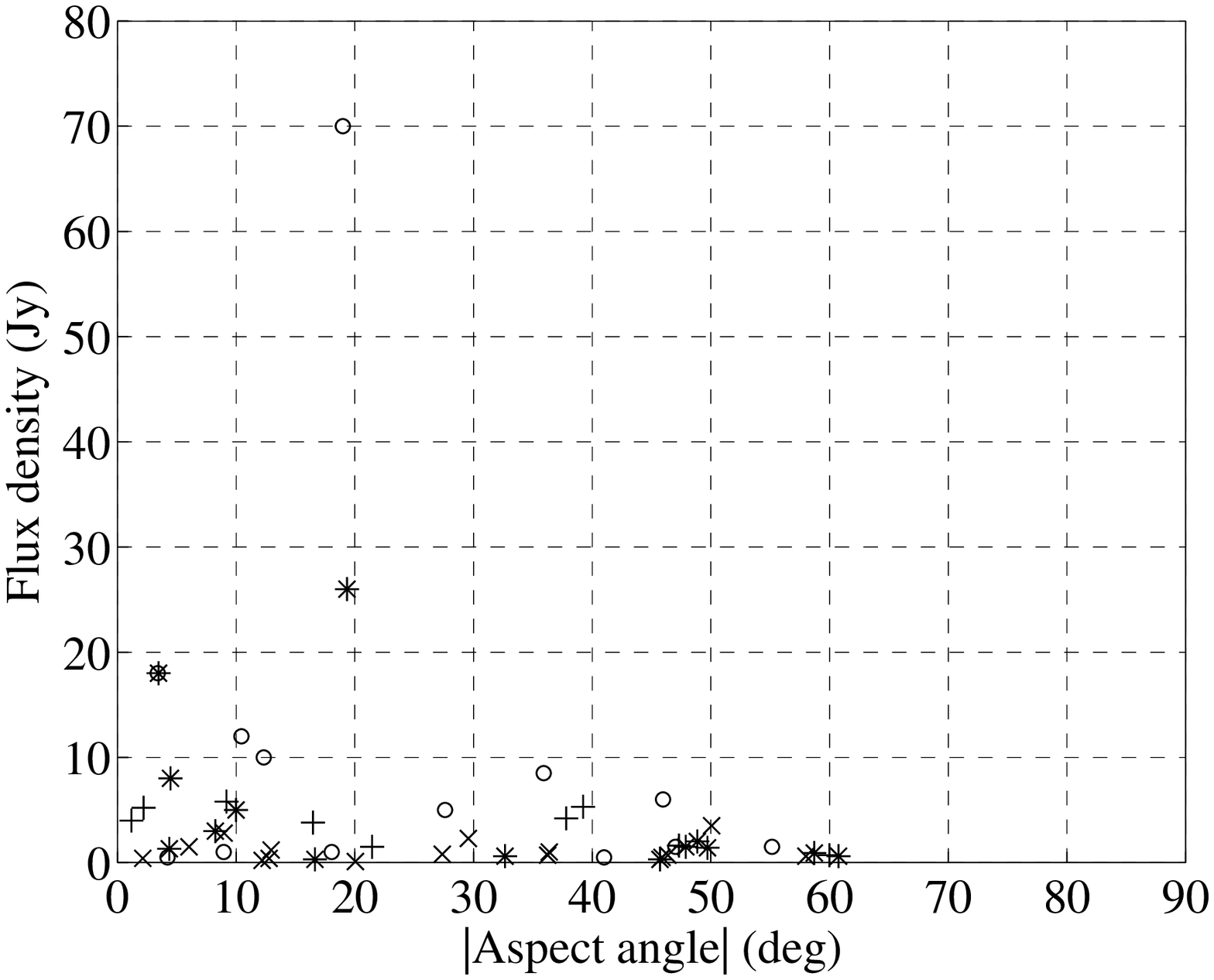}
\caption{The observed flux density as a function of the absolute value
of the aspect angle ($\theta$).  The epochs are represented by the
same symbols as in Figs. \ref{fig:aligned_maps} and
\ref{fig:aligned_CSE}.}
\label{fig:aspect_angle}
\end{figure}

In Fig. \ref{fig:aspect_angle} the observed flux density is plotted
against the absolute value of the aspect angle.  It is obvious that
the highest fluxes are found among the smallest aspect angles.  If the
two data points with the very high fluxes close to 20\degr\ are
disregarded the highest flux indeed is found close to 0\degr.  In
addition, the upper envelope of the scatter plot is decreasing from a
maximum close to 0\degr\ to a minimum close to 60\degr.  The fact that
this correlation is disturbed by the very strong masers at 20\degr\
must be explained by the disparate intrinsic maser strengths.

Fig. \ref{fig:aspect_angle} is an additional evidence, albeit a weak
one, for the \water\ masers in the CSE of RX~Boo to be predominantly
tangentially beamed.  However, the typical beaming angle is impossible
to extract from the data as we cannot disentangle the deviation of the
line of sight from the direction of the maximum amplification and the
deviation of the direction of the maximum amplification from the
tangential direction.

\subsection{\label{asymmetry} Maser shell asymmetry}
So far, the \water\ maser shell has been assumed to be radially
symmetric.  However, there is a prominent asymmetry present in the
distribution of \water\ maser emission of RX~Boo, which manifests
itself during 1990 -- 1995 by the absence of emission in the SW part
of the shell and by the NW-SE velocity gradient
(Sect. \ref{id_MapsCompsId}).  How persistent is this asymmetry?
Interferometric observations of the \water\ masers of RX~Boo have been
made a number of times, but the first sensitive map available is from
VLA observations in 1985 (Bowers et al. \cite{Bowers93}).  An E-W
velocity gradient is evident also in this map, but in addition there
is some emission in the SW part of the shell.  VLBA observations taken
in 1995 January and July are available from Marvel (\cite{Marvel96}),
but they are difficult to compare with our VLA maps because all the
blueshifted emission with $v_{\rm los} < +0.5$~\kms\ was not detected.
This emission seems to have been resolved out.  Nevertheless, we
conclude that the incomplete ring-like structure and the velocity
gradient was present at least over eleven years between 1985 and 1995,
which is about 70\% of the gas crossing time in the \water\
maser-supporting shell of RX~Boo.

Evidence for asymmetry is coming also from the \water-maser velocity
interval, which is centred on $v_{\rm c} = 0$~\kms\ (Fig.
\ref{fig:v_t_diagram_rxboo}), while the CO-emission velocity interval
is centred on $v_* = +1$~\kms\ (Teyssier et al. \cite{Teyssier06}).
The blueshifted projected expansion velocities extend to higher values
$\vert v_{\rm los} - v_* \vert\,\approx 6$~\kms\ than the redshifted
velocities $\vert v_{\rm los} - v_* \vert\,\approx 4$~\kms.  This
velocity asymmetry has persisted over the full monitoring time range
of 20 years.

\begin{figure}[h]
\begin{minipage}[t]{8cm}
\includegraphics[width=4.5cm,angle=-90]{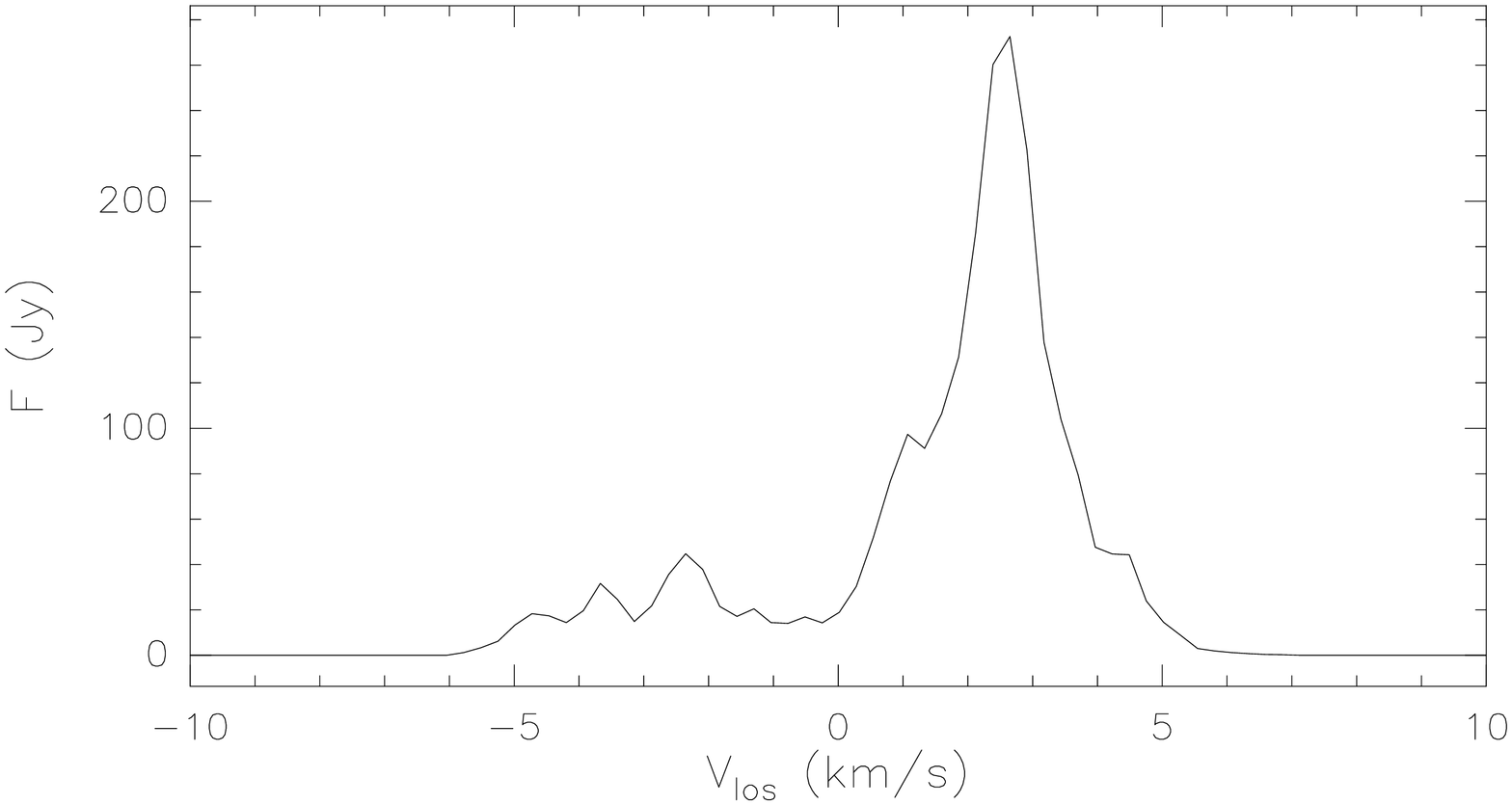}  \\[0.5cm]
\includegraphics[width=4.5cm,angle=-90]{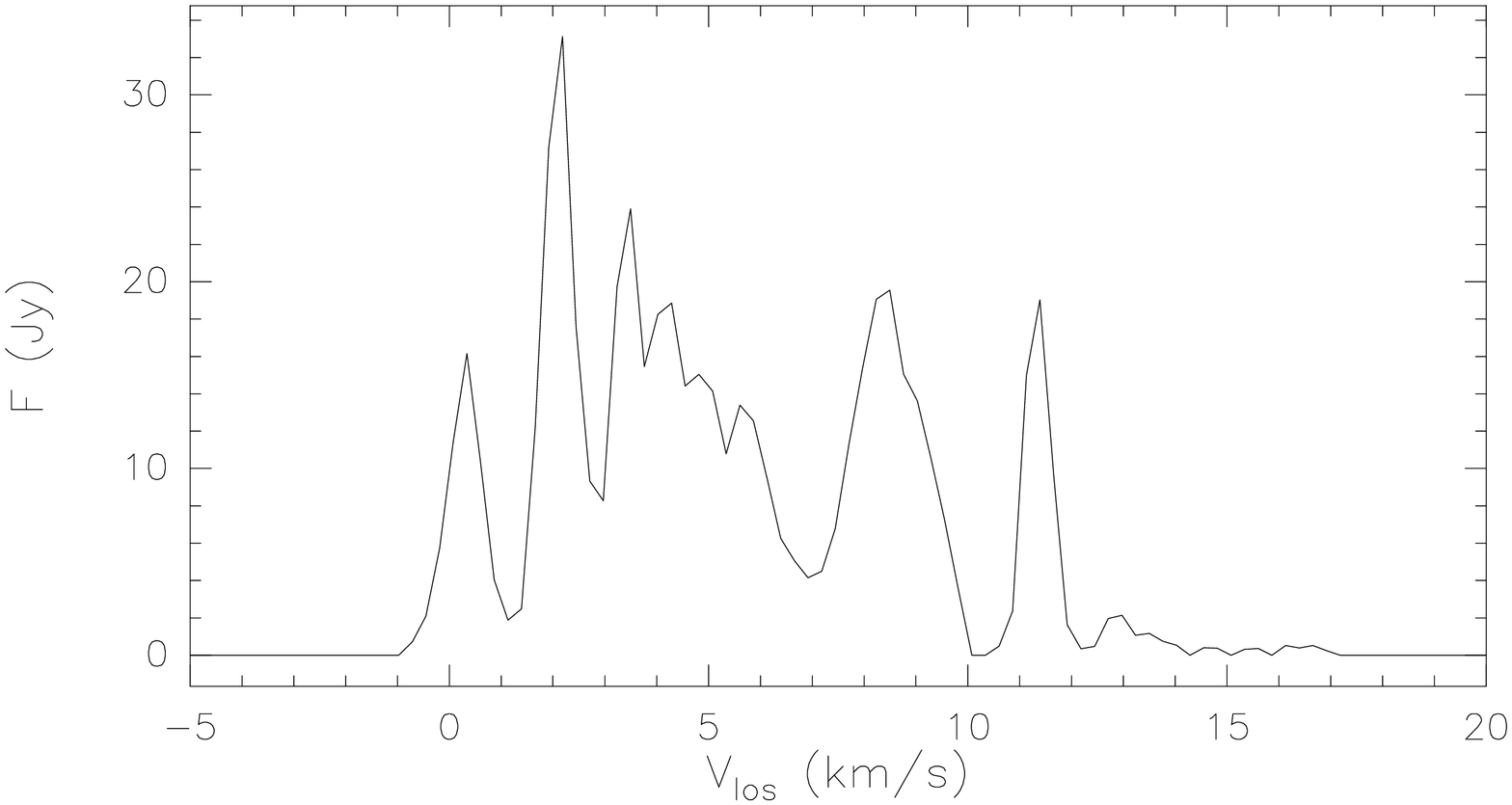}
\end{minipage}
\caption{Upper envelope of all spectral profiles of RX~Boo (top) and
SV~Peg (bottom).  In RX~Boo the spectra containing the flare (1995
January -- May) were omitted.  The flare occured at $\sim$3.5~\kms\
close to the strongest peak in the upper envelope and reached a peak
flux of 1600~Jy.  The upper envelope is defined as the highest flux
density ever achieved in velocity bins of 0.3~\kms\ during the
monitoring program.}
\label{fig:upper_envelope}
\end{figure}

In Fig. \ref{fig:upper_envelope} the upper envelopes of the spectral
profiles of RX~Boo and SV~Peg are shown.  These envelopes show the
highest flux densities measured during the monitoring period at each
velocity.  Contrary to SV~Peg, where the highest flux densities are
more or less on a similar level across the maser velocity range, the
profile of RX~Boo is dominated by emission at $\sim$2.5~\kms\
(spectral component I).  Emission close to this velocity was present
over the full monitoring period (see Table \ref{tab:compSVPegF-I} in
the Appendix), emphasizing the longevity of the asymmetries.

Asymmetries present in individual observations can easily be explained
as due to shells unevenly filled with maser clouds.  The clouds
themselves are explained by Richards et al. (\cite{Richards99}) as
clumps of enhanced density, which form as condensations in the
extended stellar atmosphere.  Due to the randomness of the formation
process of such clumps, parts of the shell might not contain clumps at
a given epoch. But we have observed the persistence of the asymmetric
emission structure and the velocity gradient over a time range more
than half of the crossing time through the shell. And, the asymmetry
of the \water\ maser velocity interval outlasted the 16.5 years
crossing time, showing that the morphology of the maser site
distribution is not due to random fluctuations.  It is therefore more
likely that the persistent asymmetries are evidence for a non
spherically-symmetric mass loss process in RX~Boo.  The asymmetric
distribution of maser emission components found in RX~Boo is not
untypical.  >From VLA-observations of \water\ masers in 13 Mira
variables and 3 SRVs, Bowers \& Johnston (\cite{Bowers94}) concluded
that the emission is best modelled by ``clumps, segments, or filaments
in an asymmetric but ellipsoidal-type geometry''.

For a strong maser to occur, some very special (and rare) physical
conditions have to be fulfilled: the pumping of the masing transition
has to win over thermal de-excitation processes and the coherent
amplification path length has to be sufficiently long for the maser
flux density to be detected.  Therefore, an alternative explanation of
the asymmetric distribution of \water\ masers in the envelope of
RX~Boo might be provided by a non-spherical distribution of {\it maser
conditions}.

A possible explanation of such a distribution might be found in the
photospheric structure of the star.  Giant and supergiant stars have
extremely deep convective layers and as a result the convection cells
of the photosphere are expected to be gigantic (cf. Schwarzschild
\cite{Schwarzschild75}).  Freytag (\cite{Freytag03}) has made a
3-dimensional radiative hydrodynamic model of the SRV Betelgeuse
($\alpha$~Orionis; spectral class M2Iab) that simulates
high-resolution observations of surface features (cf. Young et
al. \cite{Young00}) and typical irregular flux variations of this star
(cf. Goldberg \cite{Goldberg84}).  However, we are unaware of any
published simulations of pulsating giants or of AGB stars of more
moderate sizes.  Nevertheless, it is conceivable that the random
character of the convection elements is responsible for irregular
light variations or for an irregular modulation of periodic
variations.  It is conceivable too that the convection elements to a
certain degree are preserved as some kind of `blobs' in the stellar
wind.  As these blobs reach the \water\ maser region, a larger blob
probably creates longer amplification path lengths than a number of
smaller blobs.  In addition such blobs will expand considerably during
their travel outwards, not only due to the divergence but also
radially due to the acceleration.  In the case of RX~Boo a convective
blob is expected to increase its volume by a factor of $\sim$50 during
its journey from the photosphere to the inner radius of the \water\
maser region.

However, the real mystery is the longevity of the asymmetry of the
\water\ maser region of RX~Boo.  In the convection-wind picture this
implies that the lifetime of convection cells either be long or that
the occurence of large cells persists over long time periods.  In the
Betelgeuse simulation by Freytag (\cite{Freytag03}) the lifetime of
convection cells seems to be of the order of one year and the
likelihood of occurence of a cell seems to be constant over both time
and stellar surface.  If the typical lifetime of convection cells in
the RX~Boo photosphere is of the same order, it would take about one
year for the corresponding blob to getting detached from the
photosphere and becoming part of the stellar wind.  However, when this
blob reaches the \water\ maser region ($\sim$14~AU from the star), the
inflation due to acceleration and divergence would increase the time
for the blob to pass a fixed point by a factor of about 5.

\section{Conclusions}
Water vapour maser emission from CSEs is often strong enough to be
observed with small radio telescopes even in rather distant late-type
stars.  Because of its variability, however, the information to be
derived from a single maser spectrum is limited.  Neither the
intensities nor the velocities are necessarily representative.  To
understand the cause of the strong variability of the masers in SRVs
we have monitored RX~Boo and SV~Peg, two stars representative of their
class, with single-dish radio telescopes over a period of two decades.
The interpretation of the variations seen in single-dish spectra is
usually compromised by blending of the maser peaks due to the overlap
of two or more maser clouds in velocity space.  To overcome this
limitation, we obtained in addition for RX~Boo four interferometric
observations taken over a time interval of almost three years.

We find that the \water\ masers in RX~Boo are located in an incomplete
ring of thickness 22~AU, which is filled only in part.  The time that
the gas needs to cross this ring is 16.5 years. This ring-like
distribution of maser spots remained stable over at least 11 years.
Compared to these timescales, the survival time of individual maser
clouds is much shorter having a typical duration of $\sim$1 year.  The
variability of \water\ masers is then due to the emergence and
disappearance of shortlived maser clouds.  The variations in
intensities of these clouds are not dependent on variations of the
stellar luminosity, as we find no correlation between the optical and
the radio lightcurve.

It follows from the interpretation of single-dish spectra that the
strength of individual maser peaks or the details in the shape of the profile
are snap-shots reflecting the distribution of the maser clouds and their
luminosities at the time of observation.  If RX~Boo and SV~Peg are typical
SRVs, the lifetime of these clouds in SRVs is limited to
about 1 year.  Also stellar radial velocities and shell outflow velocities will
be biased, if the maser cloud distribution is not uniform in the shell.

For RX~Boo, the persistent overall distribution of the maser emission
regions indicates that the underlying mass outflow from the star or
the distribution of maser conditions are not spherically symmetric.
Whether deviations from spherical symmetry are present in the \water\
maser regions of SRVs in general, will have to be answered by new
monitoring programmes relying on interferometry for a larger sample of
stars.

\begin{acknowledgements}
We like to thank the NRAO\footnote{NRAO is operated by Associated
Universities, Inc., under cooperative agreement with the US National
Science Foundation.} staff for generous support, especially
E.\,Brinks, B.\,Clarke, P.\,Perley, R.\,Perley, and M.\,Rupen.  We are
grateful to the Arcetri Radio group for helping us to observe our
sources as part of a larger monitoring project.  The staff of the
Medicina observatory has been very supportive solving operational
problems with the telescope as they occurred; we especially thank
A. Orfei and G. Maccaferri.  R. Cesaroni (Arcetri) devised various
ingenious routines for the display of the data collected during the
monitoring.  The Max-Planck-Institut f\"ur Radioastronomie and the
Istituto di Radioastronomia supported this work by ample allocation of
observing time and computing facilities. DE was supported by the
Deutsche Forschungsgemeinschaft (DFG) through travel grants Re\,353/25
and En\,176/2.  AW was supported by the Swedish Science Research
Council (now Science Council).  This research has made use of the
SIMBAD database, operated at Centre de donn\'{e}es astronomiques de
Strasbourg (CDS), France.  We acknowledge with gratitude the variable
star observations from the AAVSO International Database contributed by
observers worldwide and used in this research. We thank the referee
A.\,Richards for providing constructive comments and help in improving
the contents of this paper.

\end{acknowledgements}

\appendix
\newpage
\section{Single-dish data of RX~Boo and SV~Peg}

\begin{longtable}{rrcrrr|rr|rr|rr|rr}

\caption{\label{tab:compRXBooA-E}Maser components of RX~Boo. The columns
give the observing date in Gregorian and Julian (JD) format, the noise of the
spectra (rms), the integrated flux \SI\ in units of \powerten{-22} W
m$^{-2}$, and the velocities \vp\ and flux densities \Sp\ of maser peaks
associated with the spectral components A -- E in Table \ref{tab:components}.}\\

\hline\noalign{\smallskip}

 & & & &\multicolumn{2}{c}{A} & \multicolumn{2}{c}{B} & \multicolumn{2}{c}{C} & 
\multicolumn{2}{c}{D} & \multicolumn{2}{c}{E} \\[0.1cm]

Date & JD & rms & \SI & \multicolumn{2}{c}{\vp\gm\Sp} &
\multicolumn{2}{c}{\vp\gm\Sp} & \multicolumn{2}{c}{\vp\gm\Sp} &
\multicolumn{2}{c}{\vp\gm\Sp} & \multicolumn{2}{c}{\vp\gm\Sp} \\[0.1cm]

 & & [Jy] &  & \multicolumn{10}{c}{\hrulefill\ [km\,s$^{-1}$, Jy] \hrulefill}\\

\noalign{\smallskip}\hline\noalign{\smallskip}

\endfirsthead

\caption{continued.} \\

\hline\noalign{\smallskip}

 & & & & \multicolumn{2}{c}{A} & \multicolumn{2}{c}{B} & \multicolumn{2}{c}{C} &
\multicolumn{2}{c}{D} & \multicolumn{2}{c}{E} \\[0.1cm]

Date & JD & rms & \SI & \multicolumn{2}{c}{\vp\gm\Sp} &
\multicolumn{2}{c}{\vp\gm\Sp} & \multicolumn{2}{c}{\vp\gm\Sp} &
\multicolumn{2}{c}{\vp\gm\Sp} & \multicolumn{2}{c}{\vp\gm\Sp} \\[0.1cm]

 & & [Jy] & & \multicolumn{10}{c}{\hrulefill\ [km\,s$^{-1}$, Jy] \hrulefill}\\

\noalign{\smallskip}\hline\noalign{\smallskip}

\endhead

\hline

\endfoot
                                                                     
 30.03.87 & 2446885 & 3.41 & 2743   & $-$    & $-$  & $-$    & $-$  & $-$   
&$-$  & $-$    & $-$  & $-$    & $-$  \\                                 
 22.06.87 & 2446969 & 1.16 & 1732   &\multicolumn{8}{c|}{\sevendash\vgm
$-3.8$\gm 4:\vgm \sevendash } & $-$    & $-$ \\                               
 05.09.87 & 2447044 & 2.30 & 2170   & $-$    & $-$  & $-$    & $-$  & $-$   
&$-$  & $-$    & $-$  & $-$    & $-$  \\                                 
 12.09.87 & 2447051 & 4.41 & 2305   & $-$    & $-$  & $-$    & $-$  & $-$   
&$-$  & $-$    & $-$  & $-$    & $-$  \\                                 
 05.11.87 & 2447105 & 5.16 & 1928   & $-$    & $-$  & $-$    & $-$  & $-$   
&$-$  & $-$    & $-$  & $-$    & $-$  \\                                 
 11.02.88 & 2447203 & 0.86 & 1927   & $-$    & $-$  & $-$    & $-$  & $-$   
&$-$  & $-$    & $-$  & $-$    & $-$  \\                                 
 19.02.88 & 2447211 & 4.27 & 2018   & $-$    & $-$  & $-$    & $-$  & $-$   
&$-$  & $-$    & $-$  & $-$    & $-$  \\                                 
 02.05.88 & 2447284 & 5.55 & 2544   & $-$    & $-$  & $-$    & $-$  & $-$   
&$-$  & $-$    & $-$  & $-$    & $-$  \\                                 
 13.07.88 & 2447356 & 2.56 & 1410  
&\multicolumn{8}{c|}{\sevendash\vgm$-3.6$\gm14:\vgm \sevendash } & $-$    & $-$
\\                                
 13.10.88 & 2447448 & 4.06 & 2009  
&\multicolumn{6}{c|}{\threedash\vgm$-4.2$\gm18.5\vgm \threedash }& $-2.4$ &23.1 
& $-$    & $-$  \\               
 31.01.89 & 2447558 & 2.53 & 1599  
&\multicolumn{6}{c|}{\threedash\vgm$-3.9$\gm16:\vgm \threedash } & $-2.4$ &42.7 
& $-$    & $-$  \\               
 21.03.89 & 2447607 & 3.79 & 1571   & $-$    & $-$  & $-$    & $-$  & $-$   
&$-$  & $-2.4$ &32.5  & $-$    & $-$  \\                                 
 04.12.89 & 2447865 & 1.44 &  374   & $-$    & $-$  & $-$    & $-$  & $-$   
&$-$  & $-$    & $-$  & $-$    & $-$  \\                                 
 06.02.90 & 2447929 & 0.95 & 1105   & $-4.8$ & 2.8  & $-$    & $-$  & $-3.0$ &
6.0 & $-$    & $-$  & $-$    & $-$  \\                                 
 17.02.90 & 2447940 & 0.25 & 1008   & $-4.6$ & 2.6  & $-$    & $-$  & $-2.9$ &
6.2 & $-$    & $-$  & $-$    & $-$  \\                                 
 31.03.90 & 2447982 & 0.14 &  842   & $-4.7$ & 3.9  & $-$    & $-$  & $-2.9$ &
5.7 & $-$    & $-$  & $-1.3$ &  1.6  \\                                
 17.04.90 & 2447999 & 0.95 &  845   & $-$    & $-$  & $-$    & $-$  & $-$    &
$-$ & $-$    & $-$  & $-$    & $-$  \\                                 
 11.05.90 & 2448023 & 0.22 &  883   & $-4.6$ & 4.9  & $-$    & $-$  & $-2.9$ &
4.5 & $-$    & $-$  & $-1.2$ &  1.7  \\                                
 15.07.90 & 2448088 & 1.84 &  528   & $-$    & $-$  & $-$    & $-$  & $-3.0$ &
3.0 & $-$    & $-$  & $-$    & $-$  \\                                 
 21.10.90 & 2448186 & 0.13 &  759   & $-4.5$ & 5.0  & $-$    & $-$  & $-2.5$ &
4.4 & $-$    & $-$  & $-1.2$ &  1.3  \\                                
 18.01.91 & 2448275 & 1.94 &  680   & $-$    & $-$  & $-$    & $-$  & $-2.7$ &
8.5 & $-$    & $-$  & $-$    & $-$  \\                                 
 29.01.91 & 2448286 & 1.82 &  587   & $-4.7$ & 6.4  & $-$    & $-$  & $-2.9$ &
5.4 & $-$    & $-$  & $-$    & $-$  \\                                 
 30.03.91 & 2448346 & 0.10 &  534   & $-4.6$ & 3.4  & $-$    & $-$  & $-2.6$ &
3.5 & $-$    & $-$  & $-$    & $-$  \\                                 
 01.05.91 & 2448378 & 0.26 &  404   & $-4.6$ & 3.0  & $-$    & $-$  & $-2.7$ &
4.7 & $-$    & $-$  & $-1.0$ &  0.8  \\                                
 17.05.91 & 2448394 & 1.33 &  354   & $-$    & $-$  & $-$    & $-$  & $-2.9$ &
9.2 & $-$    & $-$  & $-0.8$ &  1.3  \\                                
 25.10.91 & 2448555 & 0.72 &  249   & $-$    & $-$  & $-$    & $-$  & $-2.7$ &
4:  & $-$    & $-$  & $-$    & $-$  \\                                 
 18.01.92 & 2448640 & 0.19 &  426   & $-4.5$ & 1.9  & $-3.7$ & 1.2  & $-2.8$ &
3.7 & $-1.9$ & 2.6  & $-$    & $-$  \\                                 
 29.02.92 & 2448682 & 0.12 &  340   & $-4.6$ & 1.6  & $-$    & $-$  & $-2.6$ &
3.3 & $-$    & $-$  & $-$    & $-$  \\                                 
 18.04.92 & 2448731 & 0.64 &  359  
&\multicolumn{8}{c|}{\sevendash\vgm$-3.3$\gm4:\vgm \sevendash} & $-$    & $-$ 
\\                                 
 05.07.92 & 2448809 & 0.18 &  418   & $-4.5$ & 2.9  & $-$    & $-$  & $-2.5$ &
5.6 & $-$    & $-$  & $-$    & $-$  \\                                 
 01.09.92 & 2448867 & 0.10 &  234   & $-4.8$ & 4.6  & $-3.6$ & 3.9  & $-2.3$ &
4.6 & $-$    & $-$  & $-$    & $-$  \\                                 
 15.10.92 & 2448911 & 0.69 &  277  
&\multicolumn{8}{c|}{\sevendash\vgm$-3.5$\gm4.7\vgm \sevendash}& $-$    & $-$ 
\\                                 
 24.10.92 & 2448920 & 0.69 &  205   & $-$    & $-$  & $-$    & $-$  & $-$   
&$-$  & $-$    & $-$  & $-$    & $-$  \\                                 
 22.12.92 & 2448979 & 0.13 &  221   & $-4.9$ & 2.4  & $-3.7$ & 5.0  & $-$   
&$-$  & $-2.0$ & 2.5  & $-1.0$ &  1.2  \\                                
 26.01.93 & 2449014 & 0.48 &  331   & $-4.9$ & 4.2  & $-3.6$ & 7.5  & $-$   
&$-$  & $-2.3$ & 4.5  & $-1.2$ &  1.0  \\                                
 21.04.93 & 2449099 & 0.14 &  694   & $-4.6$ &11.0  & $-3.5$ & 6.2  & $-2.8$
&10.8 & $-2.0$ & 7.3  & $-1.0$ &  2.9  \\                                
 13.05.93 & 2449121 & 0.58 &  268   & $-4.5$ & 4.8  & $-$    & $-$  & $-2.9$ &
7.0 & $-$    & $-$  & $-$    & $-$  \\                                 
 03.11.93 & 2449295 & 0.54 &  264  
&\multicolumn{8}{c|}{\sevendash\vgm$-3.6$\gm5.8\vgm \sevendash}& $-$    & $-$ 
\\                                 
 30.11.93 & 2449322 & 0.23 &  257   & $-$    & $-$  & $-4.0$ & 4.0  & $-2.8$ &
3.0 & $-$    & $-$  & $-$    & $-$  \\                                 
 08.03.94 & 2449420 & 0.22 &  373   & $-$    & $-$  & $-4.2$ & 4.5  & $-3.4$ &
3.4 & $-2.1$ & 3.9  & $-$    & $-$  \\                                 
 14.04.94 & 2449457 & 0.34 &  160   & $-$    & $-$  & $-$    & $-$  & $-3.2$ &
1.7 & $-$    & $-$  & $-$    & $-$  \\                                 
 08.09.94 & 2449604 & 1.35 &  272   & $-$    & $-$  & $-$    & $-$  & $-$    &
$-$ & $-$    & $-$  & $-$    & $-$  \\                                 
 28.10.94 & 2449654 & 0.61 &  361   & $-$    & $-$  & $-4.2$ & 3.9  & $-3.4$ &
3.0 & $-2.1$ & 3.4  & $-$    & $-$  \\                                 
 18.01.95 & 2449736 & 0.85 & 2319   & $-$    & $-$  & $-$    & $-$  & $-$    &
$-$ & $-$    & $-$  & $-$    & $-$  \\                                 
 09.03.95 & 2449786 & 0.47 & 9382   & $-$    & $-$  & $-$    & $-$  & $-3.1$ &
3.3 & $-1.6$ & 5.3  & $-$    & $-$  \\                                 
 12.03.95 & 2449789 & ---  & 9408   & $-$    & $-$  & $-$    & $-$  & $-$    &
$-$ & $-$    & $-$  & $-$    & $-$  \\                                 
 13.03.95 & 2449790 & 0.33 & 6558   & $-$    & $-$  & $-$    & $-$  & $-3.2$ &
2.1 & $-1.5$ & 3.9  & $-$    & $-$  \\                                 
 27.03.95 & 2449804 & 0.30 & 2585   & $-$    & $-$  & $-$    & $-$  & $-$    &
$-$ & $-$    & $-$  & $-$    & $-$  \\                                 
 09.04.95 & 2449817 & 0.20 & 3129   & $-$    & $-$  & $-$    & $-$  & $-3$  
&$<$2 & $-1.5$ & $<$4 & $-$    & $-$  \\                                 
 03.05.95 & 2449841 & 0.13 & 1869   & $-$    & $-$  & $-$    & $-$  & $-3$  
&$<$2 & $-1.5$ & $<$5 & $-$    & $-$  \\                                 
 03.06.95 & 2449872 & 0.21 & 1021   & $-$    & $-$  & $-3.9$ & 1.7  & $-$   
&$-$  &\multicolumn{4}{c}{{------}\gm$-1.0$\gm5.9\gm {------} }  \\      
 04.06.95 & 2449873 & 0.23 &  739   & $-$    & $-$  & $-3.8$ & 1.2  & $-$   
&$-$  &\multicolumn{4}{c}{{------}\gm$-1.1$\gm4.2\gm {------} }  \\      
 05.06.95 & 2449874 & 0.19 &  832   & $-$    & $-$  & $-3.7$ & 1.6  & $-$   
&$-$  &\multicolumn{4}{c}{{------}\gm$-1.2$\gm5.0\gm {------} }  \\      
 23.06.95 & 2449892 & 0.22 &  947   & $-$    & $-$  & $-$    & $-$  & $-3.2$ &
2.7 & $-$    & $-$  & $-1.0$ &  9.1 \\                                 
 18.09.95 & 2449982 & 0.19 &  728   & $-$    & $-$  & $-$    & $-$  & $-$   
&$-$  & $-2.1$ &10.0  & $-$    & $-$  \\                                 
 18.01.96 & 2450101 & 0.08 &  453   & $-$    & $-$  & $-$    & $-$  & $-2.6$ &
1.8 & $-$    & $-$  & $-0.9$ &  2.9 \\                                 
 01.03.96 & 2450144 & 0.93 &  447   & $-$    & $-$  & $-$    & $-$  & $-$    &
$-$ & $-$    & $-$  & $-0.9$ &$<$5  \\                                 
 03.09.96 & 2450330 & 0.17 &  560   & $-$    & $-$  & $-$    & $-$  & $-2.8$ &
2.6 &\multicolumn{4}{c}{{------}\gm $-1.4$\gm2.2\gm {------} }  \\     
 14.03.97 & 2450522 & 0.13 & 1316   & $-$    & $-$  & $-$    & $-$  & $-2.7$
&10.7 & $-$    & $-$  & $-$    & $-$  \\                                 
 06.05.97 & 2450575 & 0.71 & 1277   & $-$    & $-$  & $-$    & $-$  & $-2.6$ &
9.4 & $-$    & $-$  & $-$    & $-$  \\                                 
 12.10.97 & 2450734 & 0.14 &  650   & $-$    & $-$  & $-$    & $-$  & $-2.3$
&13.3 & $-$    & $-$  & $-0.5$ &  5.9 \\                                 
 21.10.97 & 2450743 & 0.58 &  420   & $-$    & $-$  & $-$    & $-$  & $-2.3$ &
7.9 & $-$    & $-$  & $-0.5$ & $<$4 \\                                 
 15.12.97 & 2450798 & 0.47 &  567   & $-$    & $-$  & $-$    & $-$  & $-2.2$
&11.2 & $-$    & $-$  & $-$    & $-$  \\                                 
 29.01.98 & 2450843 & 0.58 &  616   & $-$    & $-$  & $-$    & $-$  & $-2.2$
&16.0 & $-$    & $-$  & $-$    & $-$  \\                                 
 19.03.98 & 2450892 & 0.53 &  431   & $-$    & $-$  & $-$    & $-$  & $-2.3$ &
6.0 &\multicolumn{4}{c}{{------}\gm $-1.2$\gm4.1\gm{------} }  \\      
 10.04.98 & 2450914 & 0.80 &  281   & $-$    & $-$  & $-$    & $-$  & $-$   
&$-$  &\multicolumn{4}{c}{{------}\gm $-1.5$\gm$<$5\gm{------} } \\      
 11.05.98 & 2450945 & 0.74 &  327   & $-$    & $-$  & $-$    & $-$ 
&\multicolumn{4}{c|}{{------}\gm $-1.8$\gm5.3\gm{------} }      & $-$    & $-$ 
\\
 02.02.99 & 2451212 & 0.38 &  567   & $-$    & $-$  & $-$    & $-$  & $-3.4$ &
9.5 &\multicolumn{4}{c}{{------}\gm $-1.2$\gm4.6\gm{------} }  \\      
 24.06.99 & 2451354 & 0.12 &  824   & $-$    & $-$  & $-$    & $-$  & $-3.6$
&31.7 &\multicolumn{4}{c}{{------}\gm $-0.9$\gm$<$9\gm{------} } \\      
 27.10.00 & 2451845 & 1.00 &  437   & $-4.4$ & 9.9  & $-$    & $-$  & $-$   
&$-$  & $-$    & $-$  & $-0.6$ & 16.4 \\                                 
 20.12.00 & 2451899 & 0.79 &  304   & $-4.2$ & 3.3  & $-$    & $-$  & $-$   
&$-$  & $-$    & $-$  & $-0.6$ &  7.2 \\                                 
 07.05.01 & 2452037 & 0.65 &  343   & $-4.3$ & 2.2  & $-$    & $-$  & $-$   
&$-$  & $-$    & $-$  & $-0.9$ &  4.8 \\                                 
 18.09.01 & 2452171 & 0.66 &  442   & $-$    & $-$  & $-$    & $-$  & $-$   
&$-$  & $-$    & $-$  & $-1.0$ & 14.5 \\                                 
 24.10.01 & 2452207 & 0.76 &  469   & $-$    & $-$  & $-$    & $-$  & $-$   
&$-$  & $-$    & $-$  & $-1.0$ & 12.6 \\                                 
 19.03.02 & 2452353 & 1.89 &  922   & $-$    & $-$  & $-$    & $-$  & $-$   
&$-$  & $-$    & $-$  & $-$    & $-$  \\                                 
 18.04.02 & 2452383 & 0.80 & 1107   &
\multicolumn{10}{c}{{----------------------}\vgm $-6.0 -
0.0$\gm$\approx2-4$\vgm{----------------------} }  \\  
 20.06.02 & 2452446 & 0.15 &  512   & $-$    & $-$  &
\multicolumn{8}{c}{{--------------}\vgm $-4.0 -
0.0$\gm$\approx2$\vgm{--------------} }  \\     
 26.06.02 & 2452452 & 1.11 &  434   & $-$    & $-$  & $-$    & $-$  & $-$   
&$-$  & $-$    & $-$  & $-$    & $-$  \\                                 
 01.10.02 & 2452549 & 0.96 &  483   & $-$    & $-$  & $-$    & $-$  & $-$   
&$-$  & $-$    & $-$  & $-$    & $-$  \\                                 
 24.10.02 & 2452572 & 0.99 &  287   & $-$    & $-$  & $-$    & $-$  & $-$   
&$-$  & $-$    & $-$  & $-$    & $-$  \\                                 
 13.12.02 & 2452622 & 1.47 &  511   & $-$    & $-$  & $-$    & $-$  & $-$   
&$-$  & $-$    & $-$  & $-$    & $-$  \\                                 
 19.12.02 & 2452628 & 1.33 &  480   & $-$    & $-$  & $-$    & $-$  & $-$   
&$-$  & $-$    & $-$  & $-$    & $-$  \\                                 
 15.01.03 & 2452655 & 1.55 &  488   & $-$    & $-$  & $-$    & $-$  & $-$   
&$-$  & $-$    & $-$  & $-$    & $-$  \\                                 
 02.04.03 & 2452732 & 1.05 &  425   & $-$    & $-$  & $-$    & $-$  & $-$   
&$-$  & $-$    & $-$  & $-$    & $-$  \\                                 
 18.11.03 & 2452962 & 0.86 &  757   & $-$    & $-$  & $-$    & $-$  & $-$   
&$-$  & $-$    & $-$  & $-$    & $-$  \\                                 
 28.01.04 & 2453033 & 0.80 &  416   & $-5.1$ & 3.5: & $-$    & $-$  & $-$   
&$-$  & $-$    & $-$  & $-1.4$ & 22.9 \\                                 
 31.03.04 & 2453096 & 0.83 &  395   & $-$    & $-$  & $-$    & $-$  & $-$   
&$-$  & $-$    & $-$  & $-1.3$ & 21.1 \\                                 
 11.05.04 & 2453137 & 0.62 &  246   & $-$    & $-$  & $-$    & $-$  & $-$   
&$-$  & $-$    & $-$  & $-1.3$ &  4.6 \\                                 
 18.06.04 & 2453175 & 0.54 &  216   & $-$    & $-$  & $-$    & $-$  & $-$   
&$-$  & \multicolumn{4}{c}{{ }\gm $-2.0 -0.0$\gm$\approx 2$\gm { } }  \\ 
 17.09.04 & 2453266 & 1.00 &  362   & $-$    & $-$  & $-$    & $-$  & $-$   
&$-$  & $-$    & $-$  & $-$    & $-$  \\                                 
 18.12.04 & 2453358 & 1.14 &  325   & $-$    & $-$  & $-3.6$ & 7.3  & $-$   
&$-$  & \multicolumn{4}{c}{{ }\gm $-2.0 -0.0$\gm$\approx 2$\gm { } }  \\ 
 12.01.05 & 2453383 & 0.96 &  447   & $-$    & $-$  & $-3.6$ & 8.3  & $-$   
&$-$  & \multicolumn{4}{c}{{ }\gm $-2.0 -0.0$\gm$\approx 2$\gm { } }  \\ 
 15.02.05 & 2453417 & 0.74 &  548   & $-$    & $-$  & $-3.9$ & 3.5  & $-2.8$ &
2.9 & $-$    & $-$  & $-0.4$ &  3.2 \\                                 
 13.04.05 & 2453474 & 1.21 &  750   & $-$    & $-$  & $-$    & $-$  & $-$   
&$-$  & $-$    & $-$  & $-$    & $-$  \\                                 
 20.06.05 & 2453542 & 1.56 &  394   & $-$    & $-$  & $-$    & $-$  & $-$   
&$-$  & $-$    & $-$  & $-$    & $-$  \\                                 
 11.07.05 & 2453563 & 1.13 &  375   & $-$    & $-$  & $-$    & $-$  & $-$   
&$-$  & $-$    & $-$  & $-$    & $-$  \\                                 
 22.11.05 & 2453697 & 1.26 &  703   & $-$    & $-$  & $-$    & $-$  & $-2.6$ &
4.0:& $-$    & $-$  & $-$    & $-$  \\                                 
 14.02.06 & 2453781 & 1.32 &  940   & $-$    & $-$  & $-$    & $-$  & $-$   
&$-$  & $-$    & $-$  & $-$    & $-$  \\                                 
 07.04.06 & 2453833 & 1.22 &  527   & $-$    & $-$  & $-$    & $-$  & $-$   
&$-$  & $-$    & $-$  & $-$    & $-$  \\                                 
 05.07.06 & 2453922 & 1.01 &  380   & $-$    & $-$  & $-$    & $-$  & $-$   
&$-$  & $-$    & $-$  & $-$    & $-$  \\                                 
 01.09.06 & 2453980 & 1.33 &  436   & $-$    & $-$  & $-$    & $-$  & $-$   
&$-$  & $-$    & $-$  & $-$    & $-$  \\                                 
 17.10.06 & 2454026 & 0.62 &  436   & $-$    & $-$  & $-$    & $-$  & $-$   
&$-$  & $-$    & $-$  & $-$    & $-$  \\                                 
 01.12.06 & 2454071 & 1.01 &  542   & $-$    & $-$  & $-$    & $-$  & $-$   
&$-$  & $-$    & $-$  & $-$    & $-$  \\                                 
 17.01.07 & 2454118 & 0.94 &  320   & $-$    & $-$  & $-$    & $-$  & $-$   
&$-$  & $-$    & $-$  & $-$    & $-$  \\

\noalign{\smallskip}\hline                                                      
                                                                     
\end{longtable}                                                                 
\begin{longtable}{rrrr|rr|rr|rr|rr|rr}

\caption{\label{tab:compRXBooF-K} Maser components of RX~Boo. The columns
give the observing date in Gregorian and Julian (JD) format, and the velocities
\vp\ and flux densities \Sp\ of maser peaks associated with the spectral
components F -- K in Table \ref{tab:components}.} \\

\hline\noalign{\smallskip}

 & & \multicolumn{2}{c}{F} & \multicolumn{2}{c}{G} & \multicolumn{2}{c}{H} &
\multicolumn{2}{c}{I} & \multicolumn{2}{c}{J} & \multicolumn{2}{c}{K} \\[0.1cm]

Date & JD & \multicolumn{2}{c}{\vp\gm\Sp} & \multicolumn{2}{c}{\vp\gm\Sp} &
\multicolumn{2}{c}{\vp\gm\Sp} & \multicolumn{2}{c}{\vp\gm\Sp} &
\multicolumn{2}{c}{\vp\gm\Sp} & \multicolumn{2}{c}{\vp\gm\Sp} \\[0.1cm]

 & & \multicolumn{12}{c}{\hrulefill\ [km\,s$^{-1}$, Jy] \hrulefill} \\

\noalign{\smallskip}\hline\noalign{\smallskip}

\endfirsthead

\caption{continued.} \\

\hline\noalign{\smallskip}

 & & \multicolumn{2}{c}{F} & \multicolumn{2}{c}{G} & \multicolumn{2}{c}{H} &
\multicolumn{2}{c}{I} & \multicolumn{2}{c}{J} & \multicolumn{2}{c}{K} \\[0.1cm]
                                                                     
Date & JD & \multicolumn{2}{c}{\vp\gm\Sp} & \multicolumn{2}{c}{\vp\gm\Sp}  &
\multicolumn{2}{c}{\vp\gm\Sp} & \multicolumn{2}{c}{\vp\gm\Sp}  &
\multicolumn{2}{c}{\vp\gm\Sp} & \multicolumn{2}{c}{\vp\gm\Sp} \\[0.1cm]

 & & \multicolumn{12}{c}{\hrulefill\ [km\,s$^{-1}$, Jy] \hrulefill} \\

\noalign{\smallskip}\hline\noalign{\smallskip}

\endhead

\hline

\endfoot

 30.03.87 & 2446885 & $-$    &  $-$ & $-$    &  $-$  & 1.4    & 71.0 & 2.4   & 
212.8 & 4.1    & 40.2 & $-$    &  $-$ \\                              
 22.06.87 & 2446969 & $-$    &  $-$ & $-$    &  $-$  & 1.3    & 24:  & 2.5   & 
173.6 & 4.0    & 18.6 & $-$    &  $-$ \\                              
 05.09.87 & 2447044 & $-$    &  $-$ & $-$    &  $-$  & 1.3    & 27:  & 2.5   & 
261.9 & 4.0    & 18:  & $-$    &  $-$ \\                              
 12.09.87 & 2447051 & $-$    &  $-$ & $-$    &  $-$  & 1.5    & 32:  & 2.5   & 
294.0 & $-$    & $-$  & $-$    &  $-$ \\                              
 05.11.87 & 2447105 & $-$    &  $-$ & $-$    &  $-$  & 1.3    & 36.3 & 2.5   & 
172.4 & $-$    & $-$  & $-$    &  $-$ \\                              
 11.02.88 & 2447203 & $-$    &  $-$ & $-$    &  $-$  & 1.6    & 41:  & 2.6   & 
215.8 & $-$    & $-$  & $-$    &  $-$ \\                              
 19.02.88 & 2447211 & $-$    &  $-$ & $-$    &  $-$  & 1.6    & 44:  & 2.7   & 
232.4 & $-$    & $-$  & $-$    &  $-$ \\                              
 02.05.88 & 2447284 & $-$    &  $-$ & $-$    &  $-$  & 1.2    & 39:  & 2.8   & 
238.7 & $-$    & $-$  & $-$    &  $-$ \\                              
 13.07.88 & 2447356 & $-$    &  $-$ & $-$    &  $-$  & 1.7    & 34:  & 2.9   & 
108.0 & $-$    & $-$  & $-$    &  $-$ \\                              
 13.10.88 & 2447448 & $-$    &  $-$ & 0.9    & 22.7  & 1.9    & 61.3 & 3.0   & 
136.8 & $-$    & $-$  & $-$    &  $-$ \\                              
 31.01.89 & 2447558 & $-$    &  $-$ & 1.4    & 22.5  & 1.9    & 35.6 & 2.9   &  
93.2 & $-$    & $-$  & $-$    &  $-$ \\                              
 21.03.89 & 2447607 & $-$    &  $-$
&\multicolumn{4}{c|}{------\gm$1.6$\gm32.2\gm------}     & 2.9  & 92.6 & $-$   
& $-$  & $-$    & $-$ \\          
 04.12.89 & 2447865 & $-$    &  $-$
&\multicolumn{4}{c|}{------\gm$1.6$\gm9.7\gm------}      & 2.9  & 34.5 & $-$   
& $-$  & $-$    & $-$ \\          
 06.02.90 & 2447929 & 0.1    &  8.5 & $-$    &  $-$  & 1.4    & 35.3 & 2.7    & 
89   & $-$    & $-$  & 4.9    &  5.6 \\                              
 17.02.90 & 2447940 & $-$    &  $-$ & 1.3    &  8.8  & 1.6    & 35.8 & 2.9    & 
64.7 & $-$    & $-$  & 5.1    &  9.5 \\                              
 31.03.90 & 2447982 & $-$    &  $-$ & 0.6    & 11.0  & 1.6    & 30.7 & 2.8    & 
52.1 & $-$    & $-$  & 5.1    &  4.3 \\                              
 17.04.90 & 2447999 & 0.1    &  9.7 & $-$    &  $-$  & 1.5    & 39.4 & 2.7    & 
48.9 & $-$    & $-$  & $-$    &  $-$ \\                              
 11.05.90 & 2448023 & $-$    &  $-$ & 0.4    & 11.9  & 1.6    & 46.3 & 2.8    & 
47.3 & $-$    & $-$  & 4.5    &  2.3 \\                              
 15.07.90 & 2448088 & $-$    &  $-$ & 0.5    &  9.9  & 1.5    & 13.2 & 2.5    & 
28.7 & $-$    & $-$  & $-$    &  $-$ \\                              
 21.10.90 & 2448186 & 0.1    &  6.2 & $-$    &  $-$  & 1.8    & 35.4 & 2.8    & 
16.4 & $-$    & $-$  & 5.1    &  2.8 \\                              
 18.01.91 & 2448275 &\multicolumn{8}{c|}{\sevendash\gm$1.7$\gm27.7\gm\sevendash}
 & $-$    & $-$  & 5.1    & 8.1 \\                                   
 29.01.91 & 2448286 &\multicolumn{8}{c|}{\sevendash\gm$1.6$\gm25.1\gm\sevendash}
 & $-$    & $-$  & 4.9    & 5.1 \\                                   
 30.03.91 & 2448346 & $-$    &  $-$ & 1.0    &  9.2  & 1.7    & 27.1 & 2.7    & 
 6.9 & $-$    & $-$  & 5.0    &  1.8 \\                              
 01.05.91 & 2448378 & $-$    &  $-$ & 1.0    &  5.7  & 1.6    & 16.1 & 2.9    & 
 4.6 & $-$    & $-$  & 4.8    &  1.9 \\                              
 17.05.91 & 2448394 & $-$    &  $-$ & 1.0    & 13.1  & 1.9    &  9.7 & 2.7    & 
 8.7 & $-$    & $-$  & $-$    & $-$ \\                               
 25.10.91 & 2448555 &\multicolumn{8}{c|}{\sevendash\gm$1.7$\gm8.9\gm\sevendash} 
 & 4.6    & 4:   & $-$    & $-$ \\                                   
                                                                     
 18.01.92 & 2448640 & 0.0    &  4.2 & 0.8    &  5.2  & 2.1    & 13.7 & $-$   
&$-$    & 4.4    &  8.8 & $-$    & $-$ \\                               
 29.02.92 & 2448682 & $-0.1$ &  2.6 & 0.8    &  3.4  & 2.0    & 12.5 & $-$   
&$-$    & 4.3    &  4.0 & $-$    & $-$ \\                               
 18.04.92 & 2448731 & $-$    &  $-$
&\multicolumn{6}{c|}{\fivedash\gm$2.3$\gm9.9\gm\fivedash} & 4.2    & 7.9  & $-$ 
  & $-$ \\                      
 05.07.92 & 2448809 & $-$    &  $-$ & 0.6    &  7.0  & 2.1    &  6.4 & 3.2    & 
 5.7 & 4.0    &  6.1 & $-$    & $-$ \\                               
 01.09.92 & 2448867 & $-$    &  $-$ & 0.4    &  2.1  & 2.1    &  5.1 & $-$   
&$-$    & 3.9    &  1.3 & $-$    & $-$ \\                               
 15.10.92 & 2448911 & $-$    &  $-$
&\multicolumn{6}{c|}{\fivedash\gm$2.1$\gm4.2\gm\fivedash}  & $-$    & $-$  & $-$
   & $-$ \\                      
 24.10.92 & 2448920 & $-$    &  $-$
&\multicolumn{6}{c|}{\fivedash\gm$1.7$\gm28.8\gm\fivedash} & $-$    & $-$  & $-$
   & $-$ \\                      
 22.12.92 & 2448979 & 0.2    &  2.6 & $-$    &  $-$  & 1.9    &  5.2 & $-$   
&$-$    & 3.8    &  3.5 & $-$    & $-$ \\                               
 26.01.93 & 2449014 & 0.2    &  3.4 & $-$    &  $-$  & 1.9    &  5.8 & $-$   
&$-$    & 3.8    &  6.4 & $-$    & $-$ \\                               
 21.04.93 & 2449099 & 0.2    &  6.8 & $-$    &  $-$  & 2.1    &  7.0 & $-$   
&$-$    & 3.8    & 35.6 & $-$    & $-$ \\                               
 13.05.93 & 2449121 & 0.2    &  $<$2& $-$    &  $-$  & 2.1    & $<$2 & $-$   
&$-$    & 3.9    &  7.4 & $-$    & $-$ \\                               
 03.11.93 & 2449295 & $-0.5$ &  3.1
&\multicolumn{6}{c|}{\fivedash\gm$2.4$\gm5.6\gm\fivedash}  & $-$    & $-$  & $-$
   & $-$ \\                      
 30.11.93 & 2449322 & $-0.4$ &  3.0 & $-$    &  $-$  & 1.7    &  1.7 & 2.6    & 
 5.0 & $-$    & $-$  & $-$    & $-$ \\                               
 08.03.94 & 2449420 & $-0.4$ &  4.5 & 1.1    &  4.4  & $-$    &  $-$ & 2.6    & 
 9.4 & 4.0    &  2.1 & $-$    & $-$ \\                               
 14.04.94 & 2449457 & $-$    &  $-$ & 1.4    &  1.9  & $-$    &  $-$ & 2.5    & 
 4.4 & 4.0    & $<$3 & $-$    & $-$ \\                               
 08.09.94 & 2449604 & $-$    &  $-$ & $-$    &  $-$  & $-$    &  $-$ & 2.5    & 
13.4 & $-$    & $-$  & $-$    & $-$ \\                               
 28.10.94 & 2449654 & $-$    &  $-$ & 1.3    &  5.8  & 2.4    & 18.1 & 3.2    & 
 9.6 & $-$    & $-$  & $-$    & $-$ \\                               
 18.01.95 & 2449736 & $-$    &  $-$ & $-$    &  $-$  & 2      &$<$13 & 3.4    &
365.1 & $-$    & $-$  & $-$    & $-$ \\                               
 09.03.95 & 2449786 & $-0.3$ &  5.0 & $-$    &  $-$  & 2      &$<$30 & 3.4   
&1600.7 & $-$    & $-$  & $-$    & $-$ \\                               
 12.03.95 & 2449789 & $-$    &  $-$ & $-$    &  $-$  & $-$    &  $-$ & 3.4   
&1552.0 & $-$    & $-$  & $-$    & $-$ \\                               
 13.03.95 & 2449790 & $-0.2$ &  3.1 & $-$    &  $-$  & $-$    &  $-$ & 3.4   
&1100.7 & $-$    & $-$  & $-$    & $-$ \\                               
 27.03.95 & 2449804 & $-$    &  $-$ & $-$    &  $-$  & 2      &$<$10 & 3.3    &
429.3 & $-$    & $-$  & $-$    & $-$ \\                               
 09.04.95 & 2449817 & 0.0    &  $<$4& $-$    &  $-$  & 2      &$<$20 & 3.3    &
440.3 & $-$    & $-$  & $-$    & $-$ \\                               
 03.05.95 & 2449841 & 0.0    &  $<$5& $-$    &  $-$  & 2      &$<$20 & 3.3    &
220.9 & $-$    & $-$  & $-$    & $-$ \\                               
 03.06.95 & 2449872 & $-$    &  $-$ & 1.4    &  9.4  & $-$    & $-$  & 3.0    & 
88.9 & $-$    & $-$  & $-$    & $-$ \\                               
 04.06.95 & 2449873 & $-$    &  $-$ & 1.9    &  7.1  & $-$    & $-$  & 3.0    & 
63.5 & $-$    & $-$  & $-$    & $-$ \\                               
 05.06.95 & 2449874 & $-$    &  $-$ & 1.4    &  8.1  & $-$    & $-$  & 3.0    & 
70.0 & $-$    & $-$  & $-$    & $-$ \\                               
 23.06.95 & 2449892 & $-$    &  $-$ & 1.1    & 10.4  & $-$    & $-$  & 3.0    & 
69.3 & $-$    & $-$  & $-$    & $-$ \\                               
 18.09.95 & 2449982 & 0.4    & 23.3 & $-$    &  $-$  & 1.7    &  9.0 & 3.1    & 
23.8 & $-$    & $-$  & $-$    & $-$ \\                               
 18.01.96 & 2450101 & 0.6    & 10.5 & 1.5    &  7.1  & $-$    & $-$  & 3.0    & 
12.7 & $-$    & $-$  & $-$    & $-$ \\                               
 01.03.96 & 2450144 &\multicolumn{4}{c|}{\twodash\gm$0.9$\gm15.2\gm\twodash} &
$-$    & $-$  & 2.8   &13.8 & $-$    & $-$  & $-$    & $-$ \\          
 03.09.96 & 2450330 & 0.0    & 15.7 & 1.4    & 13.0  & $-$    & $-$  & 2.9    & 
16.4 & $-$    & $-$  & $-$    & $-$ \\                               
 14.03.97 & 2450522 & 0.0    &  4.4 & 1.1    & 94.6  & 2.3    &  6.5 & 3.2    & 
23.0 & $-$    & $-$  & $-$    & $-$ \\                               
 06.05.97 & 2450575 & $-$    &  $-$ & 1.3    & 88.5  & $-$    & $-$  & 3.2    & 
36.5 & $-$    & $-$  & $-$    & $-$ \\                               
 12.10.97 & 2450734 & $-$    &  $-$ & 0.9    & 20.6  & 2.1    &  6.5 & 3.4    & 
17.7 & $-$    & $-$  & $-$    & $-$ \\                               
 21.10.97 & 2450743 & $-$    &  $-$ & 1.0    & 14.1  & 2.1    & $<$6 & 3.4    & 
13.3 & $-$    & $-$  & $-$    & $-$ \\                               
 15.12.97 & 2450798 & \multicolumn{4}{c|}{\twodash\gm$0.5$\gm8.5\gm\twodash}   &
2.3    & 10.4 & 3.5    &17.2 & $-$    & $-$  & $-$    & $-$ \\       
 29.01.98 & 2450843 & \multicolumn{4}{c|}{\twodash\gm$0.2$\gm10.7\gm\twodash}  &
2.2    & 11.3 & 3.5    &11.8 & $-$    & $-$  & $-$    & $-$ \\       
 19.03.98 & 2450892 & $-$    &  $-$ & 0.9    &  5.7  & 2.3    &  5.8 & 3.5    & 
12.0 & $-$    & $-$  & $-$    & $-$ \\                               
 10.04.98 & 2450914 & $-$    &  $-$ & 0.9    & $<$4  & $-$    & $-$  & 3.3    & 
 9.2 & $-$    & $-$  & $-$    & $-$ \\                               
 11.05.98 & 2450945 & $-$    &  $-$ & 0.6    &  4.0  & 2.2    &  4.5 & 3.5    & 
14.9 & $-$    & $-$  & $-$    & $-$ \\                               
 02.02.99 & 2451212 & 0.1    &  8.6 & $-$    & $-$   & 1.5    &  3.3 & 3.2    & 
20.5 & $-$    & $-$  & 4.9    & 16.5\\                               
 24.06.99 & 2451354 & $-$    &  $-$ & $-$    & $-$   & $-$    & $-$  & 3.0    & 
51.4 & 4.1    & $<$7 & $-$    & $-$ \\                               
 27.10.00 & 2451845 & $-$    &  $-$
&\multicolumn{6}{c|}{\fivedash\gm$2.6$\gm\,8.8\gm\fivedash} & $-$    & $-$  &
$-$    & $-$ \\                      
 20.12.00 & 2451899 & $-$    &  $-$
&\multicolumn{6}{c|}{\fivedash\gm$2.6$\gm\,7.9\gm\fivedash}& $-$    & $-$  & $-$
   & $-$ \\                      
 07.05.01 & 2452037 & $-$    &  $-$
&\multicolumn{6}{c|}{\fivedash\gm$2.6$\gm10.7\gm\fivedash} & $-$    & $-$  & $-$
   & $-$ \\                      
 18.09.01 & 2452171 & $-$    &  $-$ & 1.2    &  5.8  & 2.2    & 17.4 & 3.0    & 
10.7 & $-$    & $-$  & $-$    & $-$ \\                               
 24.10.01 & 2452207 & $-$    &  $-$ & 1.0    &  4.9: & 2.2    & 27.1 & 3.3    & 
10.3 & $-$    & $-$  & $-$    & $-$ \\                               
 19.03.02 & 2452353 & $-$    &  $-$ & $-$    & $-$   & 2.3    & 26.7 & 3.6    & 
84.6 & $-$    & $-$  & $-$    & $-$ \\                               
 18.04.02 & 2452383 & $-$    &  $-$ & $-$    & $-$   & 2.2    & 31.5 & 3.5    & 
89.3 & $-$    & $-$  & $-$    & $-$ \\                               
 20.06.02 & 2452446 & $-$    &  $-$ & 1.5    &  6.2: & 2.3    & 12.2 & 3.5    & 
30.8 & $-$    & $-$  & $-$    & $-$ \\                               
 26.06.02 & 2452452 & $-$    &  $-$ & $-$    & $-$   & 2.4    & 11.8 & 3.5    & 
25.5 & $-$    & $-$  & $-$    & $-$ \\                               
 01.10.02 & 2452549 & $-$    &  $-$ & $-$    & $-$   & 2.7    & 24.3 & 3.5    & 
19.4 & $-$    & $-$  & $-$    & $-$ \\                               
 24.10.02 & 2452572 & $-$    &  $-$ & $-$    & $-$   & 2.8    & 14.3 & 3.5    & 
13.3 & $-$    & $-$  & $-$    & $-$ \\                               
 13.12.02 & 2452622 & $-$    &  $-$ & $-$    & $-$   & 3.0    & 22.8 & 3.7    & 
25.2 & $-$    & $-$  & $-$    & $-$ \\                               
 19.12.02 & 2452628 & $-$    &  $-$ & $-$    & $-$   & 3.0    & 15.8 & 3.6    & 
25.8 & $-$    & $-$  & $-$    & $-$ \\                               
 15.01.03 & 2452655 & $-$    &  $-$ & $-$    & $-$   & $-$    & $-$  & 3.4    & 
39.2 & $-$    & $-$  & $-$    & $-$ \\                               
 02.04.03 & 2452732 & $-$    &  $-$ & $-$    & $-$   & $-$    & $-$  & 3.3    & 
40.9 & $-$    & $-$  & $-$    & $-$ \\                               
 18.11.03 & 2452962 & $-$    &  $-$ & $-$    & $-$   & $-$    & $-$  & 3.4    &
111.8 & $-$    & $-$  & $-$    & $-$ \\                               
 28.01.04 & 2453033 & $-$    &  $-$ & $-$    & $-$   & 2.5    &  7.4 & 3.4    & 
17.6 & $-$    & $-$  & $-$    & $-$ \\                               
 31.03.04 & 2453096 & $-$    &  $-$ & $-$    & $-$   & 2.6    &  5.9 & 3.5    & 
13.8 & $-$    & $-$  & $-$    & $-$ \\                               
 11.05.04 & 2453137 & $-$    &  $-$ & $-$    & $-$   & 2.0    &$<$4.5& 3.5    & 
12.0 & $-$    & $-$  & $-$    & $-$ \\                               
 18.06.04 & 2453175 & $-$    &  $-$ & $-$    & $-$   & 1.9    &  4.2 & 3.6    & 
 9.6 & $-$    & $-$  & $-$    & $-$ \\                               
 17.09.04 & 2453266 & $-$    &  $-$ & $-$    & $-$   & 1.9    & 15.8 & 3.7    & 
11.8 & $-$    & $-$  & $-$    & $-$ \\                               
 18.12.04 & 2453358 & $-$    &  $-$ & $-$    & $-$   & 2.0    &  7.7 & 3.6    & 
 8.8 & $-$    & $-$  & $-$    & $-$ \\                               
 12.01.05 & 2453383 & $-$    &  $-$ & $-$    & $-$   & 1.8    & 11.7 & 3.4    & 
13.4 & $-$    & $-$  & $-$    & $-$ \\                               
 15.02.05 & 2453417 & $-$    &  $-$ & $-$    & $-$   & 1.6    & 14.3 & 3.3    & 
19.1 & $-$    & $-$  & $-$    & $-$ \\                               
 13.04.05 & 2453474 & $-$    &  $-$ & $-$    & $-$   & $-$    & $-$  & 3.1    & 
31.3 & 4.4    & 41.7 & $-$    & $-$ \\                               
 20.06.05 & 2453542 & $-$    &  $-$ & $-$    & $-$   & $-$    & $-$  & 3.2    & 
19.5 & 4.4    &  8.5 & $-$    & $-$ \\                               
 11.07.05 & 2453563 & $-$    &  $-$ & $-$    & $-$   & $-$    & $-$  & 3.3    & 
16.5 & 4.0    &  8.4 & $-$    & $-$ \\                               
 22.11.05 & 2453697 & $-$    &  $-$ & $-$    & $-$   &
\multicolumn{4}{c|}{\twodash\gm$2.6$\gm38.3\gm\twodash} & $-$  & $-$  & $-$    &
$-$ \\        
 14.02.06 & 2453781 & $-$    &  $-$ & $-$    & $-$   &
\multicolumn{4}{c|}{\twodash\gm$2.6$\gm86.3\gm\twodash} & 3.9  & 18.6 & $-$    &
$-$ \\        
 07.04.06 & 2453833 & $-$    &  $-$ & $-$    & $-$   & 2.6    & 34.0 & 3.1:   & 
14.3:& $-$    &  $-$ & $-$    & $-$ \\                               
 05.07.06 & 2453922 & $-$    &  $-$ & $-$    & $-$   & 2.5    & 18.1 & 3.4:   & 
10.6:& $-$    &  $-$ &$-$    & $-$ \\                                
 01.09.06 & 2453980 & $-$    &  $-$ & $-$    & $-$   &
\multicolumn{4}{c|}{\twodash\gm$2.7$\gm20.1\gm\twodash} &4.5  & 15.5 & $-$    &
$-$ \\         
 17.10.06 & 2454026 & $-$    &  $-$ & $-$    & $-$   & $-$    & $-$  & 2.9    & 
20.5 & 4.6    & 17.8 & $-$    & $-$ \\                               
 01.12.06 & 2454071 & $-$    &  $-$ & $-$    & $-$   & $-$    & $-$  & 2.9    & 
27.5 & 4.6    & 18.4 & $-$    & $-$ \\                               
 17.01.07 & 2454118 & $-$    &  $-$ & $-$    & $-$   & $-$    & $-$  & 2.6    & 
13.7 & 4.8    & 13.3 & $-$    & $-$ \\                               
\noalign{\smallskip}\hline                                                      
                                                                     
\end{longtable}

\newpage

\begin{longtable}{rrcrrr|rr|rr|rr}

\caption{\label{tab:compSVPegA-E} Maser components of SV~Peg. The columns
give the observing date in Gregorian and Julian (JD) format, the noise of the
spectra (rms), the integrated flux \SI\ in units of \powerten{-22} W
m$^{-2}$, and the velocities \vp\ and flux densities \Sp\ of maser peaks
associated with the spectral components A -- D as described in Section
\ref{SdDataSVPeg}.} \\
\hline\noalign{\smallskip}

 & & & & \multicolumn{2}{c}{A} & \multicolumn{2}{c}{B} & \multicolumn{2}{c}{C} &
\multicolumn{2}{c}{D} \\[0.1cm]                                                 

Date & JD & rms & \SI & \multicolumn{2}{c}{\vp\gm\Sp} &
\multicolumn{2}{c}{\vp\gm\Sp} & \multicolumn{2}{c}{\vp\gm\Sp} &
\multicolumn{2}{c}{\vp\gm\Sp} \\[0.1cm]                         

 & & [Jy] & & \multicolumn{8}{c}{\hrulefill\ [km\,s$^{-1}$, Jy] \hrulefill} \\

\noalign{\smallskip}\hline\noalign{\smallskip}

\endfirsthead

\caption{continued.} \\

\hline\noalign{\smallskip}

 & & & & \multicolumn{2}{c}{A} & \multicolumn{2}{c}{B} & \multicolumn{2}{c}{C} &
\multicolumn{2}{c}{D} \\[0.1cm]

Date & JD & rms & \SI & \multicolumn{2}{c}{\vp\gm\Sp} &
\multicolumn{2}{c}{\vp\gm\Sp} & \multicolumn{2}{c}{\vp\gm\Sp} &
\multicolumn{2}{c}{\vp\gm\Sp} \\[0.1cm]

 & & [Jy] & & \multicolumn{8}{c}{\hrulefill\ [km\,s$^{-1}$, Jy] \hrulefill} \\

\noalign{\smallskip}

\noalign{\smallskip}\hline\noalign{\smallskip}

\endhead

\hline

\endfoot

 17.02.90 & 2447940 & 0.21 & 142   & $-$    & $-$  & 2.2    &22.9  & $-$    &
$-$& $-$    & $-$  \\                          
 01.04.90 & 2447983 & 0.19 &  82   & $-$    & $-$  & 2.0    & 3.2  & $-$    &
$-$& $-$    & $-$  \\                          
 13.05.90 & 2448025 & 0.15 &  95   & $-$    & $-$  & 2.2    & 3.8  & $-$    &
$-$& $-$    & $-$  \\                          
 25.10.90 & 2448190 & 0.20 &  92   & 0.5    & 0.7  & 2.3    & 4.3  & $-$    &
$-$& 4.5    & 0.9  \\                          
 19.01.91 & 2448276 & 1.42 & $<$50 & $-$    & $-$  & 2.4    & 3.4  & $-$    &
$-$& $-$    & $-$  \\                          
 30.03.91 & 2448346 & 0.16 & 122   & 0.3    & 0.4  & 2.3    & 1.7  & $-$    &
$-$& 4.5    & 0.9  \\                          
 01.05.91 & 2448378 & 0.22 & 136   & 0.3    & 0.5  & 2.3    & 1.8  & $-$    &
$-$& 4.3    & 1.4  \\                          
 22.05.91 & 2448399 & 1.67 & 145   & $-$    & $-$  & $-$    & $-$  & $-$    &
$-$& $-$    & $-$  \\                          
 25.10.91 & 2448555 & 1.99 & 219   & $-$    & $-$  & 2.1    & 5.3  & $-$    &
$-$& 4.5    & 7.3  \\                          
 18.01.92 & 2448640 & 0.19 &  89   & 0.3    & 0.8  & 2.0    & 1.9  & $-$    &
$-$& 4.7    & 3.9  \\                          
 29.02.92 & 2448682 & 0.15 &  89   & 1.0    & 0.3: & 2.1    & 0.9  & $-$    &
$-$& 4.7    & 4.5  \\                          
 19.04.92 & 2448732 & 1.01 & 327   & $-$    & $-$  & $-$    & $-$  & $-$    &
$-$& 4.4    &18.6 \\                           
 02.09.92 & 2448868 & 0.20 & 102   & 0.5    & 1.2  & $-$    & $-$  & 3.6   
&12.2& $-$    & $-$  \\                          
 16.10.92 & 2448912 & 0.89 & 252   & 0.6    & 6.0  & $-$    & $-$  & 3.4   
&24.4& $-$    & $-$   \\                         
 22.12.92 & 2448979 & 0.20 & 180   & 0.3    &16.1  & 1.7    & 1.5  & 3.5    &
3.7& 4.9    & 0.7: \\                          
 25.01.93 & 2449013 & 0.48 & 101   & 0.4    &11.7  & 1.9    & 2.0:
&\multicolumn{4}{c}{\twodash\gm4.2\gm1.4\gm\twodash} \\   
 20.04.93 & 2449098 & 0.13 &  84   & 0.3    & 4.1  & 1.5    & 2.6  & 3.5    &
2.2& 4.8    & 2.2  \\                          
 02.11.93 & 2449294 & 0.65 &  74   & 0.7    & 1.7  & $-$    & $-$  & 3.3    &
2.4& 4.8    & 2.5   \\                         
 29.11.93 & 2449321 & 0.33 &  59  
&\multicolumn{4}{c}{\twodash\gm1.2\gm1.0\gm\twodash}   & 3.0    & 2.6  & 4.7   
&4.3  \\  
 17.01.94 & 2449370 & 0.47 &  64   & $-$    & $-$  & $-$    & $-$  & $-$    &
$-$& 4.5    & 6.0   \\                         
 08.03.94 & 2449420 & 0.22 &  30   & $-$    & $-$  & 1.6    & 0.7  & 3.3    &
0.9& 4.5    & 2.3   \\                         
 15.04.94 & 2449458 & 1.16 & $<$50 & $-$    & $-$  & $-$    & $-$  & $-$    &
$-$& $-$    & $-$   \\                         
 18.01.95 & 2449736 & 0.52 & $<$50 & $-$    & $-$  & 2.0    & 1.7: & $-$    &
$-$& $-$    & $-$   \\                         
 09.03.95 & 2449786 & 0.09 &  19   & $-$    & $-$  & 2.0    & 1.0 
&\multicolumn{4}{c}{\twodash\gm4.1\gm0.4\gm\twodash}  \\  
 24.06.95 & 2449893 & 0.23 &  17   & $-$    & $-$  & 1.9    & 0.7 
&\multicolumn{4}{c}{\twodash\gm4.0\gm1.0\gm\twodash}   \\ 
 27.10.00 & 2451845 & 1.15 & $<$50 & $-$    & $-$  & $-$    & $-$  & $-$    &
$-$& $-$    & $-$   \\                         
 20.12.00 & 2451899 & 0.94 & $<$50 & $-$    & $-$  & $-$    & $-$  & $-$    &
$-$& $-$    & $-$   \\                         
 07.05.01 & 2452037 & 1.09 & $<$50 & $-$    & $-$  & $-$    & $-$  & $-$    &
$-$& $-$    & $-$   \\                         
 18.09.01 & 2452171 & 0.87 & $<$50 & $-$    & $-$  & $-$    & $-$  & $-$    &
$-$& $-$    & $-$   \\                         
 24.10.01 & 2452207 & 1.21 & $<$50 & $-$    & $-$  & $-$    & $-$  & $-$    &
$-$& $-$    & $-$   \\                         
 20.03.02 & 2452354 & 1.42 & $<$50 & $-$    & $-$  & $-$    & $-$  & $-$    &
$-$& $-$    & $-$   \\                         
 24.04.02 & 2452385 & 1.24 &  65   & $-$    & $-$  & $-$    & $-$  & $-$    &
$-$& $-$    & $-$   \\                         
 27.06.02 & 2452453 & 2.87 & $<$50 & $-$    & $-$  & $-$    & $-$  & $-$    &
$-$& $-$    & $-$   \\                         
 01.10.02 & 2452549 & 0.94 &  75   & $-$    & $-$  & 2.8    & 1.8: & $-$    &
$-$& $-$    & $-$   \\                         
 29.10.02 & 2452577 & 1.62 & $<$50 & $-$    & $-$  & $-$    & $-$  & $-$    &
$-$& $-$    & $-$   \\                         
 19.12.02 & 2452628 & 1.11 & $<$50 & $-$    & $-$  & 3.2    & 1.7: & $-$    &
$-$& $-$    & $-$   \\                         
 14.01.03 & 2452654 & 1.05 & $<$50 & $-$    & $-$  & $-$    & $-$  & $-$    &
$-$& $-$    & $-$   \\                         
 03.04.03 & 2452733 & 1.47 & $<$50 & $-$    & $-$  & $-$    & $-$  & $-$    &
$-$& $-$    & $-$   \\                         
 19.11.03 & 2452963 & 1.37 & $<$50 & $-$    & $-$  & 3.1    & 2.1: & $-$    &
$-$& $-$    & $-$   \\                         
 28.01.04 & 2453033 & 0.76 & $<$50 & $-$    & $-$  & 2.9    & 1.5: & $-$    &
$-$& $-$    & $-$   \\                         
 02.04.04 & 2453099 & 0.80 & $<$50 & $-$    & $-$  & 2.3    & 1.9: & $-$    &
$-$& $-$    & $-$   \\                         
 14.05.04 & 2453140 & 0.99 & $<$50 & $-$    & $-$  & $-$    & $-$  & $-$    &
$-$& $-$    & $-$   \\                         
 20.06.04 & 2453177 & 2.29 & $<$50 & $-$    & $-$  & $-$    & $-$  & $-$    &
$-$& $-$    & $-$   \\                         
 17.09.04 & 2453266 & 1.88 & $<$50 & $-$    & $-$  & $-$    & $-$  & $-$    &
$-$& $-$    & $-$   \\                         
 20.12.04 & 2453360 & 0.77 &  54   & $-$    & $-$  & 2.8    & 2.0  & $-$    &
$-$& $-$    & $-$   \\                         
 11.01.05 & 2453382 & 0.90 & $<$50 & $-$    & $-$  & $-$    & $-$  & $-$    &
$-$& $-$    & $-$   \\                         
 12.03.05 & 2453442 & 1.08 &  56   & $-$    & $-$  & $-$    & $-$  & $-$    &
$-$& $-$    & $-$   \\                         
 15.04.05 & 2453476 & 1.02 & 104   & $-$    & $-$  & 2.9    & 3.0  & $-$    &
$-$& $-$    & $-$   \\                         
 21.06.05 & 2453543 & 1.94 & $<$50 & $-$    & $-$  & $-$    & $-$  & $-$    &
$-$& $-$    & $-$   \\                         
 12.07.05 & 2453564 & 1.46 &  81   & $-$    & $-$  & $-$    & $-$  & $-$    &
$-$& $-$    & $-$   \\                         
 22.11.05 & 2453697 & 1.33 & 175   & $-$    & $-$  & 2.6    & 6.2  & $-$    &
$-$& $-$    & $-$   \\                         
 15.02.06 & 2453782 & 1.61 & 193   & $-$    & $-$  & 2.4    & 4.7  & $-$    &
$-$& $-$    & $-$   \\                         
 09.04.06 & 2453835 & 2.57 & 179   & $-$    & $-$  & 2.5    & 6.0  & $-$    &
$-$& $-$    & $-$   \\                         
 07.07.06 & 2453924 & 2.49 & 216   & $-$    & $-$  & 2.1    & 5.3  & $-$    &
$-$& 6.1    & 4.6   \\                         
 01.09.06 & 2453980 & 1.00 & 402   & $-$    & $-$  & 2.2    & 6.6  & $-$    &
$-$& 5.7    & 8.2   \\                         
 17.10.06 & 2454026 & 0.71 & 518   & $-$    & $-$  & 2.1    &15.5  & $-$    &
$-$& 5.7    &13.7   \\                         
 30.11.06 & 2454070 & 1.47 & 260   & $-$    & $-$  & 2.0    &15.0  & $-$    &
$-$& 5.5:   & 3.4:  \\                         
 16.01.07 & 2454117 & 1.29 & 322   & $-$    & $-$  & 2.1    &35.8  & $-$    &
$-$& $-$    & $-$   \\                         
\noalign{\smallskip}\hline                                                      
                                            
\end{longtable}                                                                 

\begin{longtable}{rrrr|rr|rr|rr}

\caption{\label{tab:compSVPegF-I} Maser components of SV~Peg. The columns
give the observing date in Gregorian and Julian (JD) format, and the velocities
\vp\ and flux densities \Sp\ of maser peaks associated with the spectral
components E -- H as described in Section \ref{SdDataSVPeg}.} \\

\hline\noalign{\smallskip}

 & & \multicolumn{2}{c}{E} & \multicolumn{2}{c}{F} & \multicolumn{2}{c}{G} &
\multicolumn{2}{c}{H} \\[0.1cm]

Date & JD & \multicolumn{2}{c}{\vp\gm\Sp} & \multicolumn{2}{c}{\vp\gm\Sp} &
\multicolumn{2}{c}{\vp\gm\Sp} & \multicolumn{2}{c}{\vp\gm\Sp} \\[0.1cm]

 & & \multicolumn{08}{c}{\hrulefill\ [km\,s$^{-1}$, Jy] \hrulefill} \\

\noalign{\smallskip}\hline\noalign{\smallskip}

\endfirsthead

\caption{continued.} \\

\hline\noalign{\smallskip}

 & & \multicolumn{2}{c}{E} & \multicolumn{2}{c}{F} & \multicolumn{2}{c}{G} &
\multicolumn{2}{c}{H} \\[0.1cm]

Date & JD & \multicolumn{2}{c}{\vp\gm\Sp} & \multicolumn{2}{c}{\vp\gm\Sp}  &
\multicolumn{2}{c}{\vp\gm\Sp} & \multicolumn{2}{c}{\vp\gm\Sp} \\[0.1cm]

 & & \multicolumn{08}{c}{\hrulefill\ [km\,s$^{-1}$, Jy] \hrulefill} \\

\noalign{\smallskip}
       
\noalign{\smallskip}\hline\noalign{\smallskip}

\endhead

\hline

\endfoot

 17.02.90 & 2447940 &  7.7    & 0.6  & $-$    & $-$  & 12.9   & 0.9  & $-$    &
$-$  \\ 
 01.04.90 & 2447983 &  7.7    & 1.8  & $-$    & $-$  & 12.9   & 2.3  & $-$    &
$-$  \\ 
 13.05.90 & 2448025 &  $-$    & $-$  & $-$    & $-$  & $-$    & $-$  & $-$    &
$-$  \\ 
 25.10.90 & 2448190 &  7.7    & 2.0  & $-$    & $-$  & 13.4   & 0.4: & $-$    &
$-$  \\ 
 19.01.91 & 2448276 &  7.7    & 4.4  & $-$    & $-$  & $-$    & $-$  & $-$    &
$-$  \\ 
 30.03.91 & 2448346 &  7.4    & 1.6  & 11.4   & 12.7 & 13.4   & 0.9  & $-$    &
$-$  \\ 
 01.05.91 & 2448378 &  7.3    & 1.2  & 11.4   & 14.2 & 13.4   & 1.0  & $-$    &
$-$  \\ 
 22.05.91 & 2448399 &  $-$    & $-$  & 11.3   & 11.8 & $-$    & $-$  & $-$    &
$-$  \\ 
 25.10.91 & 2448555 &  $-$    & $-$  & 11.2   &  5.7 & 13.7   & 2.6  & $-$    &
$-$  \\ 
 18.01.92 & 2448640 &  $-$    & $-$  & 11.3   &  0.5 & 13.7   & 0.5  & 16.4   &
0.4  \\ 
 29.02.92 & 2448682 &  $-$    & $-$  & $-$    & $-$  & 13.8   & 0.5  & 16.6   &
0.4  \\ 
 19.04.92 & 2448732 &  $-$    & $-$  & $-$    & $-$  & $-$    & $-$  & $-$    &
$-$  \\ 
 02.09.92 & 2448868 &  7.5    & 1.2  & $-$    & $-$  & $-$    & $-$  & $-$    &
$-$  \\ 
 16.10.92 & 2448912 &  7.5    & 7.2  & $-$    & $-$  & $-$    & $-$  & $-$    &
$-$  \\ 
 22.12.92 & 2448979 &  7.3    & 0.4  & $-$    & $-$  & $-$    & $-$  & $-$    &
$-$  \\ 
 25.01.93 & 2449013 &  $-$    & $-$  & $-$    & $-$  & $-$    & $-$  & $-$    &
$-$  \\ 
 20.04.93 & 2449098 &  $-$    & $-$  & $-$    & $-$  & $-$    & $-$  & $-$    &
$-$  \\ 
 02.11.93 & 2449294 &  $-$    & $-$  & $-$    & $-$  & $-$    & $-$  & $-$    &
$-$  \\ 
 29.11.93 & 2449321 &  $-$    & $-$  & $-$    & $-$  & $-$    & $-$  & $-$    &
$-$  \\ 
 17.01.94 & 2449370 &  $-$    & $-$  & $-$    & $-$  & $-$    & $-$  & $-$    &
$-$  \\ 
 08.03.94 & 2449420 &  $-$    & $-$  & $-$    & $-$  & $-$    & $-$  & $-$    &
$-$  \\ 
 15.04.94 & 2449458 &  $-$    & $-$  & $-$    & $-$  & $-$    & $-$  & $-$    &
$-$  \\ 
 18.01.95 & 2449736 &  $-$    & $-$  & $-$    & $-$  & $-$    & $-$  & $-$    &
$-$  \\ 
 09.03.95 & 2449786 &  $-$    & $-$  & $-$    & $-$  & $-$    & $-$  & $-$    &
$-$  \\ 
 24.06.95 & 2449893 &  $-$    & $-$  & $-$    & $-$  & $-$    & $-$  & $-$    &
$-$  \\ 
 27.10.00 & 2451845 &  $-$    & $-$  & $-$    & $-$  & $-$    & $-$  & $-$    &
$-$  \\ 
 20.12.00 & 2451899 &  $-$    & $-$  & $-$    & $-$  & $-$    & $-$  & $-$    &
$-$  \\ 
 07.05.01 & 2452037 &  $-$    & $-$  & $-$    & $-$  & $-$    & $-$  & $-$    &
$-$  \\ 
 18.09.01 & 2452171 &  $-$    & $-$  & $-$    & $-$  & $-$    & $-$  & $-$    &
$-$  \\ 
 24.10.01 & 2452207 &  $-$    & $-$  & $-$    & $-$  & $-$    & $-$  & $-$    &
$-$  \\ 
 20.03.02 & 2452354 &  $-$    & $-$  & $-$    & $-$  & $-$    & $-$  & $-$    &
$-$  \\ 
 24.04.02 & 2452385 &  8.7    & 3.7  & $-$    & $-$  & $-$    & $-$  & $-$    &
$-$  \\ 
 27.06.02 & 2452453 &  $-$    & $-$  & $-$    & $-$  & $-$    & $-$  & $-$    &
$-$  \\ 
 01.10.02 & 2452549 &  7.9    & 1.3  & $-$    & $-$  & $-$    & $-$  & $-$    &
$-$  \\ 
 29.10.02 & 2452577 &  $-$    & $-$  & $-$    & $-$  & $-$    & $-$  & $-$    &
$-$  \\ 
 19.12.02 & 2452628 &  8.7    & 1.8: & $-$    & $-$  & $-$    & $-$  & $-$    &
$-$  \\ 
 14.01.03 & 2452654 &  $-$    & $-$  & $-$    & $-$  & $-$    & $-$  & $-$    &
$-$  \\ 
 03.04.03 & 2452733 &  $-$    & $-$  & $-$    & $-$  & $-$    & $-$  & $-$    &
$-$  \\ 
 19.11.03 & 2452963 &  $-$    & $-$  & $-$    & $-$  & $-$    & $-$  & $-$    &
$-$  \\ 
 28.01.04 & 2453033 &  $-$    & $-$  & $-$    & $-$  & $-$    & $-$  & $-$    &
$-$  \\ 
 02.04.04 & 2453099 &  $-$    & $-$  & $-$    & $-$  & $-$    & $-$  & $-$    &
$-$  \\ 
 14.05.04 & 2453140 &  $-$    & $-$  & $-$    & $-$  & $-$    & $-$  & $-$    &
$-$  \\ 
 20.06.04 & 2453177 &  $-$    & $-$  & $-$    & $-$  & $-$    & $-$  & $-$    &
$-$  \\ 
 17.09.04 & 2453266 &  $-$    & $-$  & $-$    & $-$  & $-$    & $-$  & $-$    &
$-$  \\ 
 20.12.04 & 2453360 &  $-$    & $-$  & $-$    & $-$  & $-$    & $-$  & $-$    &
$-$  \\ 
 11.01.05 & 2453382 &  $-$    & $-$  & $-$    & $-$  & $-$    & $-$  & $-$    &
$-$  \\ 
 12.03.05 & 2453442 &  $-$    & $-$  & $-$    & $-$  & $-$    & $-$  & $-$    &
$-$  \\ 
 15.04.05 & 2453476 &  $-$    & $-$  & $-$    & $-$  & $-$    & $-$  & $-$    &
$-$  \\ 
 21.06.05 & 2453543 &  $-$    & $-$  & $-$    & $-$  & $-$    & $-$  & $-$    &
$-$  \\ 
 12.07.05 & 2453564 &  $-$    & $-$  & $-$    & $-$  & $-$    & $-$  & $-$    &
$-$  \\ 
 22.11.05 & 2453697 &  8.0:   & 3.0: & $-$    & $-$  & $-$    & $-$  & $-$    &
$-$  \\ 
 15.02.06 & 2453782 &  7.1    & 5.1  & $-$    & $-$  & $-$    & $-$  & $-$    &
$-$  \\ 
 09.04.06 & 2453835 &  $-$    & $-$  & $-$    & $-$  & $-$    & $-$  & $-$    &
$-$  \\ 
 07.07.06 & 2453924 &  8.3    & 7.0: & $-$    & $-$  & $-$    & $-$  & $-$    &
$-$  \\ 
 01.09.06 & 2453980 &  8.8    & 8.0  & $-$    & $-$  & $-$    & $-$  & $-$    &
$-$  \\ 
 17.10.06 & 2454026 &  8.5    &16.7  & $-$    & $-$  & $-$    & $-$  & $-$    &
$-$  \\ 
 30.11.06 & 2454070 &  8.6    & 4.1  & $-$    & $-$  & $-$    & $-$  & $-$    &
$-$  \\ 
 16.01.07 & 2454117 &  $-$    & $-$  & $-$    & $-$  & $-$    & $-$  & $-$    &
$-$  \\ 
\noalign{\smallskip}\hline                                                      
       
\end{longtable}

\begin{figure*}[tp]
\centering
\resizebox{16cm}{!}{\includegraphics{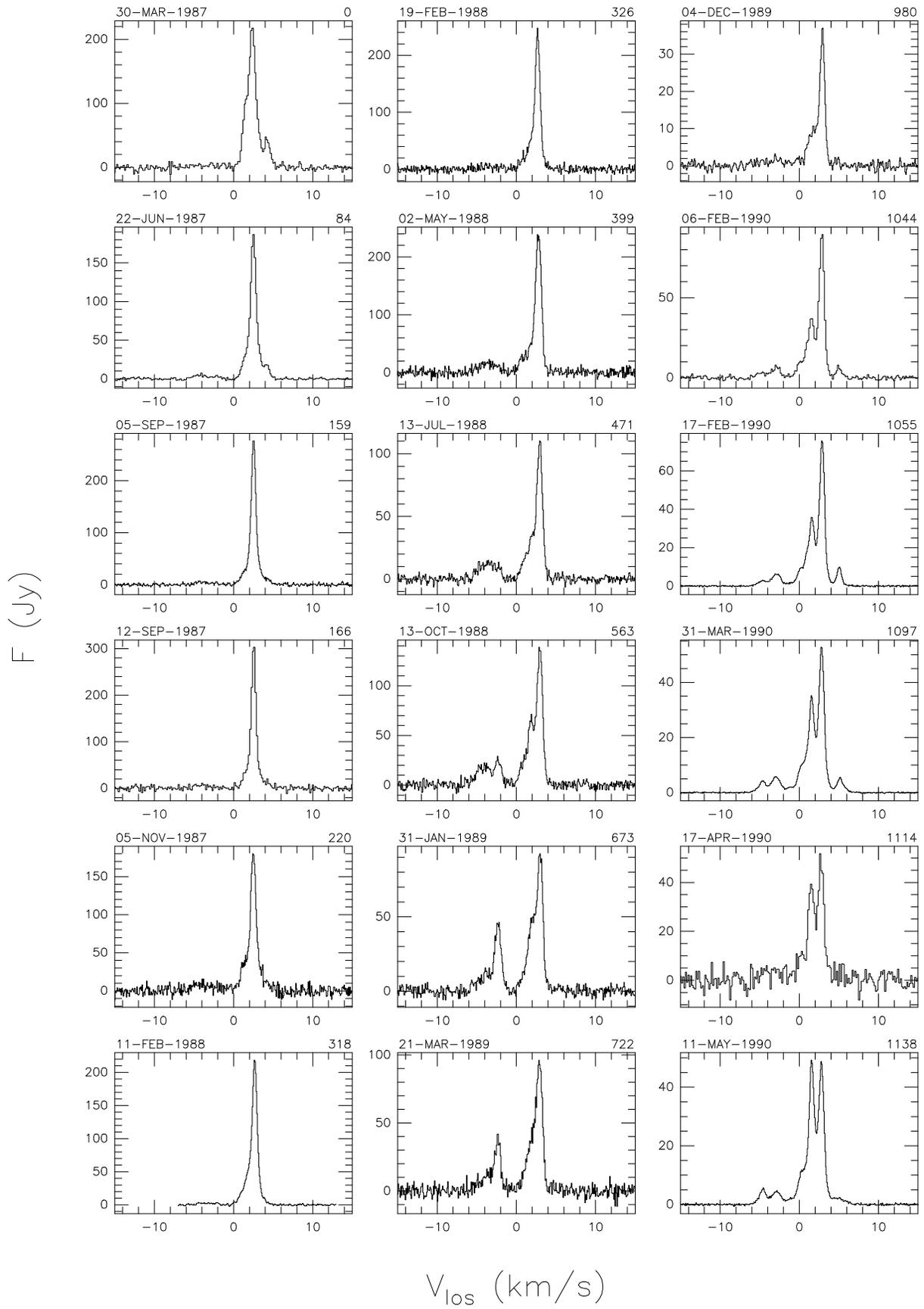}}
\caption{\water\ maser spectra of RX~Boo.  Note the varying flux density scale.}
\label{fig:spectra_rxboo_1}
\end{figure*}

\begin{figure*}[tp]
\centering
\resizebox{16cm}{!}{\includegraphics{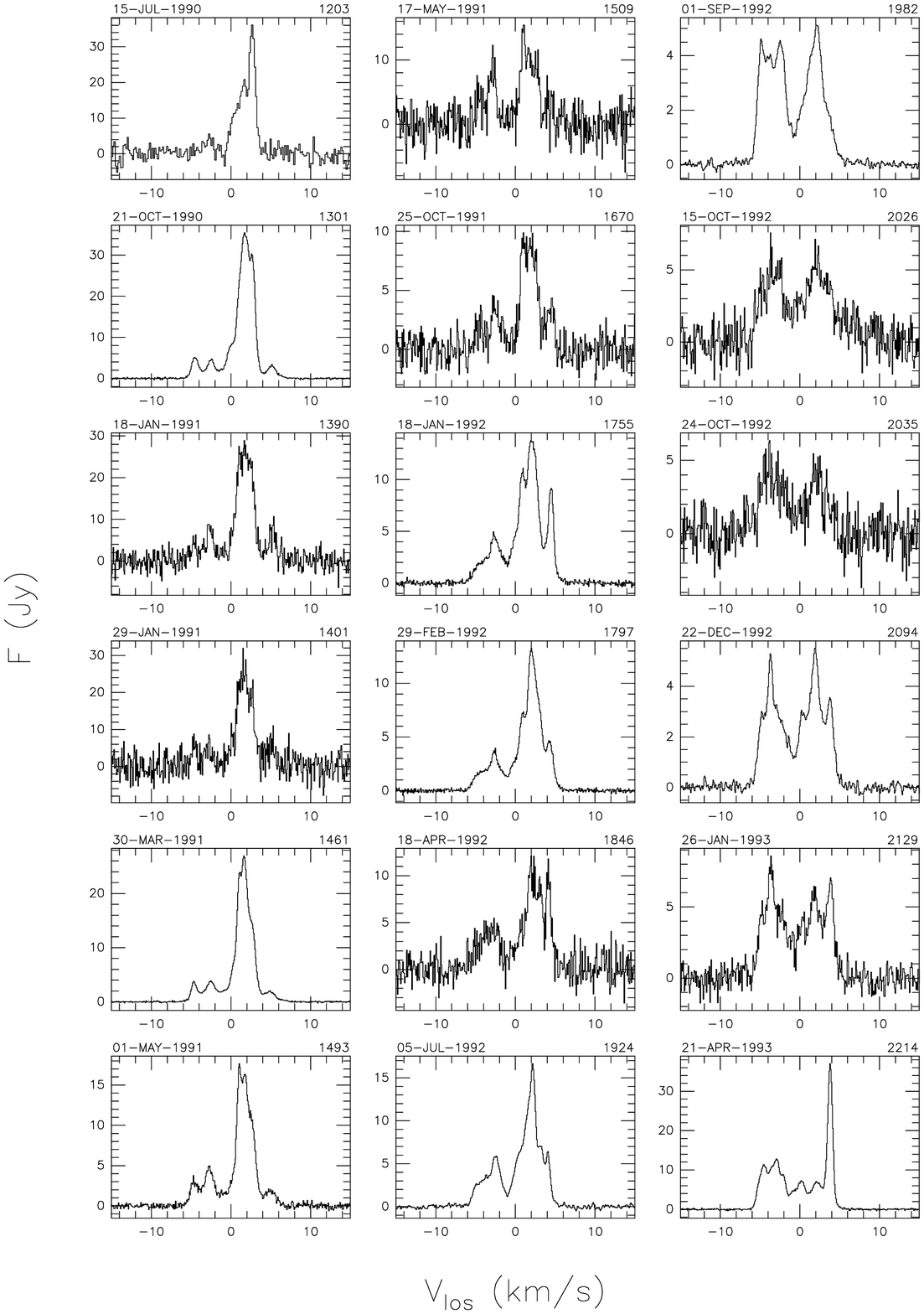}}
\caption{\water\ maser spectra of RX~Boo (continued).  Note the varying flux
density scale.}
\label{fig:spectra_rxboo_2}
\end{figure*}

\begin{figure*}[tp]
\centering
\resizebox{16cm}{!}{\includegraphics{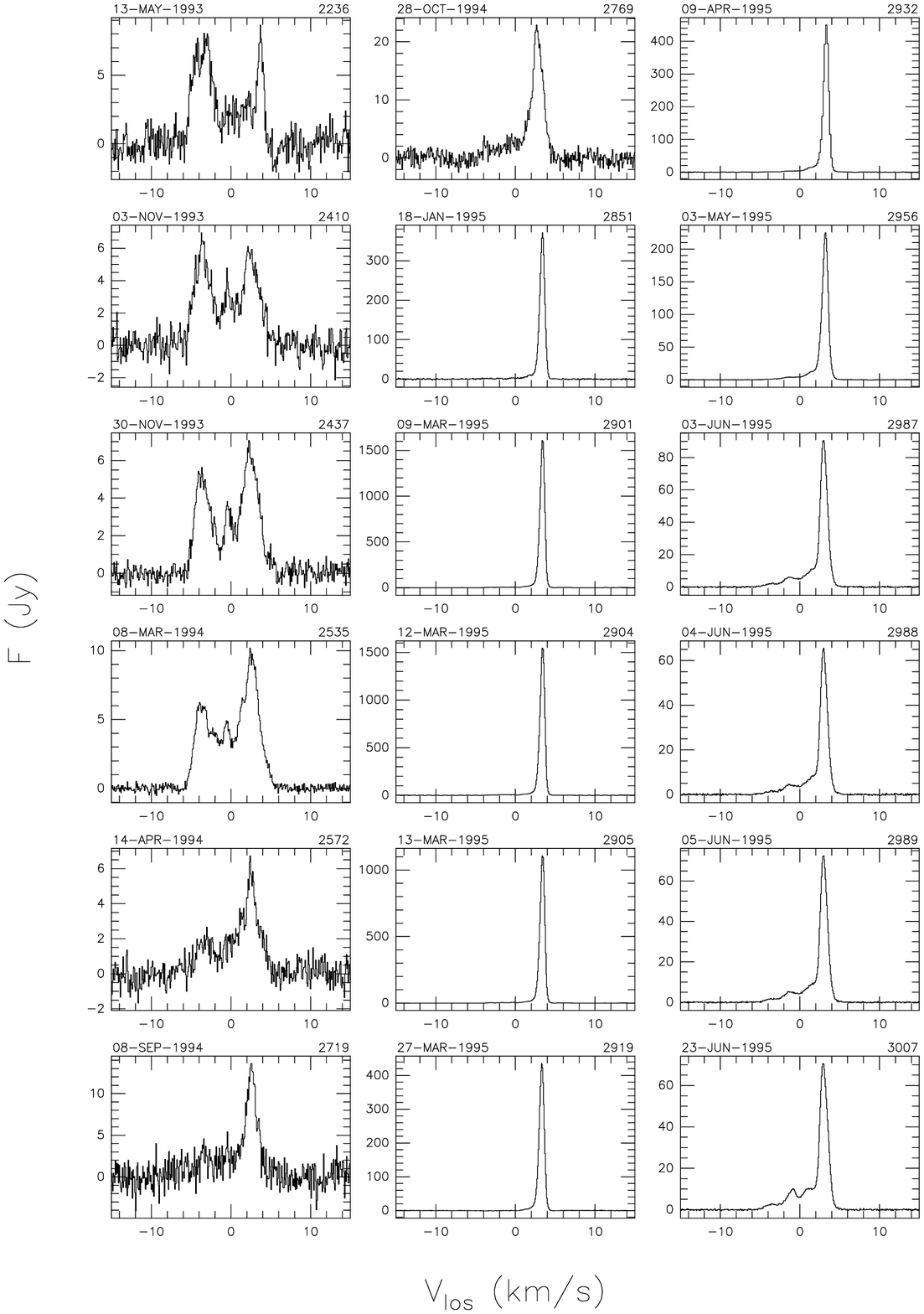}}
\caption{\water\ maser spectra of RX~Boo (continued).  Note the varying flux
density scale.}
\label{fig:spectra_rxboo_3}
\end{figure*}

\begin{figure*}[tp]
\centering
\resizebox{16cm}{!}{\includegraphics{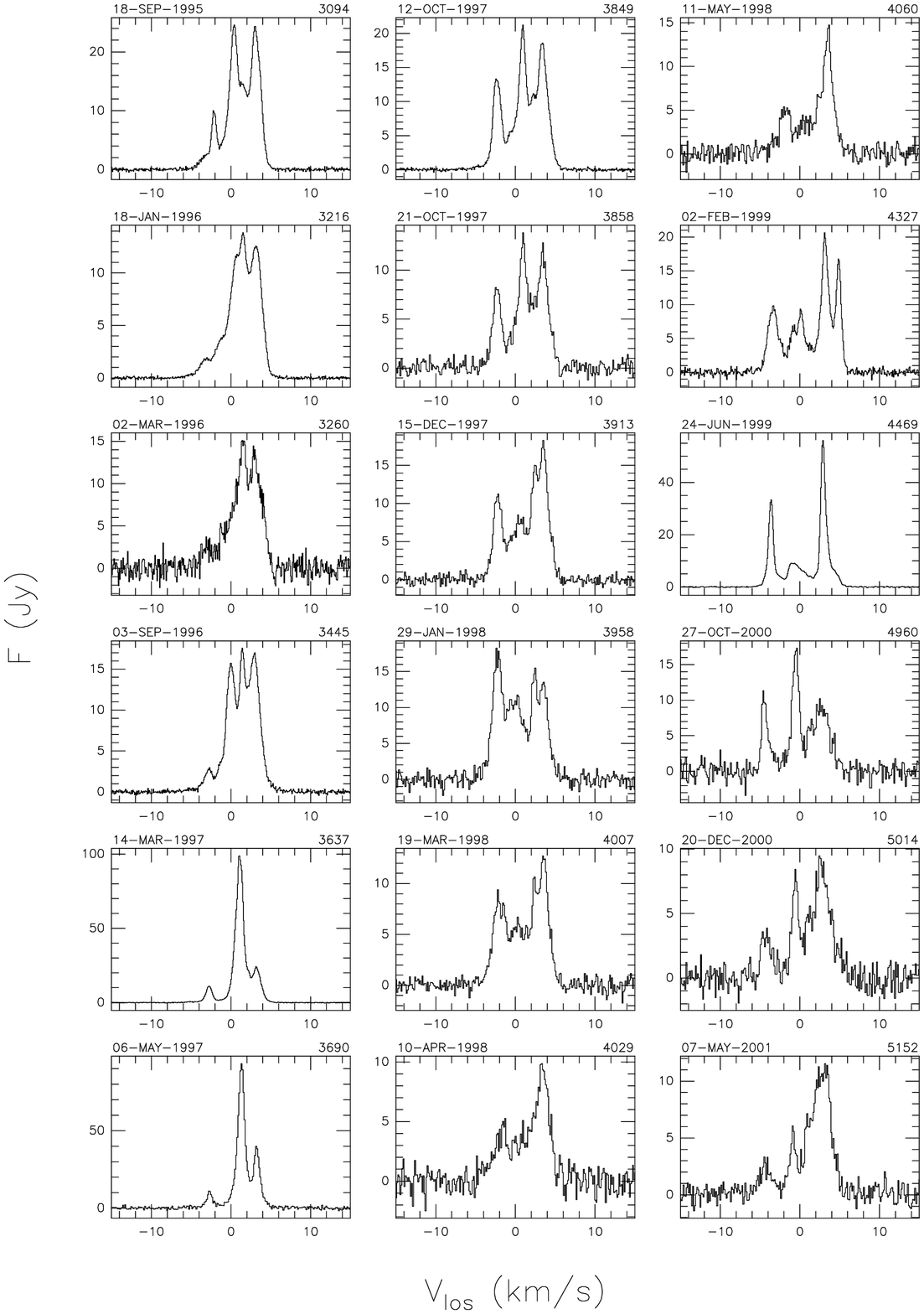}}
\caption{\water\ maser spectra of RX~Boo (continued).  Note the varying flux
density scale.}
\label{fig:spectra_rxboo_4}
\end{figure*}

\begin{figure*}[tp]
\centering
\resizebox{16cm}{!}{\includegraphics{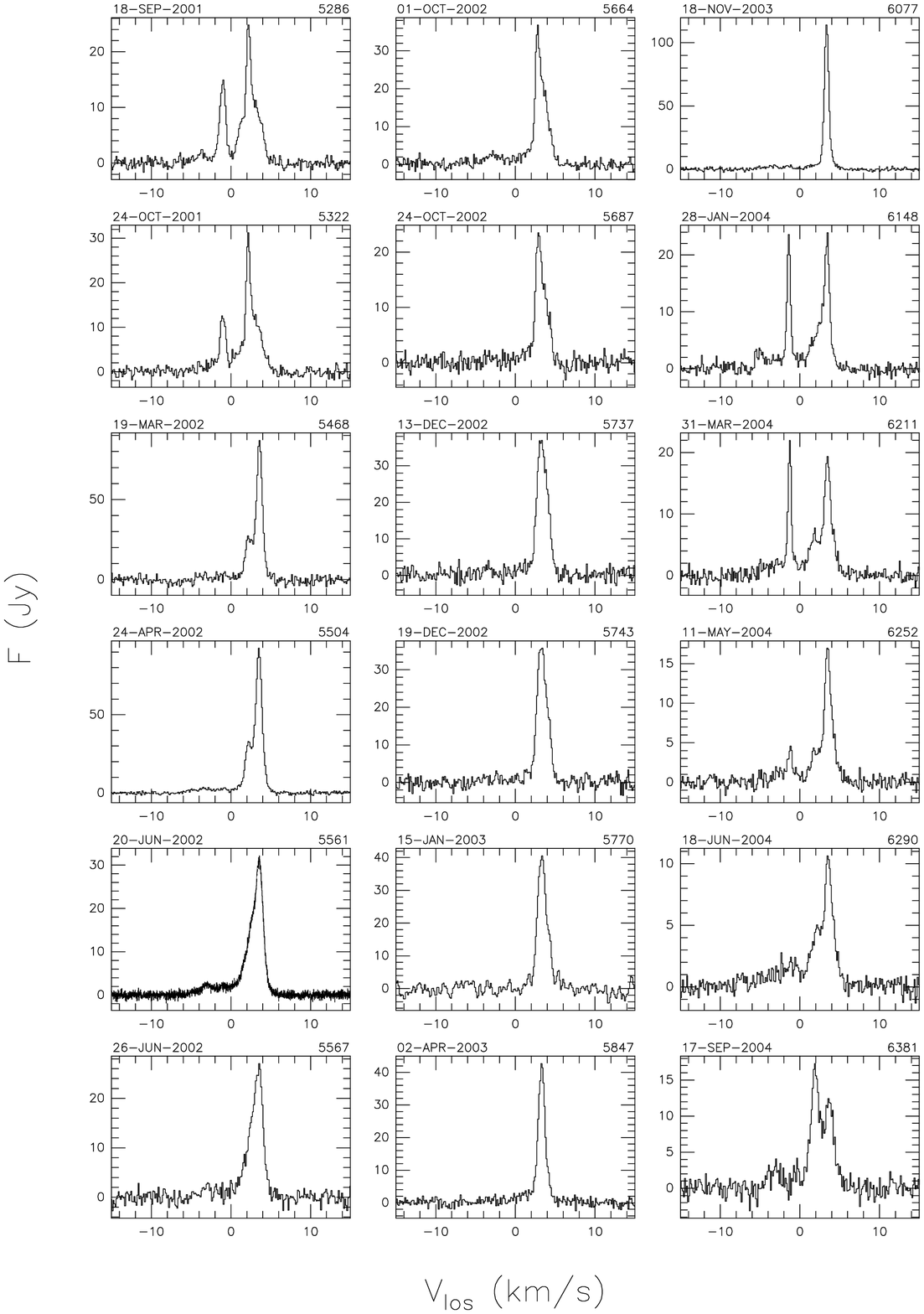}}
\caption{\water\ maser spectra of RX~Boo (continued).  Note the varying flux
density scale.}
\label{fig:spectra_rxboo_5}
\end{figure*}

\begin{figure*}[tp]
\centering
\resizebox{16cm}{!}{\includegraphics{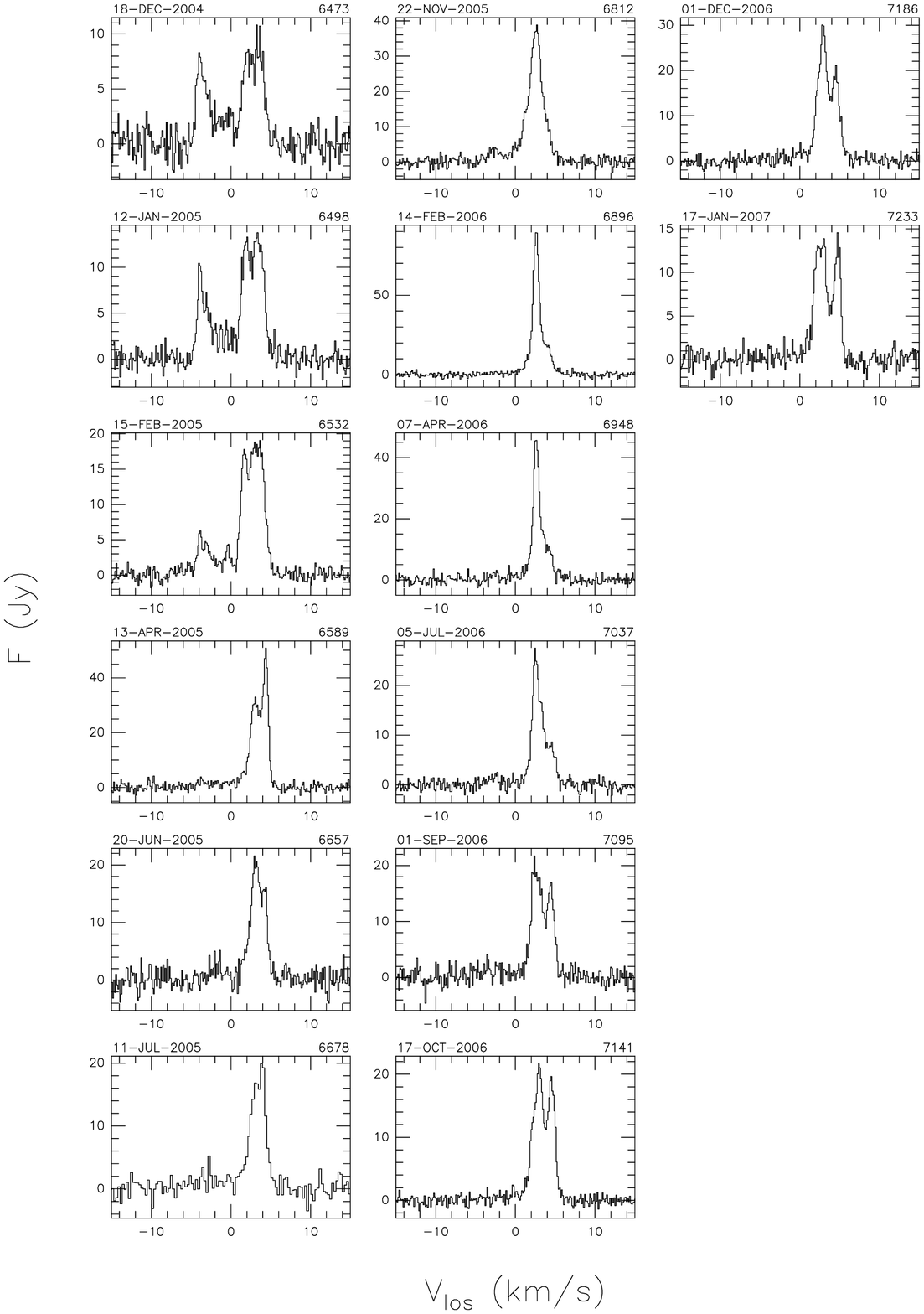}}
\caption{\water\ maser spectra of RX~Boo (continued).  Note the varying flux
density scale.}
\label{fig:spectra_rxboo_6}
\end{figure*}

\begin{figure*}[tp]
\centering
\resizebox{16cm}{!}{\includegraphics{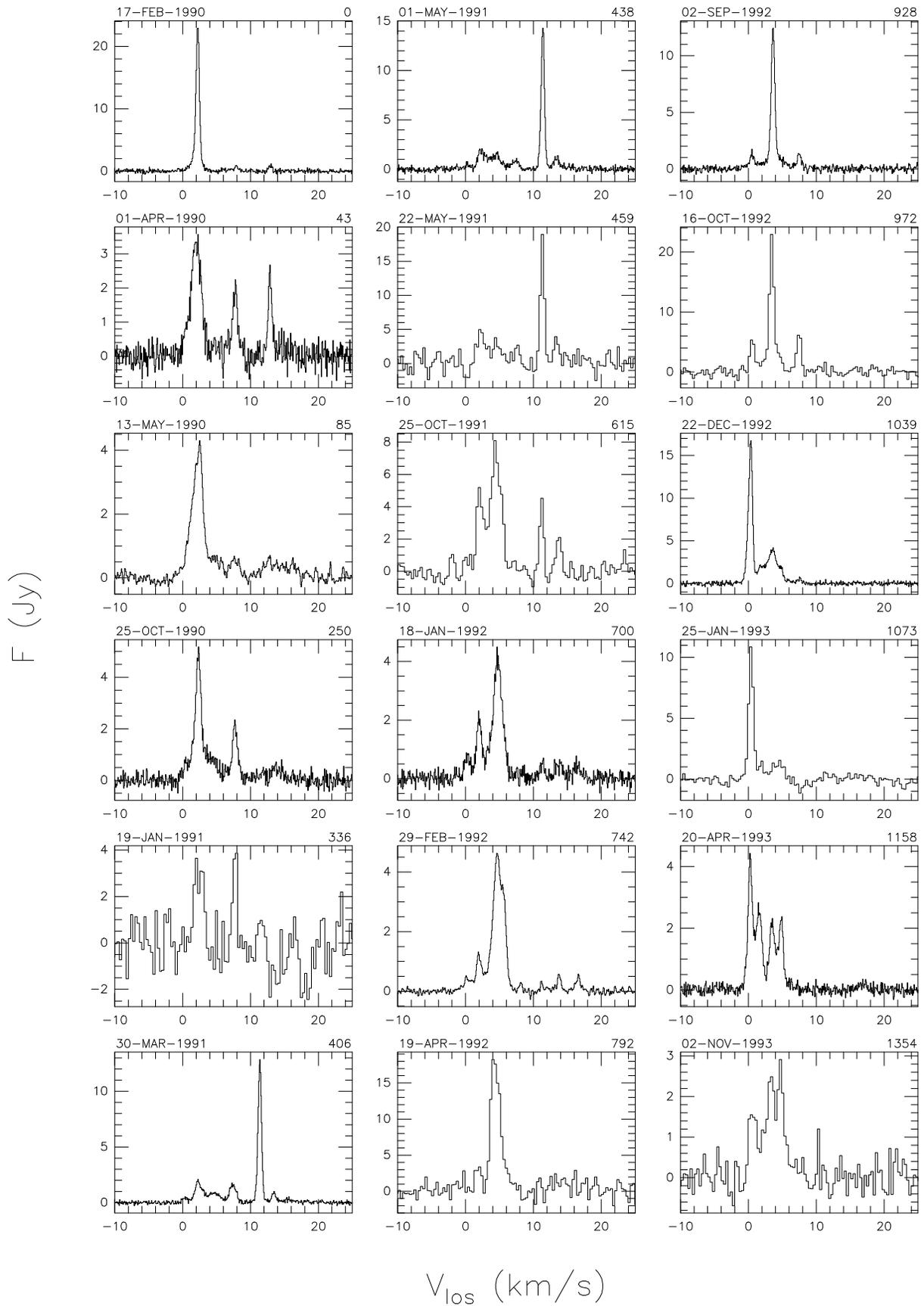}}
\caption{\water\ maser spectra of SV~Peg.  Note the varying flux density scale.}
\label{fig:spectra_svpeg_1}
\end{figure*}

\begin{figure*}[tp]
\centering
\resizebox{16cm}{!}{\includegraphics{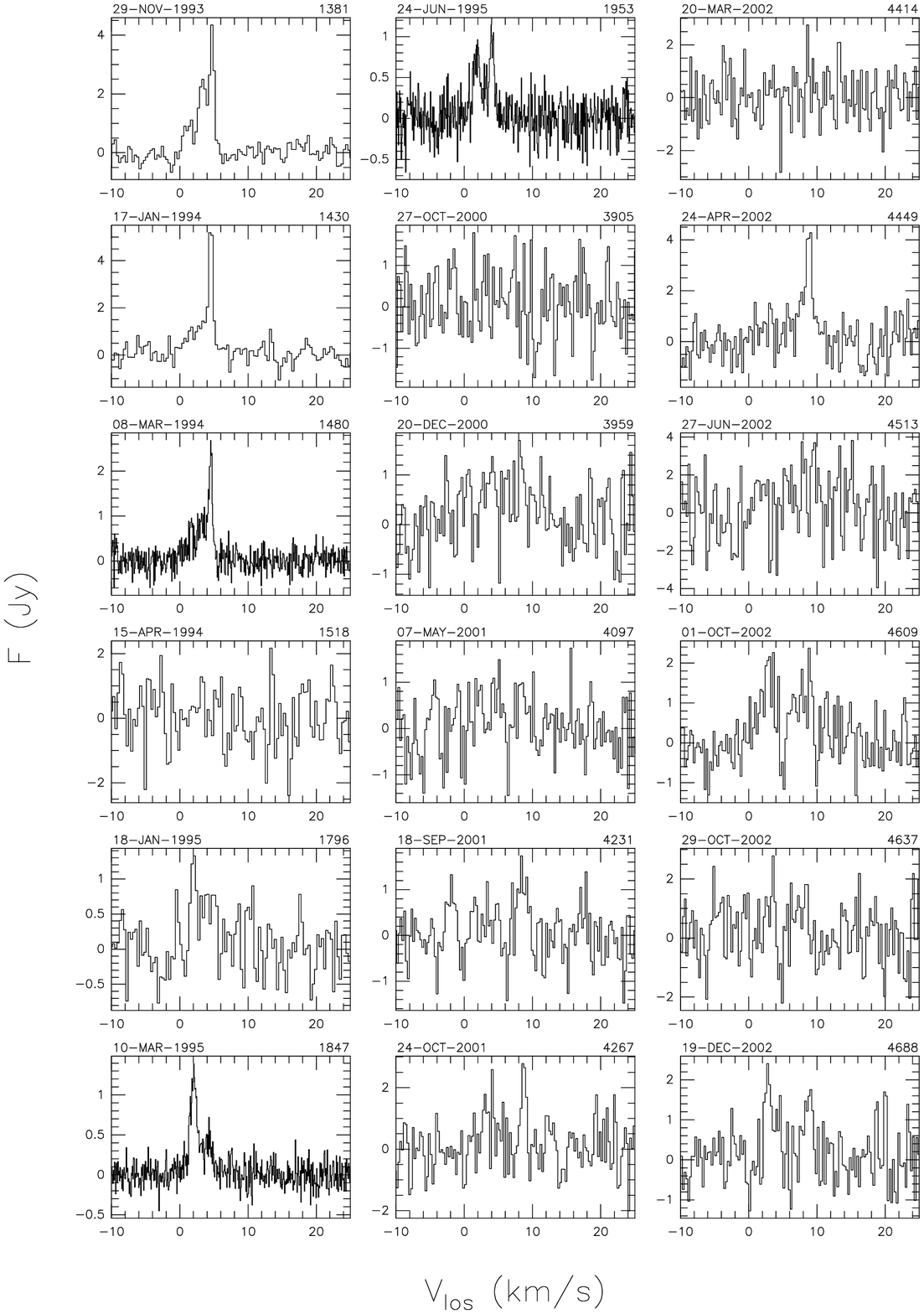}}
\caption{\water\ maser spectra of SV~Peg (continued).  Note the varying flux
density scale.}
\label{fig:spectra_svpeg_2}
\end{figure*}

\begin{figure*}[tp]
\centering
\resizebox{16cm}{!}{\includegraphics{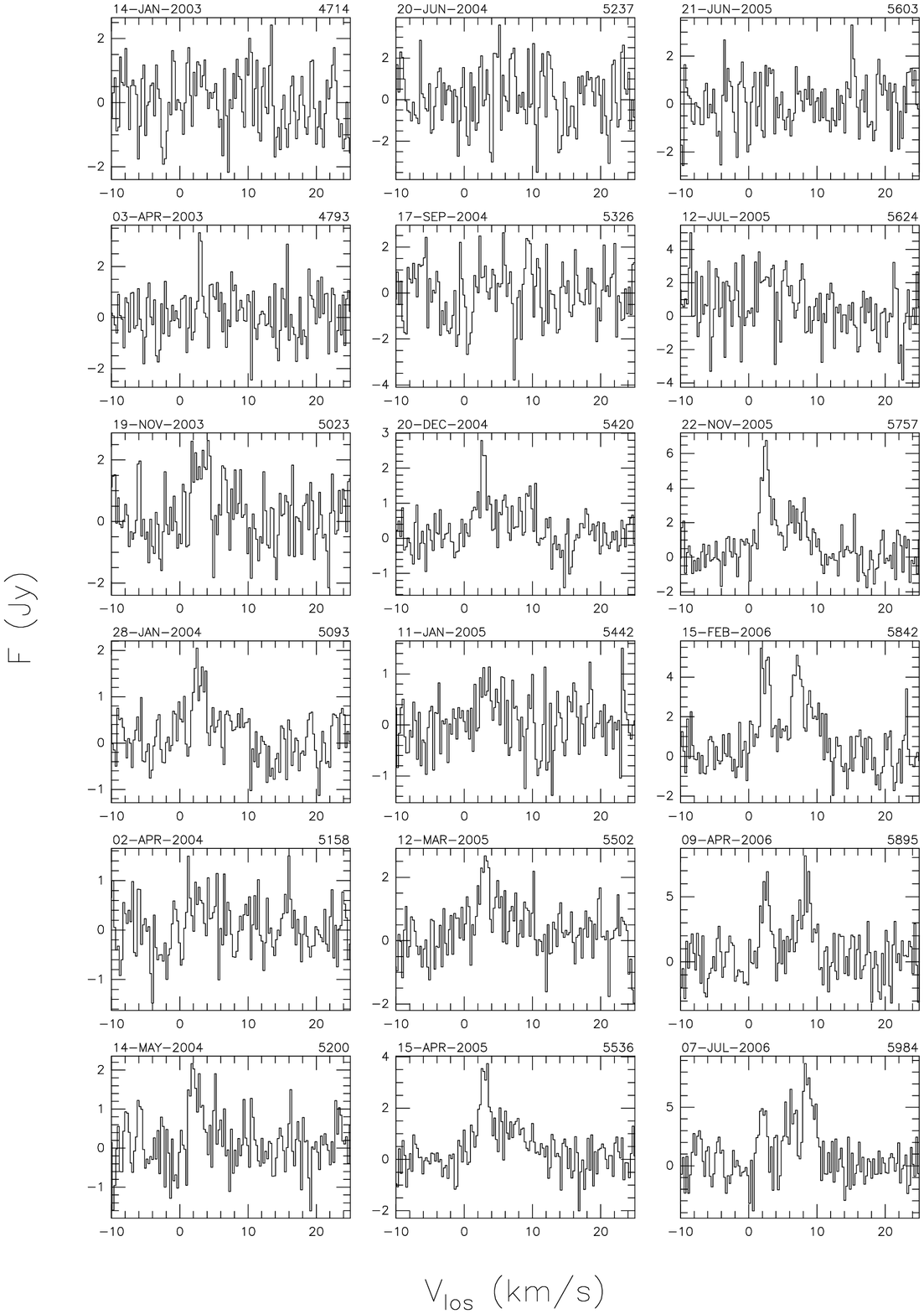}}
\caption{\water\ maser spectra of SV~Peg (continued).  Note the varying flux
density scale.}
\label{fig:spectra_svpeg_3}
\end{figure*}

\begin{figure}[tp]
\centering
\resizebox{6.03cm}{!}{\includegraphics{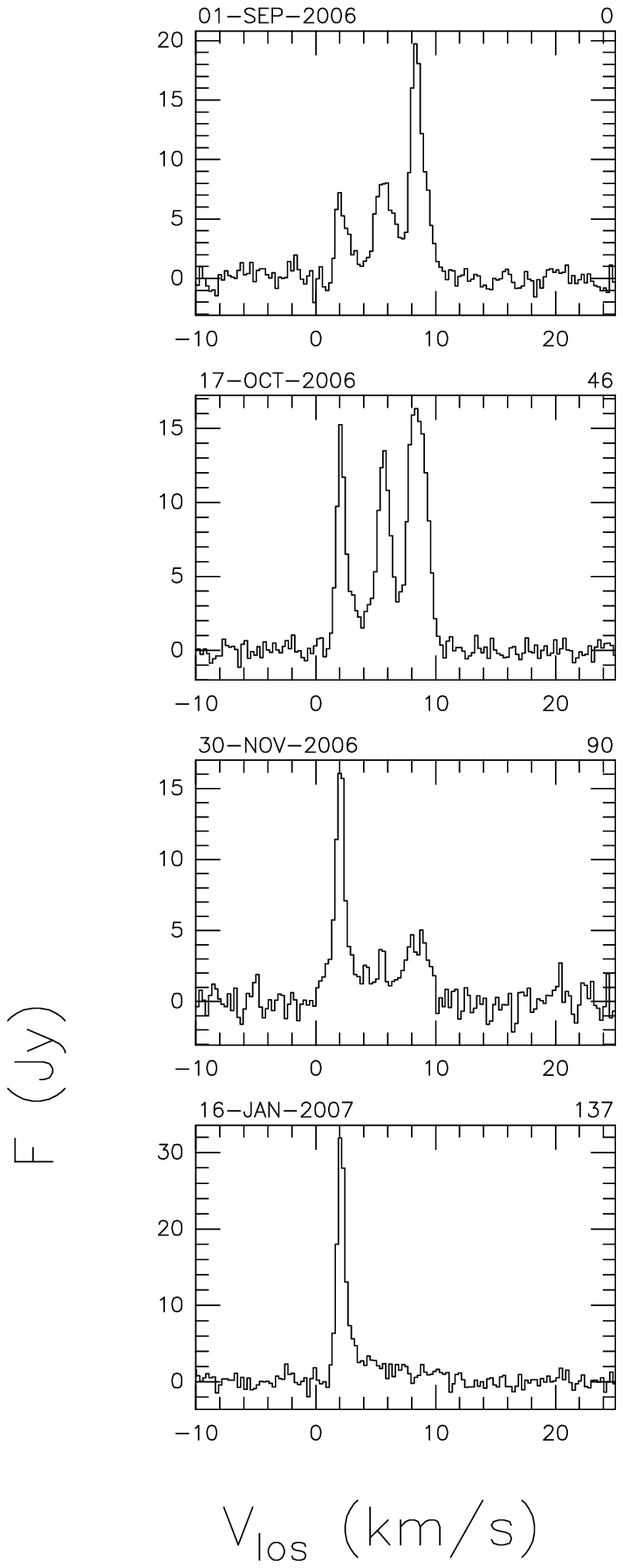}}
\caption{\water\ maser spectra of SV~Peg (continued).  Note the varying flux
density scale.}
\label{fig:spectra_svpeg_4}
\end{figure}

\end{document}